\documentclass[prc,preprint,eqsecnum,showpacs,showkeys,nofootinbib,amsmath,amssymb,tightenlines,dvips]{revtex4}
\usepackage{epsfig}
\usepackage{amssymb,latexsym,amsmath}
\newcommand{\eq}[1]{\begin{align} #1 \end{align}}

\newcommand{\p}{\partial}
\newcommand{\abs}[1]{\left\vert#1\right\vert}
\newcommand{\mc}[1]{\mathcal{#1}}
\newcommand{\comment}[1]{}

\begin{document}
\title{Analytical formulas, general properties and calculation of transport coefficients in the hadron gas:
shear and bulk viscosities.}

\author{Oleg Moroz}
\affiliation{Bogolyubov Institute for Theoretical Physics, Kiev,
 Ukraine}

\email{ moroz@mail.bitp.kiev.ua }

\date {\today}

\begin{abstract}
Elaborated calculations of the shear and the bulk viscosities in the hadron gas, using the ultrarelativistic
quantum molecular dynamics (UrQMD) model cross sections, are made. These cross sections are analyzed and
improved. A special treatment of the resonances is implemented additionally. All this allows for better
hydrodynamical description of the experimental data. The previously considered approximation of one constant
cross section for all hadrons is justified. It's found that the bulk viscosity of the hadron gas is much larger
than the bulk viscosity of the pion gas while the shear viscosity is found to be less sensitive to the hadronic
mass spectrum. The maximum of the bulk viscosity of the hadron gas is expected to be approximately in the
temperature range ${T=150-190~MeV}$ with zero chemical potentials. This range covers the critical temperature
values found from lattice calculations. We comment on some important aspects of calculations of the bulk
viscosity, which were not taken into account or were not analyzed well previously. Doing this, a generalized
Chapman-Enskog procedure, taking into account deviations from the chemical equilibrium, is outlined. Some general
properties, features, the physical meaning of the bulk viscosity and some other comments on the deviations from
the chemical equilibrium supplement this discussion. Analytical closed-form expressions for the transport
coefficients and some related quantities within a quite large class of cross sections can be obtained. Some
examples are explicitly considered. Comparisons with some previous calculations of the viscosities in the hadron
gas and the pion gas are done.
\end{abstract}



\maketitle

\section{ Introduction }
The bulk and the shear viscosity coefficients are transport
coefficients which enter in the hydrodynamic equations, and thus
are important for studying of nonequilibrium evolution of any
thermodynamic system.

There are two more additional reasons to study the shear viscosity. The first one is the experimentally observed
minimum of the ratio of the shear viscosity to the entropy density $\eta/s$ near the liquid-gas phase transition
for different substances, which may help in studying of the quantum chromodynamics phase diagram and finding of
the location of the critical point \cite{Csernai:2006zz,lacey}\footnote{Fireballs, created in heavy ion
collisions, have finite sizes and finite times of existence of their thermalized part. This puts important
restrictions on detection of the critical fluctuations of thermodynamic functions \cite{Stephanov:1999zu}.
Because of this it's also important to consider nonequilibrium dissipative corrections and nonequilibrium
phenomenons like critical slow down/speed up.}. Such a minimum was observed in theoretical results in several
models, see e. g. \cite{chakkap,Dobado:2009ek}. For a counterexample see \cite{Chen:2010vf} and references
therein. The second reason is the calculation of the $\eta/s$ in strongly interacting systems, preferably real
ones, to compare physical inputs which provide small values of the $\eta/s$. The conjectured lowest
bound\footnote{In \cite{Danielewicz:1984ww} the bound coming from the Heisenberg uncertainty principle was
obtained for the $\eta/s$. However, it was obtained using a formula, which is justified in rarified gases with
short-range interactions. It's well known already from the nonrelativistic kinetic theory that dense gases get
corrections over the particle number densities (see e. g. \cite{landau10}, Section 18), corresponding to more
than binary collisions, and in very dense gasses this bound can be quite inaccurate. In liquids and other
substances the mechanism of appearance of the shear viscosity may be different (see \cite{Schafer:2009dj} for a
review). In particular, the shear viscosity of water can be very well described by a phenomenological formula
with an exponential dependence on the inverse temperature, see e. g. \cite{Sengers}. } $\eta/s=\frac1{4\pi}$
\cite{adscftbound} was violated with different counterexamples. For some reasonable ones see
\cite{Buchel:2008vz,Sinha:2009ev}. Also see the recent review \cite{Cremonini:2011iq}. The bulk viscosity, being
very sensitive to violation of the equation of state and being connected with fluctuations through the
fluctuation-dissipation theorem \cite{Callen:1951vq}, can have a maximum near a phase transition
\cite{Kharzeev:2007wb, Karsch:2007jc, chakkap, Dobado:2012zf}. In \cite{chakkap} and \cite{Dobado:2012zf} sharp
maxima were observed in the bulk viscosity $\xi$ and the ratio $\xi/s$ in the linear $\sigma$-model for the
vacuum $\sigma$ mass $900~MeV$. Decreasing the vacuum $\sigma$ mass the maximum eventually disappears. Any
maximum of the $\xi/s$ was not observed in the large-N limit of the linear $\sigma$-model in the
\cite{Dobado:2012zf}. Any maximum of the $\xi$ was not observed in the large-N limit of the ${1+1}$-dimensional
Gross-Neveu model \cite{FernandezFraile:2010gu} (see also Sec. \ref{physmeansec} for comments).

Whether one uses the Kubo\footnote{The Kubo formulas are
distinguished from the Greet-Kubo formulas e. g. in
\cite{Muronga:2003tb, Kadanoff}.} formula or the Boltzmann
equation one faces nearly the same integral equation for the
transport coefficients \cite{jeon, jeonyaffe, Arnold:2002zm}. The
preferable way to solve it is the variational (or Ritz) method.
Due to its complexity the relaxation time approximation is used
often in the framework of the Boltzmann equation. Though this
approximation is inaccurate, does not allow to control precision
of approximation and can potentially lead to large deviations. The
main difficulty in the variational method is in calculation of
collision integrals. To calculate any transport coefficient in the
lowest order approximation in a mixture with a very large number
of components $N'$ (like in the hadron gas) one would need to
calculate roughly ${N'}^2$ 12-dimensional integrals if only the
elastic collisions are considered. Fortunately, it's possible to
simplify these integrals considerably and perform these
calculations in a reasonable time.

This paper contains calculations of the shear and the bulk
viscosity coefficients for the hadron gas using the (corrected,
see Sec. \ref{hardcorsec}) UrQMD cross sections. The calculations
are done in the framework of the Boltzmann equation with the
classical Maxwell-Boltzmann statistics, without medium effects and
with the ideal gas equation of state. The Maxwell-Boltzmann
statistics approximation allows one to obtain some relatively
simple analytical closed-form expressions. Originally the
calculations in the same approximations for the hadron gas but
with one constant cross section for all hadrons were done in
\cite{Moroz:2011vn}. The deviations in the worst cases are
relatively small. In that paper some analytical formulas of the
viscosities for 1-, 2- (explicitly) and N-component (up to
solution of the matrix equation) gases with constant cross
sections were obtained. Analogical formulas can be written down
for quite a large class of non-constant cross sections, in
particular, for the ones which appear in the chiral perturbation
theory. The final expressions may become somewhat more cumbersome;
anyway this is better than numerical integration at least in the
speed of the computation. Explicit formulas for the viscosities
with the elastic pion-pion isospin averaged cross section and
somewhat more general one are obtained in the present paper. The
results of the \cite{Moroz:2011vn} are partially reproduced in the
present paper, improving the text and adding more detailed
explanations. The presented calculations can be considered as
quite precise ones at low temperatures where the elastic
collisions dominate and the equation of state is close to the
ideal gas equation of state. At higher temperatures the
calculations with the total cross section are expected to give the
qualitative description.

\begin{figure}[h!]
\begin{center}
\epsfig{file=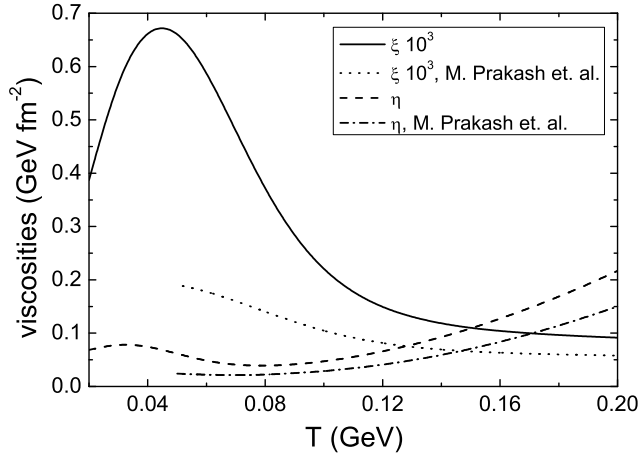,width=8.6cm} \caption{The viscosities of the
pion gas as functions of the temperature at zero chemical
potentials. The dotted and the dash-dotted lines designate
correspondingly the bulk viscosity time $10^3$ and the shear
viscosity, calculated in the 3-rd order in the \cite{prakash}. The
solid and the dashed lines designate correspondingly the bulk
viscosity time $10^3$ and the shear viscosity, calculated at the
same orders using the (corrected, see Sec. \ref{hardcorsec}) UrQMD
cross sections. \label{CompViscosPions} }
\end{center}
\end{figure}

For comparison the calculations of the viscosities are performed
for the pion gas (throughout the paper the chemical potentials are
equal to zero if else is not stated). The results are relatively
close to the results in \cite{prakash, davesne}. There the
calculations are made in the same approximations except for the
\cite{davesne}, where the Bose-Einstein statistics is used instead
of the Maxwell-Boltzmann one. The discrepancies from the used
classical statistics are not large at zero chemical potential and
become larger as the chemical potential grows (see Sec.
\ref{condappl} for the errors and comments). The comparison is
made with the results of the \cite{prakash}, see fig.
\ref{CompViscosPions}. The discrepancies up to a factor of $3.5$
for the bulk viscosity and up to a factor of $2.5$ for the shear
viscosity come most probably from somewhat different ${\pi\pi}$
elastic plus the quasielastic (through the intermediate
$\rho$-resonance) cross section of the \cite{prakash}\footnote{The
author could not reproduce this plotted total cross section by its
formula. In fact it was approximately 2.6 times larger. But the
plotted total cross section is quite close to the isospin averaged
(corrected) UrQMD $\pi\pi$ total cross section. A notable
deviation is only at $\sqrt{s}<0.5~GeV$, when the UrQMD cross
section becomes $1.5-2$ times smaller. At
$1.2~GeV<\sqrt{s}<1.9~GeV$ the UrQMD cross section is a little
larger instead.} (the averaging over the scattering angle is
expected to give small errors; see also comments below the formula
(\ref{xica})). The minima of the shear viscosities near
${T=60-70~MeV}$ are attributed to the peaks from the
$\rho$-resonances in the ${\pi\pi}$ cross sections. It's not
noticeable in the figure for the dash-dotted line. Nonzero values
of the bulk viscosities and theirs maxima are solely due to the
masses of the pions. The paper \cite{prakash} implements also the
isospin averaged current algebra elastic cross sections. These
cross sections can be reproduced in the lowest order in the chiral
perturbation theory \cite{Scherer:2002tk}. They obviously have
quite large deviations from the experimental data at high enough
energies and wrong $\sqrt{s}$ asymptotic dependence, which can be
seen from the comparison of them with the isospin averaged elastic
plus the quasielastic experimental cross sections in the
\cite{prakash}. The elastic $\pi\pi$ cross sections are rather
close to the constant $5~mb$ \cite{Bass:1998ca, Bleicher:1999xi}.

In several papers the bulk viscosity $\xi$ was calculated for the
pion gas, using the chiral perturbation theory (or the unitarized
chiral perturbation theory) and some other approaches, with quite
large discrepancies between the quantitative results. In
\cite{fernnicola} the calculations were done by the Kubo formula
in a rough approximation. There the number-changing
${2\leftrightarrow4}$ processes were neglected too, and the
non-vanishing value of the bulk viscosity is obtained due to a
trace anomaly and the pions' masses. At small temperatures, where
the effects of the trace anomaly are small, the magnitude of the
bulk viscosity is large in compare to the results of this paper
and the \cite{prakash, davesne}. For example, at ${T=20~MeV}$
(${60~MeV}$) it's larger approximately in 39 (8) times than the
bulk viscosity in this paper. The maximal values differ in
approximately 24 times. In \cite{lumoore} the calculations are
done in the framework of the Boltzmann equation and have a
divergent dependence of the $\xi$ for ${T\rightarrow0}$ because of
remained weak ${2\leftrightarrow4}$ number-changing processes (at
${T=140~MeV}$ the bulk viscosity is nearly 57 times larger than
the bulk viscosity calculated in this paper). This dependence
should change at low enough temperatures, $T=180~MeV$ or higher
ones for the pion gas, see Sec. \ref{physmeansec} for
explanations. Joining the results of the calculations at low and
high temperatures, the function $\xi(T)$ may turn out to be not
continuous at the middle temperatures (which is not a physical
effect, see Sec. \ref{physmeansec}), and the smooth function
$\xi(T)$ is to be obtained through some interpolation. In
\cite{dobado} the bulk viscosity was calculated in the framework
of the Boltzmann equation with the ideal gas equation of state and
only the elastic collisions taken into account. The Inverse
Amplitude Method was used to get the scattering amplitudes of the
pions. The quantitative results are close to the results in this
paper (discrepancies up to a factor of $2.7$). In \cite{chenwang}
the calculations are done in the framework of the Boltzmann
equation for the massless pions. There the bulk viscosity
increases rapidly so that the ratio $\xi/s$ increases with the
temperature.

Calculations of the shear viscosity in the hadron gas with a large
number of components were done in \cite{Gorenstein:2007mw}, using
some approximate phenomenological formula, and in \cite{toneev},
using the relaxation time approximation. These results are in good
agreement with the calculations of this paper. Hence, as long as
the ratio $\xi/\eta$ calculated in the \cite{toneev} for the free
massive pion gas is $8-58$ times larger (in the temperature range
$0.02~GeV<T<0.14~GeV$ with the deviations growing as the
temperature decreases) than the one calculated in this paper, one
can suspect that the difference comes from the bulk viscosity
because of the used relaxation time approximation\footnote{In the
relaxation time approximation the bulk viscosity source term is
treated somewhat differently: the $\xi$ becomes proportional to
the integral of the squared source term (times some functions of
momentum) and not to the square of the integrated source term
(times some functions of momentum). Note that in the
\cite{fernnicola} the used formula has this relaxation time
approximation form. Also there the source term is the one of a
system with the inelastic processes. These facts could help to
understand the enlarged values of the bulk viscosity. Not small
quantitative discrepancies can be noticed between the calculations
of the \cite{chakkap} and the \cite{Dobado:2012zf}.} and likely
not conserved particle numbers at low temperatures, provided that
the SHMC model's cross sections, used in the \cite{toneev}, don't
have large deviations from the UrQMD cross sections or the
experimental data, which seems to be the case. Also note that the
results in the \cite{toneev} for the free particles and the SHMC
model don't differ very much. These facts may explain why the
$\xi/s$ of the hadron gas in the \cite{toneev} is $3.65-11.3$
times larger (in the temperature range $0.1~GeV<T<0.18~GeV$) than
the $\xi/s$ calculated in this paper. At the low temperature
${T=100~MeV}$ and the vanishing chemical potentials it is 11.3
times more (at the same temperature the factor is 8.2 for the case
of the pion gas). In \cite{nhngr} the calculation of the bulk
viscosity is done for the hadron gas (with an excluded-volume
equation of state) with the masses less than $2~GeV$ using some
special formula, obtained though some ansatz
\cite{Kharzeev:2007wb}. Its quantitative accuracy has not been
clarified. The ratio $\xi/s$ in the \cite{nhngr} deviates from the
$\xi/s$ of this paper up to a factor of 1.8 in the temperature
range $0.14~GeV<T<0.18~GeV$ and is different on $4\%$ at
$T=0.14~GeV$.

Also the shear viscosity has been calculated using the Kubo
formula (or the Green-Kubo formula) in a gas of mesons and their
resonances \cite{Muronga:2003tb}. There the UrQMD simulations are
performed to calculate the energy-momentum tensor, used in the
calculations by the Kubo formula. The $\eta$ in the
\cite{Muronga:2003tb} is $1.14-1.77$ times smaller then the $\eta$
for the hadron gas in this paper. At $T=0.15~GeV$ it is $1.77$
times smaller. In \cite{Muroya:2004pu} similar calculations, using
the URASiMA event generator, are done for the shear viscosity with
close results.

The structure of the paper is the following. A misleading
viewpoint on the bulk viscosity, connected with the inelastic
processes, is commented on in Sec. \ref{physmeansec} together with
some properties, features and physical meaning of the bulk
viscosity. In that section some questions concerning the
deviations from the chemical equilibrium are addressed too. Sec.
\ref{hardcorsec} contains some comments on the constant cross
sections, which are used in approximating calculations, and some
other general comments on cross sections. Also it contains a
description of the UrQMD cross sections, which are used in the
main calculations, together with their analysis, corrections and
the consequences of the corrections for the freeze-out
temperatures. The applicability of the used through the paper
approximations is discussed in Sec. \ref{condappl}. The system of
the Boltzmann equations, its solution and formal expressions of
the transport coefficients can be found in Sec. \ref{CalcSec}. The
numerical calculations for the hadron gas are presented in Sec.
\ref{numcalc}. In Sec. \ref{singcomsec} analytical results for the
single-component gas are presented. In particular, an analytical
expression for the first order single-component shear viscosity
coefficient with constant cross section, found before in
\cite{anderson}, is corrected while the bulk viscosity coefficient
remains the same. The nonequilibrium distribution function in the
same approximation is written down. Also the viscosities with some
non-constant cross sections are written down. Some analytical
results for the binary mixture with constant cross sections are
considered in Sec. \ref{binmixsec}. Integrals of source terms
needed for the calculation of the transport coefficients can be
found in Appendix \ref{appA}. The general entropy density formula
can be found in Appendix \ref{appTherm}. It is used in the
numerical calculations for the hadron gas. Transformations of
collision brackets, being the 12-dimensional integrals which enter
in the viscosities, and some analytical formulas for them can be
found in Appendix \ref{appJ}. The closed-form expressions for
collision rates, mean free paths and mean free times are included
in Appendix \ref{appmfp}.

\section{ Some features and properties of the bulk viscosity \label{physmeansec}}

First, it should be reminded that the transport coefficients are
defined as coefficients next to their gradients in the formal
expansion of the energy-momentum tensor and the charge density
flows over the gradients of the thermodynamic functions or the
flow velocity (see e. g. \cite{landau6}, Section 136). The Kubo
formulas are not definitions of the transport coefficients, as one
might think. They may introduce some assumptions. In particular,
the Kubo formulas in the form as in the \cite{jeon} have zero
frequency and zero momentum limits, which neglect finite size and
finite time effects. Zero momentum limit implies the
thermodynamical limit. This limit is needed to avoid possible
nonphysical contributions from inappropriate choice of a current
and an ensemble \cite{Kadanoff}. The Kubo formulas in the form as
in \cite{Muroya:2004pu, Kubo} suppose thermal equilibrium in the
initial moment of time $t=-\infty$. So that any infinite
space-time scale cannot be connected with the transport
coefficients by their \emph{definitions}.

The Boltzmann equations will be used in what follows. In the case
of the elastic collisions they can be derived from the Liouville
equation\footnote{The Boltzmann equations can also be derived for
the case of the inelastic collisions from some physical
considerations, see \cite{groot} (Chap. I, Sec. 2). } in the
approximations $n_k r_{kl}^3 \ll 1$ ($r_{kl}$ is the effective
radius of two particle interactions between the particles of the
species $k$ and $l$) that is for rarified gases with short-range
interactions\footnote{This is the case of interest. Coulomb
interactions can be neglected in heavy ion collisions at all the
considered energies in this paper.}. Also the linear integral
equations for the viscosities and other transport coefficients,
derivable from the Boltzmann equation, can be obtained (with some
corrections) from the perturbative calculations for quantum field
theories at finite temperature (including the inelastic processes)
using the Kubo formulas \cite{jeon, Gagnon:2006hi, Gagnon:2007qt},
which justifies application of the Boltzmann equation when the
inelastic processes are present.

The bulk viscosity can reveal itself only when there is a nonzero
divergence of the flow velocity. This nonequilibrium perturbation
should not be confused with another possible \emph{independent}
perturbation (as was done in several papers, some of which are
mentioned below; the roots of the misleading viewpoint, perhaps,
can be found in \cite{landau6}, Section 81). Namely, this is the
homogeneous perturbation. It can be both the chemical and the
kinetic one\footnote{The inclusion of this kinetic perturbation is
similar to the inclusion of the chemical one so that it is omitted
for simplicity below. Usually this perturbation should fade first
because also the inelastic processes are responsible for the
relaxation of the momentum spectra. However, see comments for
$1+1$-dimensional systems below.}. Then it can be generalized and
made dependent on the coordinate. It just should not be
proportional to any gradient. Then the constraints of the local
conservation laws should be imposed on these perturbations. The
perturbations for the particle numbers should be such that don't
violate conservation of all charges. Considering the case of
homogeneous chemical perturbation in a gas with fixed volume, one
concludes that the temperature should change with time, being some
energy per particle. So that energy conservation should be
obtained varying the temperature. Adding an infinitesimal
correction to the temperature one gets a perturbation of the form
$C p_k^\mu U_\mu$. Such perturbations don't contribute to all
collision integrals. To describe purely chemical perturbations
they have to be chosen in the form of the momentum-independent
terms (except for the $C p_k^\mu U_\mu$ terms), otherwise there
will be contributions from the elastic collision integrals. Such
perturbations can be considered as chemical potentials-like ones
(being small, one can expand the distribution functions over them
and get these momentum-independent terms) with the arguments for
maximization of the entropy. To find the evolution of these terms
they should be separated. Let's write this in some formulas.
Multicomponent gas distribution functions with the leading
perturbations can be represented in the form (detailed definitions
can be found in Sec. \ref{CalcSecA})
 \eq{\label{distfunc}
 f_k=f_k^{(0)}(1+\tilde \varphi_k)(1+\varphi_k)\approx f_k^{(0)}(1+\tilde \varphi_k+\varphi_k),
 \quad |\tilde \varphi_k|\ll 1, \quad |\varphi_k|\ll 1,
 }
where $\varphi_k$ are the perturbation due to the gradients and
$\tilde \varphi_k$ are the chemical perturbations\footnote{Note
that if the $k$-th species have conserved particle numbers, then
the nonequilibrium chemical potential is nonphysical or redefining
the usual (thermodynamic) chemical potential.}. Following steps of
Sec. \ref{CalcSecA}, one can get the following linearized
equations from the Boltzmann equations:
 \eq{\label{generlinbeqn}
 p_k^\mu(U_\mu D+\nabla_\mu)f^{(0)}_k +
 f^{(0)}_k p_k^\mu (U_\mu D + \nabla_\mu )\tilde \varphi_k
 \approx -f_k^{(0)}\mc L_k[\varphi_k] - f_k^{(0)}\mc L_k^{inel}[\tilde \varphi_k],
 }
where $\mc L_k$ and $\mc L_k^{inel}$ is the sum of the linearized
collision integrals (divided on the $-f_k^{(0)}$) of all the
processes and of the inelastic processes correspondingly. The 2-nd
order gradients and the squared 1-st order gradients are neglected
in the l. h. s. because they are of the next order\footnote{The
question of validity of this expansion over the gradients (which
coincides with the usual order counting in the formal expansion
over the gradients in the hydrodynamics) for some profiles is not
discussed in this paper.} and should be cancelled in the next
iteration by the next corrections to the distribution functions.
Also the smallness of the $\tilde \varphi_k$ is used. If the
spatial covariant gradients $\nabla_\mu \tilde \varphi_k(t=0)$ (at
the initial moment of time) are of the same order as the gradients
of the thermodynamic functions or the flow velocity, then the
$\nabla_\mu \tilde \varphi_k$ terms in the l. h. s. should be
retained\footnote{It's a reasonable assumption in the case when
the hydrodynamical description is applicable. For example, the
chemical perturbations can be a result of a fast previous
expansion (faster than the chemical equilibration). Then the
inhomogeneities of the chemical perturbations should correlate
with the inhomogeneities of the thermodynamic functions, the flow
velocity or it's divergence.}. The covariant temporal derivatives
$D \tilde \varphi_k$ are needed to describe the temporal evolution
of the $\tilde \varphi_k$. Then the equations (\ref{generlinbeqn})
can be split onto the separate equations for the $\tilde
\varphi_k$ and the $\varphi_k$
 \eq{\label{generlinbeqn1}
 p_k^\mu U_\mu D \tilde \varphi_k
 +p_k^\mu \nabla_\mu \tilde \varphi_k
 \approx - \mc L_k^{inel}[\tilde \varphi_k],
 }
 \eq{\label{generlinbeqn2}
 p_k^\mu U_\mu D f^{(0)}_k + p_k^\mu \nabla_\mu f^{(0)}_k
 \approx -f_k^{(0)}\mc L_k[\varphi_k].
 }
The equations (\ref{generlinbeqn}) can be split within the
framework of the perturbation theory over the gradients. Let's
consider also the condition $\varphi_k \ll \tilde \varphi_k$ in
the (\ref{distfunc}). Then neglecting the $\varphi_k$ in the
(\ref{distfunc}) and repeating the steps of Sec. \ref{CalcSecA},
one can get the following linearized equations:
 \eq{\label{fadingeq}
 p_k^\mu U_\mu D \tilde \varphi_k \approx - \mc L_k^{inel}[\tilde
 \varphi_k].
 }
The equations (\ref{fadingeq}) are precise in the homogeneous case
(the approximation is only from the linearization). The 1-st order
gradients and the $\nabla_\mu \tilde \varphi_k$ are neglected.
Then using the (\ref{fadingeq}) and the (\ref{generlinbeqn}), one
can get
 \eq{\label{generlieqn}
 (p_k^\mu U_\mu D+p_k^\mu \nabla_\mu)f^{(0)}_k
 +f^{(0)}_k p_k^\mu \nabla_\mu \tilde \varphi_k
 \approx -f_k^{(0)}\mc L_k[\varphi_k].
 }
Solving the system of equations (\ref{fadingeq}) in the local rest
frame, one gets the leading exponential fading dependencies on
time\footnote{If the expansion rate is much larger than the
collision rates of the inelastic processes (e. g. because of a
substantial decrease of the temperature), then the chemical
perturbations should enlarge instead. If the r. h. s. of the
(\ref{generlinbeqn1}) is smaller than the second term of the l. h.
s., then one can consider another approximation, when the $k$-th
species particle numbers are conserved. Then the chemical
perturbation becomes an addition to the thermodynamic chemical
potential.} (in a covariant form this should be an explicit
space-time dependence). Such dependencies were obtained in some
previous studies, see e. g. \cite{Matsui:1985eu, Song:1996ik}. The
equations (\ref{generlieqn}) are different from the ones obtained
from the common Chapman-Enskog procedure (see e. g. \cite{groot},
Chap. V) because of the $\nabla_\mu \tilde \varphi_k$ terms. The
contributions from the small chemical perturbations can be
neglected in the considered order in the transport coefficients
because they are multiplied on the 1-st order gradients. The
$\nabla_\mu \tilde \varphi_k$ terms can be cancelled, introducing
terms proportional to the $\nabla_\mu \tilde \varphi_k(t=0)$ into
the $\varphi_k$ terms. If the spatial distributions of the $\tilde
\varphi_k(t=0)$ are such that $\nabla_\mu \tilde \varphi_k(t=0)$
are of the 2-nd or a higher order, then the $\nabla_\mu \tilde
\varphi_k$ can be neglected. This assumption or approximation is
used in the calculations of this paper. In the linear response
theory one can also introduce independent small chemical
perturbations with the same conclusions for the 1-st order
transport coefficients and find evolution of the perturbations
with time.

Note that the deviation from the chemical equilibrium itself is
not necessarily a source of the bulk viscosity, as is stated in
\cite{Paech:2006st}. If the bulk viscosity is not equal to zero
only because of the particles' masses and they are tended to zero,
the bulk viscosity source term and the bulk viscosity tend to zero
even if there are inelastic processes (see the end of Sec.
\ref{CalcSecA}). In the \cite{Paech:2006st} the independent
chemical perturbations and the perturbations due to the gradients
were just connected through the perturbations of particle numbers,
and the bulk viscosity became proportional to the chemical
relaxation time. Formally infinite chemical relaxation time
doesn't imply any divergencies in the chemical perturbations
$\tilde \varphi_k$, but rather approximation of conserved particle
numbers. Note that the dependence on the strength of the inelastic
processes is different for the chemical perturbations and the
perturbations due to the gradients. Increasing the strength of the
inelastic processes the chemical relaxation time decreases. And
the gradients' relaxation time increases, because the transport
coefficients, at least in rarified gases with short-range
interactions, roughly speaking, are inversely proportional to the
integrated cross sections (in an ideal liquid the gradients'
relaxation time is infinite). What happens with the bulk viscosity
if the inelastic processes become weaker is discussed below.

Making the inelastic processes weaker in compare to the elastic
ones, the bulk viscosity eventually gets a formal dominant
contribution from them because of the approximate zero mode(s)
\cite{jeon}, connected with possible conservation of particle
number(s)\footnote{\label{footn2}If the particles involved into
the inelastic processes are massive, then the formal dominant
contribution is the exponential one over the temperature and grows
as the temperature decreases. If the particles are massless or
approximately massless, as in high-temperature QCD
\cite{Arnold:2006fz}, then a more complicated situation can occur,
and one may need to compare some differences of processes' rates
(and not just equilibrium collision rates), arising in the
collision matrix ($\tilde C^{ab}_{mn}$ in assignments of the
\cite{Arnold:2006fz}). Under the same pair of used test-functions
(indexed by $m$, $n$ in the $\tilde C^{ab}_{mn}$), and for the
same pair of particle species, smaller differences of processes'
rates can be neglected. Comparing among \emph{different pairs of
test-functions} the smallest nonzero contributions dominate, or
rather as can be obtained directly from the inverted collision
matrix. }. As long as it's clear that the bulk viscosity is not
responsible for the chemical equilibration, it's also clear that
there may be the approximation of conserved particle numbers if
the momentum spectrum, as well as the gradients, can relax by
means of only the elastic collisions (which is usually the case)
and the elastic processes make a dominant contribution to the
collision rates. The question is only at what concrete temperature
does this approximation sets in. Let's make an illustrative
example of
what nonphysical contributions one can get from formally remained
weak inelastic processes. Consider infinitely weak inelastic
processes and the perturbation of the flow velocity such that the
energy-momentum tensor gets a sizable contribution from the bulk
viscosity term, not large in compare to the pressure (cf.
(\ref{T0}), (\ref{T1})) to remain the perturbation theory
applicable. Then it's obvious that this contribution is not
physical because it is created by the practically absent processes
and the infinitesimal perturbation of the flow velocity. Instead,
this system is practically described by the equilibrium
thermodynamic functions. This also answers positively the question
whether the thermodynamic chemical potential can be introduced for
approximately conserved particle number in principle. As far as
the author knows, the first correct comment (albeit somewhat
inaccurate) on this issue can be found in the \cite{jeonyaffe}.
However, note that in fact there is no divergent mean free paths,
corresponding to the inelastic processes (IMFP) in this case. They are
cut by the mean free paths, corresponding to the elastic processes
(and the overall collision rate have the dominant contribution
from the elastic collisions). So that it may be not necessary for
the chemical relaxation time to be much larger than any relevant
time scale (like the gradients relaxation time or the time of
existence of the thermal part of the system) to switch off the
inelastic processes. That's why a criterion based on comparison of
collision rates of elastic and inelastic processes can be
considered to switch off the inelastic processes. Such a
comparison is done in the UrQMD studies of the hadron gas in
\cite{Bleicher:2002dm} (see Sec. \ref{condappl} for farther
discussions). According to \cite{Goity}, the chemical relaxation
time of the $2\leftrightarrow4$ processes in the pion gas is much
larger than the thermal relaxation time. And e. g. at $T=180~MeV$
the chemical relaxation time is equal to $40~fm$, which is larger
than the typical lifetime of the thermal part of the expanding
fireball (see e. g. \cite{Bleicher:2002dm}). So that it's the
inelastic $2\leftrightarrow4$ processes which should be neglected
in the pion gas at $T=180~MeV$ or even higher temperatures, which
wasn't done in the \cite{lumoore}.
To show importance of the gradients relaxation time, let's consider the following possible case. Let's consider
the only perturbation - propagating sound wave, perturbed in a point. It's possible for the IMFP to be much
larger than the gradients relaxation size (on which the wave can be considered as damped) and be much smaller
than the system's size at the same time. Then, the bulk viscosity cannot be defined by the IMFP in this case,
because it enters in the sound attenuation constant. Thus, the gradients relaxation size and time are cutting
parameters. Note that they exists even in infinite systems considered during infinite time interval.

The bulk viscosity source terms increases substantially if
particle numbers are not conserved (cf. (\ref{alfrac1}),
(\ref{alfrac2}); in mixtures these particle numbers should also be
not small). This reflects additional fluctuations from not
conserved particle numbers. Though the inelastic processes have to
be effective enough to consider the approximation of not conserved
particle numbers. Perhaps, the point at which the bulk viscosities
in the different approximations cross can provide a criterion for
switching on/off the inelastic processes. If this is not so, then
one would have to make some interpolation in the intermediate
region\footnote{Perhaps, the bulk viscosity calculated without
constant test-functions (except for zero modes of the inelastic
collision integrals, used to conserve charges) can provide a good
interpolation.}. Note that e. g. in the calculations by the Kubo
formulas through the direct calculations of the energy-momentum
tensor as in the \cite{Muronga:2003tb} it's not needed to use the
approximation of conserved or not conserved particle numbers
(which defines the number of independent thermodynamic chemical
potentials, through which the chemical potentials of all particles
are expressed, cf. (\ref{mukdef})). There the energy-momentum
tensor should be a smooth function of time and the thermodynamic
functions as long as the inelastic processes fade smoothly. Then
the bulk viscosity should be a smooth function of the temperature
and particles' chemical potentials regardless of the number of the
independent chemical potentials.

In the \cite{Arnold:2006fz} a bottleneck for the relaxation to
equilibrium characterized by the bulk viscosity due to the weakest
processes' rates is assumed. Instead, there are rather dominant
contributions from some test-functions\footnote{Not a bottleneck
from some perturbations, because one actually doesn't have a
choice in the form of the momentum dependence of the perturbations
corresponding to the transport coefficients. The kinetic
perturbation can be of different forms of the momentum
dependence.} (as is commented in the footnote \ref{footn2}), which
should not be specially treated though, except for the ones which
are the approximate zero modes making a dominant contribution. A
similar dominance\footnote{Another similar dominance can exist
from particle species interacting weakly with all particles.} is
present also in other transport coefficients, in particular, when
there is only one type of processes. Although in QCD at high
enough temperatures the equilibrium $2\leftrightarrow 2$ elastic
collisions rate is parametrically the largest one\footnote{The
estimate can be easily inferred from \cite{Arnold:2000dr}.},
$O(\alpha_s T)$, because of cancellations the momentum transfer
takes place with the rate $O(\alpha_s^2\ln(1/\alpha_s) T)$, which
is parametrically smaller than the particle number change rate
$O(\alpha_s^{3/2} T)$ from the effective "$1\leftrightarrow 2$"
processes. This provides an example when the equilibrium collision
rates may differ substantially from the relevant collision rates.
The "$1\leftrightarrow 2$" processes provide small chemical
relaxation time in compare to the thermal relaxation time, which
justifies the approximation of not conserved particle numbers and
the enhancement of the bulk viscosity from the source terms at
least at small enough $\alpha_s$, whereas the contributions to the
bulk viscosity from the collision integrals of the
"$1\leftrightarrow 2$" processes are suppressed at small enough
$\alpha_s$ (the inelastic $2\leftrightarrow 2$ processes are not
suppressed, but they are of the order $O(\alpha_s^2\ln(1/\alpha_s)
T)$). To avoid misunderstanding it may be mentioned that taking
the total collision rate of the "$1\leftrightarrow 2$" processes
as formally infinite by taking the corresponding matrix elements
as formally infinite ones, one gets zero bulk viscosity and zero
mean free paths as long as both the gluons and quarks take part in
these processes (see also footnote \ref{footn1}).

In the case of a ${1+1}$-dimensional single-component gas the
elastic collisions cannot result in the relaxation of the momentum
spectra and, hence, cannot stimulate the system to evolute towards
equilibrium\footnote{There are forward scatterings and momentum
interchange. As long as the particles are not distinguishable the
momentum interchange from the elastic collisions is equivalent to
the forward scatterings or absence of the elastic collisions at
all.}. As a result, the exponentially divergent bulk viscosity was
obtained in the paper \cite{FernandezFraile:2010gu}. Considering
again the example about the infinitely small perturbation of the
flow velocity and assuming also a finite size of the system, it's
again obvious that the weak inelastic processes may make
nonphysical contributions (in this case the mean free path is
formally cut by the system's size). If this is the case, then the
hydrodynamical description becomes inapplicable, and might use
simulations of particles' collisions or the Boltzmann equations in
the approximation without collisions (on a time scale much smaller
than the chemical relaxation time). If the ${1+1}$-dimensional
description is only an approximate one (that is with small angle
elastic scatterings in higher dimensions), the relaxation of the
momentum spectrum by the elastic collisions should be considered.
And if a $1+1$-dimensional gas has at least two components with
different masses, then a nontrivial momentum exchange in the
elastic collisions is possible. This results in the possibility of
the relaxation of the momentum spectra by only the elastic
collisions \cite{Cubero}.

Let's summarize this section with formulation of the physical
meaning of the bulk viscosity. The bulk viscosity reflects
deviation of the value of the pressure from its local equilibrium
value (as can be seen from the (\ref{T1})), appearing when the
system expands/compresses, because of the delay in the
equilibration. The bulk viscosity is not responsible for the
restoration of the chemical or the kinetic equilibria - it's
responsible for the relaxation of the divergence of the flow
velocity. If there are inelastic processes, then the particle
numbers also get nonequilibrium contributions (cf. (\ref{pflow}),
(\ref{fpert}), (\ref{varphi})) such that the charge is conserved
locally (cf. (\ref{cofchf}))\footnote{One should keep in mind that
while studying the chemical perturbations $\tilde \varphi_k$
through the thermodynamic functions first the contributions from
the transport coefficients' terms should be subtracted.}. Though
these contributions together with the contribution to the pressure
may become nonphysical because of the approximate zero modes (if
such ones appear in the calculations). The magnitude of the bulk
viscosity changes from theory to theory. Under some quite general
assumptions a nonzero value of the bulk viscosity can be connected
with violation of the scale invariance due to a nonzero value of
the energy-momentum tensor \cite{Coleman:1970je, Callan:1970ze}.
Of course, the beta function can contribute to the energy-momentum
tensor and the bulk viscosity too \cite{jeon}.

\section{ The hard core interaction model and the UrQMD cross sections \label{hardcorsec}}

In a non-relativistic classical theory of particle interactions
there is a widespread model, used in approximate calculations,
called the hard core repulsion model or the model of hard spheres
with some radius $r$. For its applications to the high-energy
nuclear collisions see \cite{Gorenstein:2007mw} and references
therein. The differential scattering cross section for this model
can be inferred from the problem of scattering of point particle
on the spherical potential ${U(r)=\infty}$ if ${r\leq a}$ and
${U(r)=0}$ if ${r>a}$ \cite{landau1}. In this model the
differential cross section is equal to $a^2/4$. To apply this
result to the gas of hard spheres with the radius $r$ one can
notice that the scattering of any two spheres can be considered as
the scattering of the point particle on the sphere of the radius
$2r$, so that one should take ${a=2r}$. The total cross section
$\sigma_{tot}$ is obtained after integration over the angles of
the $r^2 d\Omega$ which results in the $\sigma_{tot}=4\pi r^2$.
For collisions of hard spheres of different radiuses one should
take ${a=r_k+r_l}$ or replace the $r$ on the $\frac{r_k+r_l}2$:
 \eq{\label{hccs}
 \sigma_{tot,kl}=\pi (r_k+r_l)^2.
 }
The relativistic generalization of this model is the constant (not
dependent on the scattering energy and angle) differential cross
sections model.

The hard spheres model is classical, and connection of its cross
sections to cross sections, calculated in any quantum theory, is
needed. For particles, having a spin, the differential cross
sections averaged over the initial spin states and summed over the
final ones will be used.\footnote{It's assumed that particle
numbers of the same species but with different spin states are
equal. If this were not so then in approximation, in which the
spin interactions are neglected and probabilities to have certain
spin states are equal, the numbers of the particles with different
spin states would be approximately equal in the mean free time.
With equal particle numbers their distribution functions are equal
too. This allows one to use the summed over the final states cross
sections in the Boltzmann equations.} If colliding particles are
identical and their differential cross section is integrated over
the momentums (or the spatial angle to get the total cross
section) then it should be multiplied on the factor $\frac12$ to
cancel double counting of the momentum states. These factors are
exactly the factors $\gamma_{kl}$ next to the collision integrals
in the Boltzmann equations (\ref{boleqs}). The differential cross
sections times these factors will be called the classical
differential cross sections.

The UrQMD cross sections are used in the numerical calculations of
Sec. \ref{numcalc}\footnote{Very high energy dependence of any
used UrQMD cross section is not important because of the
exponential suppression $e^{-\sqrt{s}/T}$. The used cross sections
were cut on the ${\sqrt{s}=5~GeV}$ and were continued by a
corresponding constant continuously at higher energies. At small
enough momentums there is another somewhat weaker suppression. The
momentum space density of each particle provides $p^2$
suppression. This may (partially) suppress some deviations from
the experimental data of some UrQMD cross sections (like for the
${\Lambda p}$ pair) at ${\sqrt{s}\sim m_k+m_l}$. To estimate at
what temperatures some discrepancies in cross sections can appear
one can equate the $\sqrt{s}$ to the sum of the averaged
one-particle energies $e_k$ (\ref{epsandek}) of the two colliding
particles.}. These cross sections are described in
\cite{Bass:1998ca, Bleicher:1999xi}. More details can be found in
the UrQMD program codes. Below there is some description mainly of
what is different or new.

The UrQMD cross sections are averaged over the initial spin states
and summed over the final ones. As long as the UrQMD cross
sections are total ones (integrated over the scattering angle),
the factors $\gamma_{kl}$ are already absorbed into them (in what
follows only such cross sections will be considered in this
section tacitly). Dividing them on the $4\pi$, one gets the
classical differential cross sections, averaged over the
scattering angle.

The UrQMD codes (version 1.3) were modified to get accurately
tabulated (with a step of ${25~MeV}$) cross sections. Resonances'
masses and widths (they are tuned in their uncertainty regions to
describe the experimental data better), used in the UrQMD codes,
have somewhat different values than the ones in the
\cite{Bass:1998ca}. Influence of variation of these parameters was
studied in \cite{Gerhard:2012fj}. The UrQMD codes implement
somewhat different averaging of the c. m. momentums over the
resonances' masses\footnote{Averaged powers of the momentums are
used, not powers of the averaged momentums.} than in the papers
\cite{Bass:1998ca, Bleicher:1999xi}. It was found that using the
resonance dominating cross sections from the papers
\cite{Bass:1998ca, Bleicher:1999xi} some of these cross sections
could have a large rise at small c. m. momentums if constant
widths are used in the calculations of the averaged c. m.
momentums in the energy dependent widths. So that one should be
aware of this fact\footnote{It may be mentioned that one should be
also aware of possible differences in storing of the floating
point numbers in different programming languages or while using
different compilers.}. The UrQMD codes have a low energy cut-off
at ${\sqrt{s} \sim m_k+m_l+0.01~GeV}$ (and a similar one over the
c. m. momentum if triggered) for the resonance dominating cross
sections, and no large low energy rise was found there.

An important ingredient of the UrQMD model is the Additive Quark
Model (AQM), which is used for unknown cross sections.
Universality of hadrons, based on jet quenching arguments, is used
to support this model. This model describes the experimentally
known cross sections well at sufficiently high energies.
Application of this model is better than elimination of the
corresponding hadrons, which is the same as equating their all
cross sections to zero and, hence, exclusion of their
contributions from the thermodynamic functions (infinite mean
paths, no thermalization).

At this point an interruption should be made to consider some
important questions related to different types of the UrQMD cross
sections. These different types are used due to several reasons
and are the following: the elastic cross section(s) (ECS(s)), the
elastic plus the quasielastic cross section(s) (EQCS(s)), the
total cross section(s) (TCS(s)) and the previous two types with
enhanced in some way resonances' cross sections (index "2" is
appended in the abbreviations).

Of course, the system of the Boltzmann equations would have a
solution with any of these cross sections. Usage of the ECSs is
completely self-consistent as long as only the elastic
${2\leftrightarrow 2}$ collision integrals are used in the
calculations of the viscosities. However, there are reasons to
consider also the EQCSs. Exactly this type of cross sections,
being averaged over the isospin, is implemented in \cite{prakash}.
The quasielastic cross sections ${2\rightarrow 1\rightarrow 2}$
can be used as rightful contributions to the ECSs in the
approximation that the 4-momentum of the intermediate resonance
does not change (the effects of the exclusion of the resonances as
independent particles are considered in Sec. \ref{numcalc}). The
mean free paths of the intermediate resonances without
contributions of the decays, being not equal to zero, also
introduce some errors, which are neglected. The EQCSs conserve
particle numbers, which is consistent with the only elastic
collision integrals, implemented in the calculations. There are
also some additional arguments for the usage of these cross
sections. From the phenomenological considerations one can take
into account shortening of the mean free paths (or enlarging of
the collision rates) due to creation of the resonances. In other
words, there would have to be contributions from the inelastic
${2\leftrightarrow 1}$ collision integrals next to the elastic
collision integrals, and they are taken into account approximately
by the contributions from the quasielastic cross sections.

Resonances are not just intermediate particles, and they can
collide with other particles. They make not negligibly small
contribution to the thermodynamic functions and the viscosities.
So that they are also included in the calculations as independent
particles with their parameters and corresponding
${2\leftrightarrow 2}$ collision integrals. They would have to
have shortening of their mean free paths from their decays and
contributions from the inelastic ${2\leftrightarrow 1}$ collision
integrals too. These contributions may be taken into account from
the following collision rate considerations. A resonance's decay
rate can be approximately replaced with just its total width.
Then, given a resonance, one would have to redistribute its width
(that is not changing the whole collision rate containing the
contribution of the decay rate) in such a way that the cross
section of the collision of this resonance with a resonance of the
same species gets an addition\footnote{\label{footn1}This
enhancement leads to the shortening of the mean free paths of the
resonances of only this species, as needed. In the \emph{formal}
limit of this infinitely large enhancement other collision
integrals can be neglected and the Boltzmann equation for this
species decouples. Then from the solution of the Boltzmann
equation for a single-component gas (see Sec. \ref{singcomsec})
one concludes that the nonequilibrium perturbation to the
distribution function of this species vanishes in this limit. Note
that infinitely strong interactions also with particles of all
other particle species would result in zero transport
coefficients.}. Using an approximate expression for the collision
rates (in the nonrelativistic approximation, applicable in this
case) from Appendix \ref{appmfp}, one easily finds the addition
${\Gamma_k/(\sqrt{2} n_k \langle \abs{\vec v_k}\rangle)}$ (where
$\Gamma_k$ is the width) to the ${4\pi \sigma^{cl}_{kk}}$. Such
cross sections seem to be the most physically preferable ones
because they take into account more realistic mean free paths than
in the previous case while not violating the conservation of the
particle numbers too.

The TCSs are used to take into account even larger
shortening\footnote{Not the largest one. The effect of the
enhancements of the resonances' TCSs is of $5\%$ for the bulk
viscosity and of $50\%$ for the shear viscosity so that TCS2s are
additionally considered.} of the mean free paths than in the case
of the EQCSs. However, such cross sections introduce some
inconsistency, implying that the conservation of the particle
numbers is violated. As long as there may be contributions from
some partial cross sections to the UrQMD ECSs or the EQCSs which
were not taken into account (see below), the TCSs can be used as
the upper bounds for the ECSs and the EQCSs. However, it's
expected that these bounds are excessively high. If so, the TCSs
(rather TCS2s) can be considered not only as the approximation
taking into account real mean free paths but also as some measure
of deviation from the approximation of only the elastic and the
quasielastic collisions with the following arguments. If the TCSs
were approximately equal to the ECSs or the EQCSs, or the numbers
of particles with large inelastic cross sections were small, then
one could expect small errors due to the negligibility of the
inelastic collisions.

Continuing the discussion of the details of the UrQMD cross
sections, it should be mentioned that the UrQMD TCSs are the most
reliable ones. The sum of the partial cross sections is not always
equal to the TCSs by their construction. If this is the case, then
some partial cross sections are rescaled depending on their
reliability\footnote{This information, including some other
information about the cross sections, is stored in the array
SigmaLn of the file blockres.f}.

The magnitudes of the partial cross sections, implemented in the
UrQMD codes, are used to determine, what a partial cross section
to choose in a given collision, using a random number generator.
Among these partial cross sections there are the ECSs. Exactly
these ECSs are used in the present calculations. However, if a
partial cross section with a string excitation is chosen in a
given collision, there is a probability to end up with the elastic
collision if the $\sqrt{s}$ is too small. These contributions to
the ECSs are not calculated and are not added to the ECSs. Also
the string excitations can, possibly, end up with creation of a
resonance. Contributions to the EQCSs from the string excitations
are taken into account partially (see below).

The ECSs, if not known from the experiment, are taken in the form
of some extrapolations, discussed below, or the AQM is used. The
normalization on the corresponding TCSs can change the ECSs
notably. The meson meson (MM) ECSs are equal to $5~mb$. The meson
baryon (MB) ECSs are equal to the AQM rescaled experimental
${\pi^+ p}$ cross sections. But after the normalization they
become equal to zero in the resonances dominated energy range
(approximately below ${\sqrt{s}=1.7~GeV}$). The anti-baryon baryon
(${\bar B B}$) ECSs are equal to the AQM rescaled experimental
${\bar p p}$ cross sections. Other ECSs are equal to the AQM ECSs.

Before discussing the quasielastic cross sections first let's
write for convenience the resonance dominated cross sections
formula for a reaction ${2\rightarrow 1\rightarrow any}$.
Correcting a typo and rewriting it in a somewhat different form
than in \cite{Bass:1998ca, Bleicher:1999xi}, one gets
 \eq{\label{resdomxsec}
 \sigma^{ij}_{tot}(\sqrt s)=\sum_{R}\frac{g_R}{g_i g_j}\frac{\pi}{p_{cm}^2}
 \frac{\Gamma_{R,tot}^2 b_{R\rightarrow ij}}{(M_R-\sqrt s)^2+\Gamma_{R,tot}^2/4}, \quad
 b_{R\rightarrow ij}\equiv |\langle j_i,m_i,j_j,m_j || J_R, M_R \rangle|^2
 \frac{\Gamma_{R\rightarrow ij}}{\Gamma_{R,tot}},
 }
where $\Gamma_{R\rightarrow ij}$ is the partial energy-dependent
width of the decay of the resonance $R$ into particles of types
$i$ and $j$ without specification of their isospin projection,
$\Gamma_{R,tot}$ is the total energy-dependent width of the decay
of the resonance $R$, $g_i$ is the spin degeneracy factor,
$b_{R\rightarrow ij}$ is the energy-dependent branching ratio. The
squared Clebsch-Gordan coefficients allow to specify the branching
ratio $b_{R\rightarrow ij}$ for the pair of the particles with
concrete isospin projections. The squared Clebsch-Gordan
coefficients should be normalized in such a way that they give
unity after summation over all isospin projections in a given
multiplet. This formula represents contributions from all possible
resonances through which the reaction can take place. Now it's
easy to write down the cross sections for the quasielastic
${2\rightarrow 1\rightarrow 2}$ scatterings:
 \eq{
 \sigma^{ij}_{quasi}(\sqrt s)=\sum_{R}\frac{g_R}{g_i g_j}\frac{\pi}{p_{cm}^2}
 \frac{\Gamma_{R,tot}^2 b_{R\rightarrow ij}^2}{(M_R-\sqrt s)^2+\Gamma_{R,tot}^2/4}.
 }
One more multiplier $b_{R\rightarrow ij}$ takes into account the
fact that a resonance $R$ decays only into the $ij$ pair and
represents the probability of this decay.

The ${K^- p}$ TCS is not described by the formula
(\ref{resdomxsec}) completely, and a partial cross section,
attributed to the s-channel strings excitations, is added in the
UrQMD model to fit the TCS to the experimental data. In the UrQMD
model this s-channel strings cross section is added also to other
strange meson nonstrange baryon TCSs when annihilation is possible
due to the quark content. From comparison with the experimental
data for the ${K^- p}$ ECS \cite{Beringer} (actually it's believed
to be the EQCS because smaller peaks from the resonances are
reproduced there) it was found that the half of the s-channel
strings cross section is enough to describe well this experimental
${K^- p}$ cross section. Then the half of the s-channel strings
cross section is added to other strange meson nonstrange baryon
EQCSs when annihilation is possible. These contributions from the
strings excitations are the most low energetic ones. They are the
only contributions from the strings excitations which are added.
The next in the energy scale possible contributions to the EQCSs
may be in the ${\bar B B}$ cross sections. In other pairs the
string excitations appear approximately from ${\sqrt{s}=3~GeV}$.

There is an important omission, found in the UrQMD codes (present
also in the last version 3.3). The function fcgk returns incorrect
(two times smaller) values of the squared Clebsch-Gordan
coefficients for the resonances dominated cross sections in some
cases. The first case is for the pairs of unflavored mesons from
the same multiplet with the isospin ${I=1}$. For example, the
function fcgk returns $0.5$ for the only possible isospin
decomposition of the $\rho^+$ to the ${\pi^+ \pi^0}$ pair, because
the states ${\pi^+ \pi^0}$ and ${\pi^0 \pi^+}$ are counted as
different ones. As a result, the peak from the $\rho$-resonance
becomes two times smaller than e. g. in the $\rho^0$-resonance
isospin decomposition. The second less important case is for the
pairs of unflavored mesons with the isospin ${I=1}$ and
anti-nucleons. The third even less important case is for the pair
${\bar K K^*}$ and it's charge conjugate.

Let's make some comments on the errors what the above-mentioned
omissions cause in some quantities at zero chemical potentials,
which in turn demonstrate sensitivity to different changes in the
cross sections. The errors in the viscosities with the ECSs are
less than $2\%$. The errors in the shear viscosity with the EQCSs
(the TCSs) reach $57-63\%$ ($29-32\%$) at $T=0.07~GeV$. Outside
the temperature range $0.03~GeV \leq T \leq 0.14~GeV$ the errors
reach $11.6\%$ ($5.3\%$). The errors in the bulk viscosity with
the EQCSs (the TCSs) reach $14.4-15.4\%$ ($10.6-11.4\%$) at
$T=0.07~GeV$. Outside the temperature range $0.03~GeV \leq T \leq
0.13~GeV$ the errors reach $4.8\%$ ($2.1\%$). The errors in the
total number of collisions per unit time per unit volume (using
the TCSs and including the decay rates) reach $10.2\%$ (at
$T=0.07~GeV$). Outside the temperature range $0.04~GeV \leq T \leq
0.14~GeV$ the errors reach $5.1\%$. In view of the errors for the
total number of collisions the kinetic freeze-out temperatures
found in the UrQMD studies \cite{Bleicher:2002dm} should decrease,
becoming closer to the experimentally extracted ones (see
\cite{Heinz:2007in} and references therein). The chemical
freeze-out temperature may change in a less extent. This is
because both the inelastic and the quasielastic processes' cross
sections (like of the quasielastic collision of the $\pi^+ \pi^0$
pair and of the reaction $\pi^+ \pi^0 \rightarrow K^+ \bar K^0$)
increase, so that the temperature at which the inelastic processes
cease to be dominant may almost not change.

It's observed that some of the UrQMD detailed balance cross
sections (e. g. for the ${\Delta^+ \Delta^0}$ pair) are not
symmetric under the particle interchange. This is because the
function W3j, calculating the Wigner ${3-j}$ symbols, doesn't
return zero in some cases. Namely, the selection rule for the sum
${J_1+J_2+J_3}$ is not included. In principle, such omission could
result in negative values of the essentially non-negative
viscosities but, as long as only small fraction of cross sections
is affected, this omission has caused only negligibly small errors
in the viscosities. But e. g. the error in the ${\Delta^+
\Delta^0}$ TCS is approximately $25\%$.

Also some fixes of the UrQMD cross sections are made. It's found
that the ${K^+ p}$ UrQMD ECS has large deviations from the
experimental data \cite{Beringer} in the range
${1.6~GeV<\sqrt{s}<5~GeV}$ (the UrQMD ${K^+ p}$ cross section
reaches ${17~mb}$ in the region ${1.9~GeV<\sqrt{s}<3.1~GeV}$). To
fit this cross section to the experimental data it is replaced
with the AQM ECS in the range ${3~GeV<\sqrt{s}<5~GeV}$ and is
interpolated smoothly with the sine function in the range
${1.6~GeV<\sqrt{s}\leq 3~GeV}$ with the cross section being equal
to ${12.5~mb}$ at ${\sqrt{s}=1.6~GeV}$. This replacement is also
applied to other MB ECSs, when annihilation is not possible due to
the quark content.

The next fix is for the BB ECSs. It's found that the ${\Lambda p}$
UrQMD ECS has quite large deviations form the experimental data
\cite{Beringer} too. To fit this cross section to the experimental
data it is replaced with the AQM ECS in the range
${4~GeV<\sqrt{s}<5~GeV}$ and interpolated smoothly with the sine
function in the range ${2.2~GeV<\sqrt{s}\leq 4~GeV}$ with the
cross section being equal to the AQM TCS at ${\sqrt{s}=2.2~GeV}$.
This replacement is also applied to other BB ECSs.

Some other found lacks result in negligible errors in the
viscosities. However, errors in the corresponding mean free paths
and possible other quantities may be not negligible ones. Two of
such lacks can be mentioned. The first one is the following. The
${\pi p}$ and ${K N}$ cross sections are fitted to the
experimental data. And their charge conjugates are calculated
using general formulas and so cause deviations up to $50\%$ for
${\sqrt{s}>1.7~GeV}$. The second lack is the following. In some
not large energy regions with ${\sqrt{s}<1.7~GeV}$ the resonance
dominated cross sections are equal to zero for some small numbers
of pairs because there is no resonances which could be created by
this pair. These regions are replaced by a constant continuously.

Let's also comment on the deviations from the fixes described in
the last four preceding paragraphs. The altogether deviations in
the viscosities and the total number of collisions with the TCSs
are less than $0.1\%$. The altogether deviations in the
viscosities with the ECS or the EQCS are in the range $21-27\%$.
The largest contribution is from the MB cross sections' fixes. At
$T\leq 0.12~GeV$ the deviations are less than $5\%$ (the
temperatures above $T=0.27~GeV$ are not studied).

\section{ Conditions of applicability \label{condappl}}

Before proceeding forth first the applicability of the Boltzmann
equation and of the calculations of the transport coefficients
should be clarified.

Although the Boltzmann equations are valid for any perturbations
of the distribution functions they should be slowly varying
functions of the space-time coordinates to justify that they can
be considered as functions of macroscopic quantities like the
temperature, the chemical potentials or the flow velocity or, in
other words, that one can apply thermodynamics locally. Then one
can make the expansion over the independent gradients of the
thermodynamic functions and the flow velocity (the Chapman-Enskog
method), which vanish in equilibrium. Smallness of these
perturbations of the distribution functions in compare to their
leading parts ensures the validity of this expansion and that the
gradients are small\footnote{The magnitudes of thermodynamic
quantities can also be restricted by this condition or,
conversely, not restricted even if transport coefficients diverge.
See also Sec. \ref{singcomsec} of this paper. The smallness of the
shear and the bulk viscosity gradients can also be checked by the
condition of smallness of the $T^{(1)\mu\nu}$ (\ref{T1}) in
compare to the $T^{(0)\mu\nu}$ (\ref{T0}). Of course, the next
corrections should be small too.}. Because these perturbations are
inversely proportional to coupling constants one can say that they
are proportional to some product of particles' mean free paths and
the gradients. So that, in other words, the mean free paths should
be much smaller than characteristic lengths, on which the
macroscopic quantities change considerably\footnote{ It's clear
that the mean free paths should be smaller than the system's size
too. }.

As is discussed in Sec. \ref{physmeansec}, the inelastic processes
may need addition treatment in the calculations of the bulk
viscosity. There is a need to specify reasonable conditions when
the inelastic processes can be neglected. One could use the
following reliable criterion, which takes into account both the
particle number densities and the intensity of the interactions:
 \eq{
 \int_{t_1}^{t_2}dt \int_0^{V(t)} d^3x \sum_{n\in~\text{all channels}}\widetilde R^{inel}_{k',n}< 1,
 }
where $V(t)$ is the system's volume, $\sum_{n\in~\text{all
channels}}\widetilde R^{inel}_{k',n}$ is the number of reactions
of particles of the $k'$-th species\footnote{Primed indexes run
over the particle species without regard to their spin states.
This assignment is clarified more in Sec. \ref{CalcSecA}.} over
all channels per unit time per unit volume (analog of
(\ref{totratekl})), $t_1$ is chosen to satisfy the inequality, and
$t_2$ is equal to the moment of time at which the divergence of
the flow velocity is relaxed (if this time can be estimated
reliably with remained inelastic processes) or to the moment of
time at which the system becomes practically not interacting
(after expansion) because of large cumulative mean free path in
compare to the system's size. Though this criterion is likely to
be too strict, and at some higher temperatures the approximation
of conserved particle numbers should still work well. The main
alternative criterion is based on comparison of collision rates of
elastic and inelastic processes (as implemented in the
\cite{Bleicher:2002dm}). Using this criterion and some other ones,
the chemical freeze-out line\footnote{This is an approximation. In
fact this should be a range in which particles of different
particle species have their own freeze-out points.} in the
${T-\mu_B}$ plane can be built for the hadron gas, see e. g.
\cite{Cleymans:2005xv, Andronic:2005yp}. At zero chemical
potentials the chemical freeze-out temperature is approximately
equal to ${T_{ch. f.}=160-170~MeV}$. The remaining question is how
good is the approximation of only the elastic collisions at ${T
\lesssim 160~MeV}$. From the hydrodynamical description of the
elliptic flow at RHIC it's found that $\xi/s \lesssim 0.05$ near
the chemical freeze-out \cite{Dusling:2011fd}. The constant value
$\xi/s=0.04$ provides a good description of the elliptic flow both
at RHIC and LHC \cite{Bozek:2011ph}. It seems that the
approximation of conserved particle numbers is not implemented in
the bulk viscosity formula used in the \cite{nhngr}. The bulk
viscosity obtained from it is very close to the one of this paper.
These results support the choice of the approximation of only the
elastic collisions at ${T\lesssim 160~MeV}$ and show that the
deviations are likely no more than in 2-3 times. Anyway, the
numerical calculations by the Kubo formula through simulations of
collisions are desirable along and around the chemical freeze-out
line for more accurate calculations (though the procedure of
collisions of particles introduce some errors itself
\cite{Bass:1998ca}, which should be kept in mind).

Errors due to the Maxwell-Boltzmann statistics, used instead of
the Bose-Einstein or the Fermi-Dirac ones, were found to be small
for the vanishing chemical potentials\footnote{It should be
mentioned that if the particles of the $k$-th particle species are
bosons and if $\mu_k(x^\mu) \geq m_k$ then there is a (local)
Bose-Einstein condensation for them, which should be treated in a
special way.}. According to calculations for the pion gas in
\cite{davesne}, the bulk viscosity becomes $25\%$ larger at
${T=120~MeV}$ and $33\%$ larger at ${T=200~MeV}$ for the vanishing
chemical potential. Although the relative deviations of the
thermodynamic quantities of the pion gas at the nonvanishing
chemical potential ${\mu=100~MeV}$ are not more than
$20\%$\footnote{The relative deviations of the thermodynamic
quantities grow with the temperature for some fixed value of the
chemical potential and tend to some constant.} the bulk viscosity
becomes up to $2.5$ times more. The shear viscosity becomes $15\%$
less at ${T=120~MeV}$ and $25\%$ less at ${T=200~MeV}$ for the
vanishing chemical potential and $33\%$ less at ${T=120~MeV}$ and
$67\%$ less at ${T=200~MeV}$ for the ${\mu=100~MeV}$. The
corrections to the bulk viscosity of the fermion gas, according to
calculations of the bulk viscosity source term, not presented in
this paper, are of the opposite sign and approximately of the same
magnitude. So that for the hadron gas the error due to the used
classical statistics can be even smaller than for the pion gas.

The numerical calculations in Sec. \ref{numcalc} of the
viscosities with the total cross sections justify the choice of
one constant cross section for all hadrons. It's approximately
equal to ${20~mb}$, corresponding to the effective radius
${r=0.4~fm}$ (as given by the (\ref{hccs})), which is used in the
estimations below.

The condition of applicability of the ideal gas equation of state
is controlled by the dimensionless parameter $\upsilon n$ which
appears in the first correction from the binary collisions in the
virial expansion and should be small. Here ${\upsilon=16\pi
r^3/3}$ is the so called excluded volume parameter and $1/n$ is
the mean volume per particle. One finds ${\upsilon n\approx 0.09}$
at ${T=120~MeV}$, ${\upsilon n\approx 0.2}$ at ${T=140~MeV}$ and
${\upsilon n\approx 1}$ at ${T=180~MeV}$ for the vanishing
chemical potentials. Along the chemical freeze-out line (its
parametrization can be found in \cite{Gorenstein:2007mw}) the
$\upsilon n$ grows from $0.07$ to $0.49$ with the temperature.
From comparison with lattice calculations \cite{Borsanyi:2010cj}
one can find that the corrections to the ideal gas equation of
state are small at ${T \lesssim 140~MeV}$. One could suspect that
even small corrections to the thermodynamic quantities can result
in large corrections for the bulk viscosity, though this seems to
be not the case. The errors in the bulk viscosity from the
scale-violating contributions of the hadrons' masses are less than
the errors from the contributions to the trace of the
energy-momentum tensor (for more details see Sec. \ref{numcalc}).

One more important requirement, which one needs to justify the
Boltzmann equation approach, is that the mean free time should be
much larger than $\hbar/\Omega$ (the $\Omega$ is the
characteristic single-particle energy) \cite{danielewicz} or the
de Broglie wavelength should be much smaller than the mean free
path \cite{Arnold:2002zm} to distinguish independent acts of
collisions and for particles to have well-defined on-shell energy
and momentum. This condition gets badly satisfied for high
temperatures or densities. The mean free path of the particle
species $k'$ is given by the formula (\ref{mfp}) or the formula
(\ref{lel}) if the inelastic processes can be neglected. The
wavelength can be written as ${\lambda_{k'}\approx\frac1{\langle
\abs{\vec p_{k'}}\rangle}}$, where the averaged modulus of the
momentum of the $k'$-th species $\langle \abs{\vec p_{k'}}\rangle$
is
 \eq{\label{avmom}
 \langle\abs{\vec p_{k'}}\rangle=\frac{\int d^3p_{k'} |\vec p_{k'}|
 f^{(0)}_{k'}(p_{k'})}{\int d^3p_{k'} f^{(0)}_{k'}(p_{k'})}=\frac{2
 e^{-z_{k'}} T (3 + 3z_{k'} + z_{k'}^2)}{z_{k'}^2
 K_2(z_{k'})}=\sqrt{\frac{8 m_{k'} T}{\pi}}
 \frac{K_{5/2}(z_{k'})}{K_2(z_{k'})},
 }
where ${z_{k'}\equiv m_{k'}/T}$, $K_2(x)$ is the modified Bessel
function of the second kind. As it follows from the (\ref{avmom})
the largest wavelength is for the lightest particles, the
$\pi$-mesons. The elastic collision mean free paths are close to
each other for all particle species. Hence, the smallest value of
the ratio $\lambda_{k'}/l^{el}_{k'}$ is for the $\pi$-mesons. Its
value is close to the value of the $\upsilon n$ and is
exponentially suppressed for small temperatures too. At the
temperature ${T=140~MeV}$ (${180~MeV}$) and the vanishing chemical
potentials this ratio is equal to 0.18 (0.7). Along the chemical
freeze-out line it grows from $0.12$ to $0.37$ with the
temperature.

To go beyond these conditions one can use the Kubo (or Green-Kubo)
formulas, for instance. In the \cite{jeon} the Kubo formulas were
used to perform perturbative calculations of the viscosities in
the leading order. Basing on this result, an example of effective
weakly coupled kinetic theory of quasiparticle excitations with
thermal masses and thermal scattering amplitudes was presented in
the \cite{jeonyaffe}. There the function $U(q)$ (appearing because
of the temperature dependence of the mass) takes into account the
next in the coupling constant correction to the energy-momentum
tensor and the equation of state\footnote{In the hadron gas it's
believed that the vacuum masses are large in compare to their
thermal corrections for the most of the hadrons at temperatures
$T\lesssim 200~MeV$ or even higher ones. Then expanding over the
thermal correction in the matrix elements, one would get even
smaller corrections than the ones to the equation of state in
coupling constants (because of coupling constants next to the
matrix elements) in a perturbation theory, e. g. chiral
perturbation theory.}. For further developments see
\cite{Arnold:2002zm,Gagnon:2006hi,Gagnon:2007qt}. For some other
approaches see
\cite{Blaizot:1992gn,Calzetta:1986cq,Calzetta:1999ps} and
\cite{Arnold:1997gh} with references therein.

\section{ Details of calculations \label{CalcSec} }
\subsection{ The Boltzmann equation and its solution \label{CalcSecA} }

The calculations in this paper go close to the ones in
\cite{groot} though with some differences and generalizations.
Let's start from some definitions. Multi-indices $k,l,m,n$ will be
used to denote particle species with certain spin states. Indexes
$k',l',m',n'$ will be used to denote particle species without
regard to their spin states (and run from 1 to the number of the
particle species $N'$) and $a,b$ to denote conserved quantum
numbers\footnote{In systems with only the elastic collisions each
particle species have their own "conserved quantum number", equal
to 1.}. Quantifiers $\forall$ with respect to the indexes are
omitted in the text where they may be needed which won't result in
a confusion. Because nothing depends on spin variables one has for
every sum over the multi-indexes
 \eq{
  \sum_k ... = \sum_{k'}g_{k'}...,
 }
where $g_{k'}$ is the spin degeneracy factor. The following
assignments will be used:
 \begin{eqnarray}\label{assign1}
  \nonumber n\equiv\sum_{k}n_k&\equiv&\sum_{k'} n_{k'}, \quad n_a\equiv \sum_k q_{ak} n_k,
  \quad x_k\equiv\frac{n_k}{n}, \quad x_{k'}\equiv\frac{n_{k'}}{n}, \quad x_a\equiv\frac{n_a}{n}, \\
  \hat \mu_k&\equiv&\frac{\mu_k}{T}, \quad \hat \mu_a\equiv\frac{\mu_a}{T},
  \quad z_k\equiv\frac{m_k}{T}, \quad \pi_k^\mu\equiv \frac{p_k^\mu}{T},
  \quad \tau_k\equiv \frac{p_k^\mu U_\mu}{T},
 \end{eqnarray}
where $q_{ak}$ denotes values of conserved quantum numbers of the
$a$-th kind of the $k$-th particle species. Everywhere the
particle number densities are summed, the spin degeneracy factor
$g_{k'}$ appears and then gets absorbed into the $n_{k'}$ or the
$x_{k'}$ by the definition. All other quantities with primed and
unprimed indexes don't differ, except for rates, the mean free
times and the mean free paths defined in Appendix \ref{appmfp},
the $\gamma_{kl}$ commented below, the coefficients
$A_{k'l'}^{rs}$, $C_{k'l'}^{rs}$ and, of course, quantities, whose
free indexes set the indexes of the particle number densities
$n_k$. Also the assignment ${\int \frac{d^3p_k}{p_k^0}\equiv
\int_{p_k}}$ will be used for compactness somewhere.

The particle number flows are\footnote{The $+,-,-,-$ metric
signature is used throughout the paper.}
 \eq{\label{pflow}
 N^{\mu}_k=\int \frac{d^3p_k}{(2\pi)^3p^0_k} p^\mu_k f_k,
 }
where the assignment ${f_k(p_k)\equiv f_k}$ is introduced. The
energy-momentum tensor is
 \eq{\label{enmomten}
 T^{\mu\nu}=\sum_k \int \frac{d^3p_k}{(2\pi)^3p_k^0}p_k^\mu p_k^\nu f_k.
 }
The local equilibrium distribution functions are
 \eq{\label{loceq}
 f^{(0)}_k=e^{(\mu_k-p_k^\mu U_\mu)/T},
 }
where $\mu_k$ is the chemical potential of the $k$-th particle
species, $T$ is the temperature and $U_\mu$ is the relativistic
flow 4-velocity such that ${U_\mu U^\mu=1}$ (with a frequently
used consequence ${U_\mu\p_\nu U^\mu=0}$). The local equilibrium
is considered as perturbations of independent thermodynamic
variables and the flow velocity over a global equilibrium such
that they can depend on the space-time coordinate $x^\mu$.
Additional chemical perturbations could also be considered, but
they don't enter in the first order transport coefficients if they
are small, as is discussed in Sec. \ref{physmeansec}. The chemical
equilibrium implies that the particle number densities are equal
to their global equilibrium values. The global equilibrium is
called the time-independent stationary state with the maximal
entropy\footnote{The kinetic equilibrium implies that the momentum
distributions are the same as in the global equilibrium. Thus, a
state of a system with both the pointwise (for the whole system)
kinetic and the pointwise chemical equilibria is the global
equilibrium.}. The global equilibrium of an isolated system can be
found by variation of the total nonequilibrium entropy functional
\cite{landau5} over the distribution functions with condition of
the total energy and the total net charges conservation:
 \eq{
 U[f]=\sum_k \int \frac{d^3p_kd^3x}{(2\pi)^3}
 f_k(1-\ln f_k)-\sum_k\int \frac{d^3p_kd^3x}{(2\pi)^3}\beta
 p^0_kf_k-\sum_{a,k} \lambda_a q_{ak} \int \frac{d^3p_kd^3x}{(2\pi)^3} f_k,
 }
where $\beta, \lambda_a$ are the Lagrange coefficients. Equating
the first variation to zero, one easily gets the function
(\ref{loceq}) with ${U^\mu=(1,0,0,0)}$, ${\beta=\frac1{T}}$ and
 \eq{\label{mukdef}
 \mu_k=\sum_a q_{ak}\mu_a,
 }
where ${\mu_a=\lambda_a}$ are the independent chemical potentials
coupled to the conserved net charges.

With ${f_k=f_k^{(0)}}$, substituted in the (\ref{pflow}) and the
(\ref{enmomten}), one gets the leading contribution in the
gradients expansion of the particle number flow and the
energy-momentum tensor:
 \eq{
 N^{(0)\mu}_k=n_kU^\mu,
 }
 \eq{\label{T0}
 T^{(0)\mu\nu}=\epsilon U^\mu U^\nu - P\Delta^{\mu\nu},
 }
where the projector
 \eq{\label{proj}
 \Delta^{\mu\nu}\equiv g^{\mu\nu}-U^\mu U^\nu,
 }
is introduced. The $n_k$ is the ideal gas particle number density,
 \eq{\label{ignk}
 n_k=U_\mu N^{(0)\mu}_k=\frac{1}{2\pi^2}T^3z_k^2K_2(z_k)e^{\hat \mu_k},
 }
the $\epsilon$ is the ideal gas energy density,
 \eq{\label{epsandek}
 \epsilon=U_\mu U_\nu T^{(0)\mu\nu}=\sum_k
 \int \frac{d^3p_k}{(2\pi)^3}p_k^0f_k^{(0)}=\sum_k n_k e_k, \quad
 e_k\equiv m_k\frac{K_3(z_k)}{K_2(z_k)}-T,
 }
and the $P$ is the ideal gas pressure,
 \eq{\label{pressure}
 P=-\frac13T^{(0)\mu\nu}\Delta_{\mu\nu}=\sum_k \frac13\int
 \frac{d^3p_k}{(2\pi)^3p_k^0}\vec{p_k}^2 f_k^{(0)}=\sum_k n_k T=nT.
 }
Also the following assignments are used:
 \begin{eqnarray}\label{assign2}
  \nonumber e&\equiv&\frac{\epsilon}{n}=\sum_kx_ke_k, \quad h_k\equiv e_k+T,
  \quad h\equiv\frac{\epsilon+P}{n}=\sum_k x_kh_k, \\
  \hat e_k&\equiv&\frac{e_k}{T}=z_k\frac{K_3(z_k)}{K_2(z_k)}-1,
  \quad \hat e\equiv\frac{e}{T}, \quad \hat h_k\equiv\frac{h_k}{T}=
  z_k\frac{K_3(z_k)}{K_2(z_k)}, \quad \hat h\equiv \frac{h}{T}.
 \end{eqnarray}
Above $h$ is the enthalpy per particle, $e$ is the energy per
particle and $h_k$, $e_k$ are the enthalpy and the energy per
particle of the $k$-th particle species correspondingly, which are
well defined in the ideal gas.

In the relativistic hydrodynamics the flow velocity $U^\mu$ needs
somewhat extended definition. The most convenient condition which
can be applied to the $U^\mu$ is the Landau-Lifshitz condition
\cite{landau6} (Section 136). This condition states that in the
local rest frame (where the flow velocity is zero though its
gradient can have a nonzero value) each imaginary infinitesimal
cell of fluid should have zero momentum, and its energy density
and the charge density should be related to other thermodynamic
quantities through the equilibrium thermodynamic relations
(without a contribution of nonequilibrium dissipations). Its
covariant mathematical formulation is
 \eq{\label{lLcond}
 (T^{\mu\nu}-T^{(0)\mu\nu})U_\mu=0, \quad (N^\mu_a-N^{(0)\mu}_a)U_\mu=0.
 }
The next to leading correction over the gradients expansion to the
$T^{\mu\nu}$ can be written as an expansion over the 1-st order
Lorentz covariant gradients, which are rotationally and space
inversion invariant and satisfy the Landau-Lifshitz
condition\footnote{ Also this form of $T^{(1)\mu\nu}$ respects the
second law of thermodynamics \cite{landau6} (Section 136). }
(\ref{lLcond}):
 \eq{\label{T1}
 T^{(1)\mu\nu}\equiv2\eta \overset{\circ}{\overline{\nabla^\mu
 U^\nu}}+\xi \Delta^{\mu\nu} \nabla_\rho U^\rho=\eta\left(\Delta^\mu_\rho \Delta^\nu_\tau
 +\Delta^\nu_\rho \Delta^\mu_\tau-\frac23\Delta^{\mu\nu}\Delta_{\rho\tau}\right)\nabla^\rho
 U^\tau+\xi \Delta^{\mu\nu} \nabla_\rho U^\rho,
 }
where for any tensor $a_{\mu\nu}$ the symmetrized traceless tensor
assignment is introduced:
 \eq{\label{tracelessten}
 \overset{\circ}{\overline{a_{\mu\nu}}}\equiv \left(\frac{\Delta_{\mu\rho}
 \Delta_{\nu\tau}+\Delta_{\nu\rho} \Delta_{\mu\tau}}2-\frac13\Delta_{\mu\nu}
 \Delta_{\rho\tau}\right)a^{\rho\tau}\equiv \Delta_{\mu\nu\rho\tau}a^{\rho\tau},
 \quad \Delta^{\mu\nu}_{~~\rho\tau}\Delta^{\rho\tau}_{~~\sigma\lambda}=
 \Delta^{\mu\nu}_{~~\sigma\lambda}.
 }
The equation (\ref{T1}) is the definition of the shear $\eta$ and
the bulk $\xi$ viscosity coefficients. The $\xi \Delta^{\mu\nu}
\nabla_\rho U^\rho$ term in the (\ref{T1}) can be considered as a
nonequilibrium contribution to the pressure which enters in the
(\ref{T0}).

By means of the projector (\ref{proj}) one can split the
space-time derivative $\p_\mu$ as
 \eq{
 \p_\mu=U_\mu U^\nu \p_\nu + \Delta_\mu^\nu\p_\nu = U_\mu
 D+\nabla_\mu,
 }
where $D \equiv U^\nu \p_\nu$, $\nabla_\mu \equiv
\Delta_\mu^\nu\p_\nu$. In the local rest frame (where
${U^\mu=(1,0,0,0)}$) the $D$ becomes the time derivative and the
$\nabla_\mu$ becomes the spacial derivative. Then the Boltzmann
equations can be written in the form
 \eq{\label{boleqs}
 p_k^\mu\p_\mu f_k=(p_k^\mu U_\mu D + p_k^\mu \nabla_\mu )f_k
 =C_k^{el}[f_k]+C_k^{inel}[f_k],
 }
where $C_k^{inel}[f_k]$ represents the inelastic or
number-changing collision integrals (it is omitted in calculations
in this paper if the opposite is not stated explicitly) and
$C_k^{el}[f_k]$ is the elastic ${2\leftrightarrow2}$ collision
integral. The collision integral $C_k^{el}[f_k]$ has the form of
the sum of positive gain terms and negative loss terms. Its
explicit form is\footnote{The factor $\gamma_{kl}$ cancels double
counting in integration over momentums of identical particles. The
factor $\frac12$ comes from the relativistic normalization of the
scattering amplitudes. } (cf. \cite{jeon, Arnold:2002zm})
 \begin{eqnarray}\label{ckel}
  \nonumber C_k^{el}[f_k]&=&\sum_{l} \gamma_{kl}\frac12\int
  \frac{d^3p_{1l}}{(2\pi)^32p_{1l}^0}\frac{d^3p'_k}{(2\pi)^32{p'}_k^0}
  \frac{d^3p'_{1l}}{(2\pi)^32{p'}_{1l}^0}(f'_{k}f'_{1l}-f_{k}f_{1l})\\
  &\times& |\mc M_{kl}|^2(2\pi)^4\delta^{4}(p'_k+p'_{1l}-p_k-p_{1l}),
 \end{eqnarray}
where ${\gamma_{kl}=\frac12}$ if $k$ and $l$ denote the same
particle species without regard to the spin states and
${\gamma_{kl}=1}$ otherwise, ${|\mc M_{kl} (p'_k,p'_{1l};
p_k,p_{1l})|^2 \equiv |\mc M_{kl}|^2}$ is the square of the
dimensionless elastic scattering amplitude averaged over the
initial spin states and summed over the final ones. Index $1$
designates that $p_k$ and $p_{1k}$ are different variables.
Introducing ${W_{kl}\equiv W_{kl}(p'_k,p'_{1l};p_k,p_{1l})}$ as
 \eq{
 W_{kl}=\frac{|\mc M_{kl}|^2}{64\pi^2}\delta^{4}(p'_k+p'_{1l}-p_k-p_{1l}),
 }
one can rewrite the collision integral (\ref{ckel}) in the form as
in \cite{groot} (Chap. I, Sec. 2)
 \eq{\label{ckelgroot}
 C_k^{el}[f_k]=(2\pi)^3\sum_{l}\gamma_{kl}
 \int_{p_{1l},{p'}_k,{p'}_{1l}}\left(\frac{f'_{k}}{(2\pi)^3}\frac{f'_{1l}}{(2\pi)^3}
 -\frac{f_{k}}{(2\pi)^3}\frac{f_{1l}}{(2\pi)^3}\right)W_{kl}.
 }
The $W_{kl}$ is related to the elastic differential cross section
$\sigma_{kl}$ as \cite{groot} (Chap. I, Sec. 2)
 \eq{
 W_{kl}=s\sigma_{kl}\delta^{4}(p'_k+p'_{1l}-p_k-p_{1l}),
 }
where ${s=(p_k+p_{1l})^2}$ is the usual Mandelstam variable. The
$W_{kl}$ has properties $W_{kl}(p'_k,p'_{1l};p_k,p_{1l}) =
W_{kl}(p_k,p_{1l};p'_{k},p'_{1l}) =
W_{lk}(p'_{1l},p'_{k};p_{1l},p_k)$ (due to time reversibility and
a freedom of relabelling of order numbers of particles taking part
in reaction). And e.g. $W_{kl}(p'_k,p'_{1l};p_k,p_{1l}) \neq
W_{kl}(p'_{1l},p'_{k};p_{1l},p_{k})$ in the general case. The
elastic collision integrals have important properties which one
can easily prove \cite{groot} (Chap. II, Sec. 1):
 \eq{\label{c22prop0}
 \int \frac{d^3p_k}{(2\pi)^3p^0_k} C_k^{el}[f_k]=0,
 }
 \eq{\label{c22prop}
 \sum_k\int \frac{d^3p_k}{(2\pi)^3p^0_k} p_k^\mu C_k^{el}[f_k]=0.
 }
Also the $C^{el}_k[f_k]$ vanishes if ${f_k=f^{(0)}_k}$.

The distribution functions $f_k$ solving the system of the
Boltzmann equations approximately are sought in the form
 \eq{\label{fpert}
 f_k=f^{(0)}_k+f^{(1)}_k\equiv f^{(0)}_k+f^{(0)}_k\varphi_k(x,p_k),
 }
where it's assumed that $f_k$ depend on the $x^\mu$ entirely
through the $T$, $\mu_k$, $U^\mu$ or their space-time derivatives.
Also it is assumed that ${|\varphi_k|\ll 1}$. After substitution
of ${f_k=f^{(0)}_k}$ in the (\ref{boleqs}) the r. h. s. becomes
zero and the l. h. s. is zero only if the $T$, $\mu_k$ and $U^\mu$
don't depend on the $x^\mu$ (provided they don't depend on the
momentum $p^\mu_k$). The 1-st order space-time derivatives of the
$T$, $\mu_k$, $U^\mu$ in the l. h. s. should be cancelled by the
first nonvanishing contribution in the r. h. s. This means that
the $\varphi_k$ should be proportional to the 1-st order
space-time derivatives of the $T$, $\mu_k$, $U^\mu$. The covariant
time derivatives $D$ can be expressed through the covariant
spacial derivatives by means of approximate hydrodynamic
equations, valid at the same order in the gradients expansion.
Let's derive them. Integrating the (\ref{boleqs}) over the
$\frac{d^3p_k}{(2\pi)^3p^0_k}$ with the ${f_k=f^{(0)}_k}$ in the
l. h. s. with the inelastic collision integrals retained and using
the (\ref{c22prop0}) and the (\ref{pflow}) one would get (which
can be justified using explicit form of the inelastic collision
integrals)
 \eq{\label{conteq}
 \p_\mu N^{(0)\mu}_k=Dn_k+n_k\nabla_\mu U^\mu=I_k,
 }
where $I_k$ is the sum of the inelastic collision integrals
integrated over the momentum. It is responsible for the
nonconservation of the total particle number of the $k$-th
particle species and has the property ${\sum_k q_{ak} I_k=0}$. If
${C^{inel}_k[f_k]=0}$, then ${I_k=0}$ which results in
conservation of the total particle numbers of each particle
species. Multiplying the (\ref{conteq}) on the $q_{ak}$ and
summing over $k$ one gets the continuity equations for the net
charge flows:
 \eq{\label{conteq2}
 \p_\mu N^{(0)\mu}_a=Dn_a+n_a\nabla_\mu U^\mu=0.
 }
Then integrating the (\ref{boleqs}) over the
$p_k^\mu\frac{d^3p_k}{(2\pi)^3p^0_k}$ with the ${f_k=f^{(0)}_k}$
in the l. h. s. one gets
 \eq{\label{encons0}
 \p_\rho T^{(0)\rho\nu}=\p_\rho(\epsilon U^\rho U^\nu-P\Delta^{\rho\nu})=0.
 }
There is zero in the r. h. s. even if the inelastic collision
integrals are retained because they respect energy conservation
too. Note that the Boltzmann equations (\ref{boleqs}) (without any
thermal corrections) permit a self-consistent description only if
the energy-momentum tensor and the net charge flows of the ideal
gas are used. After the convolution of the (\ref{encons0}) with
the $\Delta^\mu_\nu$ one gets the Euler's equation:
 \eq{\label{eulereq}
 DU^\mu=\frac1{\epsilon+P}\nabla^\mu P=\frac1{hn}\nabla^\mu P.
 }
After the convolution of the (\ref{encons0}) with the $U_\nu$ one
gets equation for the energy density:
 \eq{\label{encons1}
 D\epsilon=-(\epsilon+P)\nabla_\mu U^\mu = - hn\nabla_\mu U^\mu.
 }

To proceed farther one needs to expand the l. h. s. of the
Boltzmann equations (\ref{boleqs}) over the gradients of
thermodynamic variables and the flow velocity. Let's choose the
$\mu_a$ and the $T$ as the independent thermodynamic variables.
Then for the $Df^{(0)}_k$ one can write the expansion
 \eq{\label{Dfk}
 Df^{(0)}_k=\sum_a \frac{\p f^{(0)}_k}{\p\mu_a}D\mu_a+\frac{\p f^{(0)}_k}{\p
 T}DT+\frac{\p f^{(0)}_k}{\p U^\mu}DU^\mu.
 }
Writing the expansion for the $Dn_a$ and the $D\epsilon$ one gets
from the (\ref{conteq2}) and the (\ref{encons1}):
 \eq{\label{Dna}
 Dn_a=\sum_b \frac{\p n_a}{\p \mu_b}D\mu_b+\frac{\p n_a}{\p T}DT
 =-n_a\nabla_\mu U^\mu,
 }
 \eq{\label{Depsilon}
 D\epsilon=\frac{\p\epsilon}{\p T}DT+\sum_a\frac{\p \epsilon}{\p \mu_a}D\mu_a
 =-hn\nabla_\mu U^\mu.
 }
The solution to the system of equations (\ref{Dna}),
(\ref{Depsilon}) can be found easily:
 \eq{\label{Teqn}
 DT=-R T\nabla_\mu U^\mu,
 }
 \eq{\label{mueqn}
 D\mu_a=T\sum_b \tilde{A}^{-1}_{ab}(R B_b-x_b)\nabla_\mu U^\mu,
 }
where
 \eq{\label{Rdef}
 R\equiv\frac{\hat h-\sum_{a,b} E_a \tilde{A}^{-1}_{ab}x_b}{C_{\{\mu\}}-\sum_{a,b}
 E_a\tilde{A}^{-1}_{ab}B_b},
 }
and
 \eq{
 \frac{\p n_a}{\p \mu_b}\equiv \frac{n}{T}\tilde{A}_{ab},
 \quad \frac{\p n_a}{\p T}\equiv \frac{n}{T}B_a, \quad
 \frac{\p\epsilon}{\p T}\equiv n C_{\{\mu\}},\quad
 \frac{\p\epsilon}{\p\mu_a}\equiv n E_a.
 }
Above it is assumed that the matrix $\tilde{A}_{ab}$ is not
degenerate\footnote{One can prove that the $N'' \times N''$ matrix
$\tilde{A}_{ab}$ in (\ref{ABCE}) is not degenerate if there are
$N''$ linearly independent conserved charges. Then one can prove
that the denominator in the (\ref{Rdef2}) is not zero.}, which is
related to the self-consistency of the statistical description of
the system. Using the ideal gas formulas (\ref{ignk}) and
(\ref{epsandek}) one gets
 \begin{eqnarray}\label{ABCE}
  &~&\tilde A_{ab}=\sum_k q_{ak}q_{bk}x_k, \quad
  E_a=\sum_k q_{ak} x_k \hat e_k, \quad B_a=E_a-\sum_b \tilde A_{ab}\hat\mu_b,\\
  \nonumber C_{\{\mu\}}&=&\sum_k x_k(3\hat h_k+z_k^2-\hat \mu_k\hat
  e_k)=\sum_kx_k(3\hat h_k+z_k^2)-\sum_aE_a\hat\mu_a \equiv
  \widetilde C_{\{\mu\}}-\sum_aE_a\hat\mu_a,
 \end{eqnarray}
and simplified expressions for the $R$ and the $D\hat\mu_a$
 \eq{\label{Rdef2}
 R=\frac{\hat h-\sum_{a,b} E_a \tilde{A}^{-1}_{ab}x_b}{\widetilde C_{\{\mu\}}-\sum_{a,b}
 E_a\tilde{A}^{-1}_{ab}E_b},
 }
 \eq{
 D\hat\mu_a=\sum_b\tilde{A}^{-1}_{ab}(R E_b-x_b)\nabla_\mu U^\mu.
 }
For the special case of the vanishing chemical potentials,
$\mu_a\rightarrow 0$, (for a chargeless system the result is the
same) the quantities $n_a$, $x_a$, $B_a$, $E_a$ tend to zero
because the contributions from particles and anti-particles cancel
each other and the chargeless particles don't contribute. Then
from the (\ref{Teqn}) and the (\ref{mueqn}) one finds
 \eq{\label{Teqnmu0}
 DT|_{\mu_a=0}=-\frac{h}{\widetilde C_{\{\mu\}}}\nabla_\mu U^\mu,
 }
 \eq{
 D\mu_a|_{\mu_a=0}=0.
 }
This means that for the vanishing chemical potentials one can
simply exclude them from the distribution functions (if one does
not study diffusion or thermal conductivity). In systems with only
the elastic collisions each particle has its own charge so that
one takes ${q_{ak}=\delta_{ak}}$ and gets
 \begin{eqnarray}\label{elquant}
  \nonumber \tilde A_{kl}&=&\delta_{kl}x_k, \quad B_k=x_k(\hat e_k-\hat \mu_k),
  \quad E_k=\hat e_k x_k, \quad R=\frac1{c_\upsilon}, \\
  C_{\{\mu\}}&-&\sum_{a,b}E_a \tilde A^{-1}_{ab} B_{b}=\sum_kx_k(-\hat h_k^2+5\hat
  h_k+z_k^2-1)\equiv\sum_kx_k c_{\upsilon,k}\equiv c_\upsilon.
 \end{eqnarray}
Then the equation for the $DT$ (\ref{Teqn}) remains the same with
a new $R$ from the (\ref{elquant}), and the equations
(\ref{mueqn}) become
 \eq{\label{mukeqn}
 D\mu_k=\left(\frac{T}{c_\upsilon}(\hat e_k-\hat \mu_k)-T\right)\nabla_\mu U^\mu.
 }
Note that in systems with only the elastic collisions the $D\mu_k$
does not tend to zero for the vanishing chemical potentials so
that the $\mu_k$ could not be omitted in the distribution
functions in this case. Because the heat conductivity and
diffusion are not considered in this paper their nonequilibrium
gradients are taken equal to zero, $\nabla_\nu P=\nabla_\nu
T=\nabla_\nu\mu_a=0$. Using the (\ref{Teqn}), (\ref{mueqn}) and
(\ref{eulereq}) the l. h. s. of the (\ref{boleqs}) can be
transformed as
 \eq{\label{boleqnlhs}
 (p_k^\mu U_\mu D+p_k^\mu\nabla_\mu)f_k^{(0)} = -Tf_k^{(0)}
 \pi_k^\mu \pi_k^\nu \overset{\circ}{\overline{\nabla_\mu
 U_\nu}}+Tf_k^{(0)}\hat Q_k\nabla_\rho U^\rho,
 }
where
 \eq{\label{Qsource}
 \hat Q_k\equiv\tau_k^2\left(\frac13-R\right)+\tau_k\sum_{a,b}q_{ak}
 \tilde A^{-1}_{ab}(RE_b-x_b)-\frac13z_k^2.
 }
Using the (\ref{tracelessten}) one can notice that the useful
equality ${ \pi_k^\mu \pi_k^\nu \overset{\circ}{
\overline{\nabla_\mu U_\nu} } = \overset{\circ}{ \overline{
\pi_k^\mu \pi_k^\nu } }\overset{\circ}{ \overline{ \nabla_\mu
U_\nu } } }$ holds. In systems with only the elastic collisions
the $\hat Q_k$ simplifies in agreement with \cite{groot} (Chap. V,
Sec. 1):
 \eq{\label{Qsource2}
 \hat Q_k=\left(\frac43-\gamma\right)\tau_k^2+
 \tau_k((\gamma-1)\hat h_k-\gamma)-\frac13z_k^2.
 }
where the assignments $\gamma$ from \cite{groot} is used. It can
be expressed through the $c_\upsilon$, defined in the
(\ref{elquant}), as ${\gamma\equiv \frac1{c_\upsilon}+1}$.
Introducing symmetric round brackets
 \eq{
 (F,G)_k\equiv\frac1{4\pi z_k^2K_2(z_k)T^2}\int_{p_k} F(p_k)G(p_k)e^{-\tau_k}.
 }
and assignments
 \eq{
 \alpha_k^r\equiv(\hat Q_k,\tau_k^r), \quad \gamma_k^r\equiv(\tau_k^r
 \overset{\circ}{\overline{\pi_k^\mu\pi_k^\nu}},
 \overset{\circ}{\overline{\pi_{k\mu}\pi_{k\nu}}}), \quad
 a^r_k\equiv(1,\tau_k^r)_k,
 }
and using explicit expressions of the $a_k^r$ from Appendix
\ref{appA} one finds for the $\alpha_k^0$ and the $\alpha_k^1$ in
systems with elastic and inelastic collisions
 \eq{\label{alphak0}
 \alpha_k^0=1+\sum_{a,b}q_{ak} \tilde A^{-1}_{ab}(R E_b-x_b)-\hat e_k R,
 }
 \eq{\label{alphak1}
 \alpha_k^1=\hat h_k+\sum_{a,b}\hat e_k q_{ak}\tilde A^{-1}_{ab}(R E_b-x_b)
 -(3\hat h_k+z_k^2)R.
 }
Then using the (\ref{alphak0}) and the (\ref{alphak1}) one can
show that
 \eq{\label{lhsnchcons}
 \sum_kq_{ak}x_k\alpha_k^0=0,
 }
 \eq{\label{lhsencons}
 \sum_kx_k\alpha_k^1=0.
 }
Because the gradients $\nabla_\mu U^\mu$ and
$\overset{\circ}{\overline{\nabla_\mu U_\nu}}$ are independent the
(\ref{lhsnchcons}) and the (\ref{lhsencons}) are direct
consequences of the local net charge (\ref{conteq2}) and the
energy-momentum (\ref{encons0}) conservations. Quantities
$(1,\overset{\circ}{\overline{\pi^\mu_{k}\pi^\nu_{k}}})$ and
$(p_k^\lambda,\overset{\circ}{\overline{\pi^\mu_{k}\pi^\nu_{k}}})$
vanish automatically because of the special tensorial
structure\footnote{ Direct computation gives $(1,\overset{\circ}{
\overline{\pi^\mu_k \pi^\nu_k} })_k \propto (C_1 U^\sigma U^\rho +
C_2\Delta^{\sigma\rho}) \Delta_{ ~~\sigma\rho }^{\mu\nu} = 0$,
$(p_k^\lambda, \overset{\circ}{ \overline{\pi^\mu_k \pi^\nu_k}
})_k \propto (C_1 U^\lambda U^\sigma U^\rho + C_2U^\lambda
\Delta^{\sigma\rho} + C_3U^\sigma \Delta^{\lambda\rho})
\Delta_{~~\sigma\rho}^{\mu\nu} = 0$.} of the $\overset{\circ}{
\overline{ \pi^\mu_{k} \pi^\nu_{k}} }$.

The next step is to transform the r. h. s. of the Boltzmann
equations (\ref{boleqs}). After the substitution of the
(\ref{fpert}) in the r. h. s. of the (\ref{boleqs}) the collision
integral becomes linear and one gets
 \eq{\label{boleqnrhs}
 C_k^{el}[f_k]\approx -f_k^{(0)}\sum_l \mc L_{kl}^{el}[\varphi_k],
 }
where
 \eq{
 \mc L_{kl}^{el}[\varphi_k]\equiv\frac{\gamma_{kl}}{(2\pi)^3}\int_{p_{1l},{p'}_k,{p'}_{1l}}
 f_{1l}^{(0)}(\varphi_k+\varphi_{1l}-\varphi'_k-\varphi'_{1l})W_{kl}.
 }
The unknown functions $\varphi_k$ are sought in the form
 \eq{\label{varphi}
 \varphi_k=\frac1{n\sigma(T)}\left(-A_k(p_k)\nabla_\mu U^\mu+C_k(p_k)
 \overset{\circ}{\overline{\pi^\mu_k \pi^\nu_k}}
 \overset{\circ}{\overline{\nabla_\mu U_\nu}}\right),
 }
where $\sigma(T)$ is some formal averaged cross section, used to
come to dimensionless quantities. Then using the (\ref{boleqnlhs})
and the (\ref{boleqnrhs}), and the fact that the gradients
$\nabla_\mu U^\mu$ and $\overset{\circ}{\overline{\nabla_\mu
U_\nu}}$ are independent, the Boltzmann equations can be written
as independent integral equations:
 \eq{\label{xieqn}
 \hat Q_k=\sum_l x_l L_{kl}^{el}[A_k],
 }
 \eq{\label{etaeqn}
 \overset{\circ}{\overline{\pi_{k}^\mu \pi_{k}^\nu}}=
 \sum_l x_l L_{kl}^{el}[C_k \overset{\circ}{\overline{\pi_{k}^\mu \pi_{k}^\nu}}],
 }
where the dimensionless collision integrals are introduced:
 \eq{
 L_{kl}^{el}[\chi_k]=\frac1{n_lT\sigma(T)}\mc L_{kl}^{el}[\chi_k].
 }
In the case of present inelastic processes the l. h. s. of the
(\ref{xieqn}) is set by the source term (\ref{Qsource}) and the r.
h. s. contains the linear inelastic collision integrals. After
introduction of inelastic processes the source terms in the
(\ref{xieqn}) become much larger as demonstrated in Sec.
\ref{singcomsec}. Using the equations (\ref{mueqn}) and
(\ref{Teqn}) and the ideal gas formulas (\ref{ABCE}) one can check
that in the zero masses limit the source terms $\hat Q_k$
(\ref{Qsource}) tend to zero and ${D\hat\mu_a=0}$ that is the
$\hat\mu_a$ don't scale and the distribution functions become
scale invariant. The source term of the shear viscosity in the
(\ref{etaeqn}) doesn't depend on the presence of inelastic
processes in the system and originates from the free propagation
term ${\frac{\vec p_k}{p_k^0} \frac{\p f_k}{\p \vec r}}$ in the
Boltzmann equation.

\subsection{ The transport coefficients and their properties }

After substitution of the $f_k^{(1)}$ with the $\varphi_k$
(\ref{varphi}) into the (\ref{enmomten}) and comparison with the
(\ref{T1}) one finds the formula for the bulk viscosity
 \eq{\label{bulkvisc}
 \xi=-\frac13\frac{T}{\sigma(T)}\sum_k x_k(\Delta^{\mu\nu}\pi_{\mu k}\pi_{\nu k},A_k)_k,
 }
and for the shear viscosity
 \eq{\label{shearvisc}
 \eta=\frac1{10}\frac{T}{\sigma(T)}\sum_k x_k(\overset{\circ}{\overline{\pi^\mu_k \pi^\nu_k}},
 C_k \overset{\circ}{\overline{\pi_{k\mu} \pi_{k\nu}}})_k,
 }
where the relation ${\Delta^{\mu\nu}_{ ~~\sigma\tau }
\Delta_\mu^\sigma \Delta_\nu^\tau = 5}$ is used.

In kinetics the conditions that the nonequilibrium perturbations
of the distribution functions does not contribute to the net
charge and the energy-momentum densities are used as a convenient
choice and are called matching conditions. They reproduce the
Landau-Lifshitz condition (\ref{lLcond}). The matching conditions
for the net charge densities can be written as
 \eq{\label{cofchf}
 \sum_k q_{ak} \int \frac{d^3p_k}{(2\pi)^3p_k^0}p_k^\mu U_\mu
 f_k^{(0)}\varphi_k=0,
 }
and for the energy-momentum density can be written as
 \eq{\label{cofemt}
 \sum_k \int \frac{d^3p_k}{(2\pi)^3p_k^0}p_k^\mu p_k^\nu U_\nu f_k^{(0)}\varphi_k=0.
 }
For the special tensorial functions $C_k \overset{\circ}{
\overline{\pi_{k\mu} \pi_{k\nu}}}$ in the (\ref{varphi}) they are
satisfied automatically and for the scalar functions $A_k$ they
can be rewritten in the form (the 3-vector part of the
(\ref{cofemt}) is automatically satisfied)
 \eq{\label{condfit}
 \sum_k q_{ak}x_k (\tau_k, A_k)_k=0, \quad \sum_k x_k (\tau_k^2, A_k)_k=0.
 }
The conditions (\ref{cofchf}) and (\ref{cofemt}) exclude the
nonphysical solutions ${A_k^{z.m.} = \sum_a C_a q_{ak} + C
\tau_k}$ (which cannot be solutions in inhomogeneous systems and
are produced just due to shifts in the $T$, $\mu_a$) of the
linearized Boltzmann equations for which the collision integrals
vanish ($A_k^{z.m.}$ are zero modes). From the formula
(\ref{bulkvisc}) one can see that in the framework of the
Boltzmann equation these conditions also eliminate ambiguity in
the $\xi$ due to freedom of addition of the $A_k^{z.m.}$ to the
solution $A_k$ of the (\ref{xieqn}). With help of these matching
conditions one can show explicitly essential positiveness of the
$\xi$. Namely, using the matching conditions (\ref{condfit}), the
equation (\ref{xieqn}) and the identity ${\Delta^{\mu\nu}
\pi_{\mu,k}\pi_{\nu,k} = z_k^2 - \tau_k^2}$, the bulk viscosity
(\ref{bulkvisc}) can be rewritten as
 \eq{\label{xipos}
 \xi=\frac{T}{\sigma(T)}\sum_k x_k(\hat Q_k,A_k)_k=
 \frac{T}{\sigma(T)}\sum_k x_k\left(\sum_l x_l L^{el}_{kl}[A_k],A_k\right)_k=
 \frac{T}{\sigma(T)}[\{A\},\{A\}],
 }
where the square brackets are introduced for sets of equal lengths
${\{F\}=(F_1,...,F_k,...)}$, ${\{G\}=(G_1,...,G_k,...)}$:
 \eq{\label{sqrbra}
 [\{F\},\{G\}]\equiv\frac1{n^2\sigma(T)}\sum_{k,l}
 \frac{\gamma_{kl}}{(2\pi)^6}\int_{p_k,p_{1l},{p'}_k,{p'}_{1l}}
 f^{(0)}_kf^{(0)}_{1l}(F_k+F_{1l}-{F'}_{k}-{F'}_{1l})G_kW_{kl}.
 }
Using the time reversibility property of the $W_{kl}$ one can show
that the equality
 \eq{
 (F_k+F_{1l}-{F'}_{k}-{F'}_{1l})G_k=\frac14(F_k+F_{1l}-{F'}_{k}-{F'}_{1l})
 (G_k+G_{1l}-{G'}_{k}-{G'}_{1l}),
 }
holds under the integration and the summation in the
(\ref{sqrbra}). Then one gets the direct consequence
 \eq{
 [\{F\},\{G\}]=[\{G\},\{F\}], \quad [\{F\},\{F\}]\geq 0.
 }
This proves the essential positiveness of the $\xi$. Similarly
using the (\ref{etaeqn}), the shear viscosity can be rewritten in
essentially positive form
 \begin{eqnarray}\label{etapos}
  \nonumber\eta&=&\frac1{10}\frac{T}{\sigma(T)}\sum_kx_k
  \left(\overset{\circ}{\overline{\pi_{k}^\mu \pi_{k}^\nu}},
  C_k \overset{\circ}{\overline{\pi_{k\mu} \pi_{k\nu}}}\right)_k
  =\frac1{10}\frac{T}{\sigma(T)}\sum_kx_k\left(\sum_l x_l L^{el}_{kl}
  [C_k \overset{\circ}{\overline{\pi_{k}^\mu \pi_{k}^\nu}}],
  C_k \overset{\circ}{\overline{\pi_{k\mu} \pi_{k\nu}}}\right)_k\\
  &=&\frac1{10}\frac{T}{\sigma(T)}[\{C \overset{\circ}{\overline{\pi^\mu \pi^\nu}}\},
  \{C \overset{\circ}{\overline{\pi_{\mu} \pi_{\nu}}}\}].
 \end{eqnarray}

The considered variational method allows to find an approximate
solution of the integral equations (\ref{xieqn}) and
(\ref{etaeqn}) in the form of a linear combination of
test-functions. The coefficients next to the test-functions are
found from the condition to deliver extremum to some functional.
One could take this functional in the form of some special norm,
as in \cite{groot}. Or one can take somewhat different functional,
like in \cite{Arnold:2003zc}, which is more convenient, and get
the same result. This generalized functional can be written in the
form
 \eq{\label{genfunctional}
 F[\chi]=\sum_k x_k(S_k^{\mu...\nu},\chi_{k\mu...\nu})_k-
 \frac12[\{\chi^{\mu...\nu}\},\{\chi_{\mu...\nu}\}],
 }
where ${S_k^{\mu...\nu}=\hat Q_k}$ and ${\chi_{k\mu...\nu}=A_k}$
for the bulk viscosity and ${S_k^{\mu...\nu} = \overset{\circ}{
\overline{\pi_{k}^\mu \pi_{k}^\nu} }}$, ${\chi_k^{\mu...\nu} = C_k
\overset{\circ}{ \overline{\pi_{k}^\mu \pi_{k}^\nu} }}$ for the
shear viscosity. Equating to zero the first variation of the
(\ref{genfunctional}) over the $\chi_{k\mu...\nu}$ one gets
 \eq{
 \sum_kx_k(S_k^{\mu...\nu},\delta\chi_{k\mu...\nu})_k-
 [\{\chi^{\mu...\nu}\},\{\delta\chi_{\mu...\nu}\}]=0.
 }
Because the variations $\delta\chi_{k\mu...\nu}$ are arbitrary and
independent the generalized integral equations follows then:
 \eq{\label{genvareqn}
 S_k^{\mu...\nu}=\sum_l x_l L^{el}_{kl}[\chi_{k\mu...\nu}].
 }
The second variation of the (\ref{genfunctional}) is
 \eq{
 \delta^2 F[\chi]=-[\{\delta\chi^{\mu...\nu}\},\{\delta\chi_{\mu...\nu}\}]\leq 0,
 }
which means that the solution of the integral equations
(\ref{xieqn}) and (\ref{etaeqn}) is reduced to the variational
problem of finding of the maximum of the functional
(\ref{genfunctional}). Using the (\ref{genvareqn}), the maximal
value of the (\ref{genfunctional}) can be written as
 \eq{
 F_{max}[\chi]=\frac12[\{\chi^{\mu...\nu}\},\{\chi_{\mu...\nu}\}]|_{\chi=\chi_{\max}}.
 }
Then using the (\ref{xipos}) and the (\ref{etapos}) one can write
the bulk and the shear viscosities through the maximal value of
the $F[\chi]$
 \eq{
 \xi=\left.2\frac{T}{\sigma(T)}F_{max}\right|_{S_k^{\mu...\nu}=\hat Q_k, \, \chi_{k\mu...\nu}=A_k},
 }
 \eq{
 \eta=\left.\frac15\frac{T}{\sigma(T)}F_{max}\right|_{S_k^{\mu...\nu}=
 \overset{\circ}{\overline{\pi_{k}^\mu \pi_{k}^\nu}}, \, \chi_k^{\mu...\nu}=
 C_k \overset{\circ}{\overline{\pi_{k}^\mu \pi_{k}^\nu}} }.
 }
This means that the precise solution of the (\ref{genvareqn})
delivers the maximal values for the transport coefficients.

The approximate solution of the system of the integral equations
(\ref{xieqn}) and (\ref{etaeqn}) are sought in the form
 \eq{\label{Atestf}
 A_{k}=\sum_{r=0}^{n_1} A_k^r\tau^r_k,
 }
 \eq{\label{Ctestf}
 C_{k}=\sum_{r=0}^{n_2} C_k^r\tau^r_k,
 }
where $n_1$ and $n_2$ set the number of the used test-functions.
Test-functions, used in \cite{Arnold:2003zc}, would cause less
significant digit cancellation in numerical calculations but there
is a need to reduce the dimension of the 12-dimensional integrals
from these test-functions as more as possible to perform the
calculations in a reasonable time. The test-functions in the form
of just powers of the $\tau_k$ seem to be the most convenient for
this purpose. Questions concerning the uniqueness and the
existence of the solution and the convergence of the approximate
solution to the precise one are covered in \cite{groot} (Chap. IX,
Sec. 1-2). As long as particles of the same particle species but
with different spin states are undistinguishable their functions
$\varphi_k$ (\ref{varphi}) are equal, and the variational problem
is reduced to the variation of the coefficients $A_{k'}^r$ and
$C_{k'}^r$, and the bulk (\ref{xipos}) and the shear
(\ref{etapos}) viscosities can be rewritten as
 \eq{\label{finxi}
 \xi=\frac{T}{\sigma(T)}\sum_{k'=1}^{N'}\sum_{r=0}^{n_1}x_{k'}\alpha_{k'}^r A_{k'}^r,
 }
 \eq{\label{fineta}
 \eta=\frac1{10}\frac{T}{\sigma(T)}\sum_{k'=1}^{N'}\sum_{r=0}^{n_2}x_{k'}\gamma_{k'}^rC_{k'}^r.
 }
After the substitution of the approximate functions $A_{k'}$
(\ref{Atestf}) and $C_{k'}$ (\ref{Ctestf}) into the
(\ref{genfunctional}) and equating the first variation of the
functional to zero one gets the following matrix equations (with
the multi-indexes ${(l',s)}$ and ${(k',r)}$) for the bulk and the
shear viscosities correspondingly\footnote{One can first derive
the same equations for the $A_k$ and $C_k$, treating them as
different functions for all $k$, with the coefficients
$A_{kl}^{rs}$ and $C_{kl}^{rs}$ having the same form as the
$A_{k'l'}^{rs}$ and $C_{k'l'}^{rs}$. Then after summation of the
equations over spin states of identical particles and taking
${A_k=A_{k'}}$, ${C_k=C_{k'}}$ one reproduces the system of
equations for the $A_{k'}$ and $C_{k'}$.}
 \eq{\label{ximatreq}
 x_{k'}\alpha_{k'}^r=\sum_{l'=1}^{N'} \sum_{s=0}^{n_1} A_{l'k'}^{sr}A_{l'}^s,
 }
 \eq{\label{etamatreq}
 x_{k'}\gamma_{k'}^r=\sum_{l'=1}^{N'} \sum_{s=0}^{n_2} C_{l'k'}^{sr}C_{l'}^s,
 }
where the introduced coefficients $A_{k'l'}^{rs}$ and
$C_{k'l'}^{rs}$ are
 \eq{\label{A4ind}
 A_{k'l'}^{rs}=x_{k'}x_{l'}[\tau^r,\tau_1^s]_{k'l'}+\delta_{k'l'}x_{k'}\sum_{m'=1}^{N'}
 x_{m'}[\tau^r,\tau^s]_{k'm'},
 }
 \eq{\label{C4ind}
 C_{k'l'}^{rs}=x_{k'}x_{l'}[\tau^r\overset{\circ}{\overline{\pi^{\mu} \pi^{\nu}}},
 \tau_1^s\overset{\circ}{\overline{\pi_{1\mu}\pi_{1\nu}}}]_{k'l'}+\delta_{k'l'}x_{k'}\sum_{m'=1}^{N'}
 x_{m'}[\tau^r\overset{\circ}{\overline{\pi^{\mu} \pi^{\nu}}},
 \tau^s\overset{\circ}{\overline{\pi_{\mu} \pi_{\nu}}}]_{k'm'}.
 }
They are expressed through the collision brackets
 \eq{\label{br1}
 [F,G_1]_{kl}\equiv\frac{\gamma_{kl}}{T^6(4\pi)^2z_k^2z_l^2K_2(z_k)K_2(z_l)\sigma(T)}
 \int_{p_k,p_{1l},{p'}_k,{p'}_{1l}}e^{-\tau_k-\tau_{1l}}(F_k-{F'}_k)G_{1l}W_{kl}.
 }
The collision brackets $[F,G]_{kl}$ are obtained from the last
formula by the replacement of the $G_{1l}$ on the $G_k$. Due to
the time reversibility property of the $W_{kl}$ one can replace
the $G_{1l}$ on the ${\frac12(G_{1l}-{G'}_{1l})}$ in the
(\ref{br1}). Then one can see that
 \eq{\label{br2pos}
 [\tau^r,\tau^s]_{kl}>0.
 }
Also it's easy to notice the following symmetries
 \eq{
 [F,G_1]_{kl}=[G,F_1]_{lk}, \quad [F,G]_{kl}=[G,F]_{kl}.
 }
They result in the following symmetric properties ${A_{k'l'}^{rs}
= A_{l'k'}^{sr}}$, ${C_{k'l'}^{rs} = C_{l'k'}^{sr}}$. Also the
microscopical particle number and energy conservation laws imply
for the $A_{l'k'}^{sr}$:
 \eq{\label{A4chcons}
 A_{k'l'}^{0s}=0,
 }
 \eq{\label{A4encons}
 \sum_{k'=1}^{N'}A_{k'l'}^{1s}=0.
 }
The (\ref{A4chcons}) together with the $\alpha_k^0=0$
(\ref{elal0}) means that the equations with ${r=0}$ in the
(\ref{ximatreq}) are excluded. From the (\ref{A4encons}) and
(\ref{lhsencons}) it follows that each one equation with ${r=1}$
in the (\ref{ximatreq}) can be expressed through the sum of the
other ones, reducing the rank of the matrix on 1. To solve the
matrix equation (\ref{ximatreq}) one eliminates one equation, for
example with ${k'=1}$, ${r=1}$. One of coefficients of $A_{l'}^1$
is independent; for example, let it be $A_{1'}^1$. Using the
(\ref{A4encons}), the matrix equation (\ref{ximatreq}) can be
rewritten as
 \eq{\label{ximatreq2}
 x_{k'}\alpha_{k'}^r=\sum_{l'=2}^{N'}A_{l'k'}^{1r}(A_{l'}^1-A_{1'}^1)
 +\sum_{l'=1}^{N'} \sum_{s=2}^{n_1}A_{l'k'}^{sr}A_{l'}^s.
 }
Then, using the (\ref{elal0}) and the (\ref{lhsencons}), the bulk
viscosity (\ref{finxi}) becomes
 \eq{\label{finxi2}
 \xi=\frac{T}{\sigma(T)}\sum_{k'=2}^{N'}x_{k'}\alpha_{k'}^1 (A_{k'}^1-A_{1'}^1)
 +\frac{T}{\sigma(T)}\sum_{k'=1}^{N'}\sum_{r=2}^{n_1}x_{k'}\alpha_{k'}^r A_{k'}^r.
 }
Then the coefficient $A_{1'}^1$ can be eliminated by shift of
other $A_{l'}^1$ and be implicitly used to satisfy one energy
conservation matching condition. The particle number conservation
matching conditions are implicitly satisfied by means of the
coefficients $A_{k'}^0$. The first term in the (\ref{finxi2}) is
present only in mixtures. That's why it is small in gases with
close to each other masses of particles of different species (like
in the pion gas). In gases with very different masses (like in the
hadron gas) contribution of the first term in the (\ref{finxi2})
can become dominant.

Analytical expressions for some lowest orders collision brackets,
which enter in the matrix equations (\ref{etamatreq}) and
(\ref{ximatreq2}), can be found in Appendix \ref{appJ}. Higher
orders are not presented because of their bulky form.

\section{ The numerical calculations \label{numcalc}}

The numerical calculations for the hadron gas involve roughly
$2\frac{(N' n)^2}2$ the 12-dimensional integrals, where $N'$ is
the number of particle species and $n$ is the number of the used
test-functions (called the order of the calculations). The
12-dimensional integrals (they are the collision brackets
${[F,G_1]_{kl}}$ and ${[F,G]_{kl}}$) can be reduced to
1-dimensional integrals. For constant cross sections they are
expressed through special functions, and for other ones numerical
methods are used. The details of calculations are described in
Appendix \ref{appJ}. This allows to perform the calculations with
a good precision in a reasonable time. Because the analytical
expressions for the collision brackets are bulky the Mathematica
package \cite{math} is used for symbolical and some numerical
manipulations.

The calculations of the viscosities are quite reliable at $T \leq
120-140~MeV$ (throughout the paper the chemical potentials are
equal to zero if else is not stated), as is discussed in Sec.
\ref{condappl}. The numerical calculations are done also for
temperatures up to ${T=270~MeV}$ for the future comparisons and to
show the position of the maximum of the bulk viscosity, when it is
present only due to the hadrons' masses. Introduction of the
inelastic processes should increase the bulk viscosity, though in
the approximation when its nonzero value is maintained only by the
hadrons' masses the maximum may shift not considerably. Taking
into account the non-ideal gas equation of state, the maximum may
shift to some extent too and become sharper, as can be seen from
the speed of sound of the \cite{toneev}, or a new smaller maximum
can appear.

\begin{figure}[h!]
\begin{center}
\epsfig{file=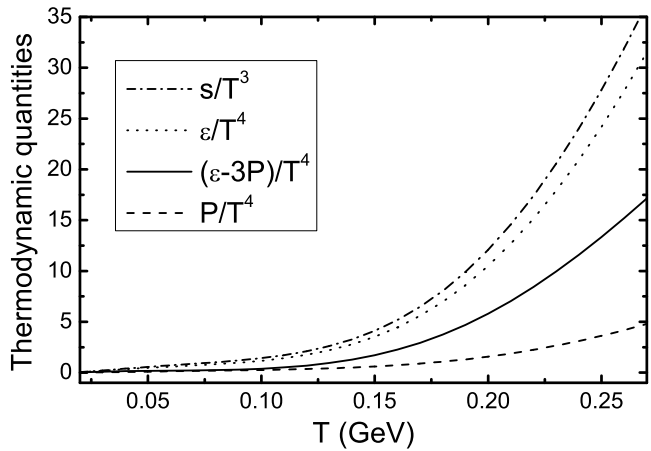,width=8.6cm} \caption{Some scaled
thermodynamic quantities (the entropy density $s$, the energy
density $\varepsilon$, the pressure $P$) and the trace of the
energy-momentum tensor of the ideal gas with the hadron gas mass
spectrum including 322 particle species. \label{ThermQuant} }
\end{center}
\end{figure}

\begin{figure}[h!]
\begin{center}
\epsfig{file=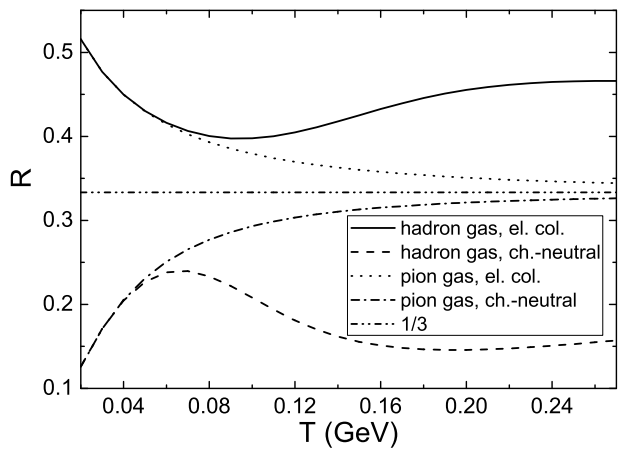,width=8.6cm} \caption{The quantity $R$
(\ref{Rdef2}), appearing in the bulk viscosity source term
(\ref{Qsource}), ate zero chemical potentials. In the case of the
charge-neutral (or chargeless) system the $R$ coincides with the
squared speed of sound $\upsilon_s^2=\frac{\p P}{\p \varepsilon}$.
In the case of only the elastic collisions it is given by the
(\ref{elquant}). It is calculated for the ideal hadron gas in the
first and the second cases (solid and dashed lines
correspondingly) and the ideal pion gas (doted and dash-dotted
lines correspondingly). The value of $1/3$ is shown for
convenience (dash-dot-dotted line). \label{Rquantity} }
\end{center}
\end{figure}

The UrQMD (version 1.3) particle list is used, which doesn't
contain charmed and bottomed particles and consists of 322
particle species including anti-particles. Some thermodynamical
quantities of the ideal hadron gas with this mass spectrum are
shown in fig. \ref{ThermQuant} and fig. \ref{Rquantity}. The
$\varepsilon$ and $P$ are given by the (\ref{epsandek}),
(\ref{ignk}), (\ref{pressure}), and the $s$ is given by
(\ref{entrden}). The quantity $R$ (\ref{Rdef2}), appearing in the
bulk viscosity source term (\ref{Qsource}) and tending to $1/3$ if
the particles' masses are tended to zero, is equal to the squared
speed of sound $R_{ch.-n.}=\upsilon_s^2=\frac{\p P}{\p
\varepsilon}$ in the case of the charge-neutral (or chargeless)
system (implying equal to zero and not developing chemical
potentials). The $R$ is given by the $R_{el. c.}$ (\ref{elquant})
in the case of only the elastic collisions. The $R_{ch.-n.}$ is
quite close to the squared speed of sound of the ideal gas in the
\cite{toneev}. For the intermediate case when there are conserved
and not conserved particle numbers it's observed that the $R$ is
above $1/3$.

The new particle list with charmed and bottomed
particles\footnote{These are the particles which are more or less
reliably detected \cite{Nakamura:2010zzi}.} (cut on $3~GeV$, which
results in negligible errors $0.01\%$ or less) from the THERMUS
package \cite{thermus} was used in the calculations at zero
chemical potentials in the \cite{Moroz:2011vn}. The errors in the
trace of the energy-momentum tensor because of neglected rhe
charmed and bottomed particles grow with the temperature and are
equal to $13\%$ ($21\%$) at ${T=140~MeV}$ ($270~MeV$). The errors
in the $R$ are less than $3.2\%$. The errors in the shear
viscosity (calculated with one constant cross section for all
hadrons) are less than $0.6\%$ and in the bulk viscosity are less
than $7.5\%$. An additional study of the mass spectrum dependence
of the viscosities can be found in the \cite{Moroz:2011vn}.

\begin{figure}[h!]
\begin{center}
\epsfig{file=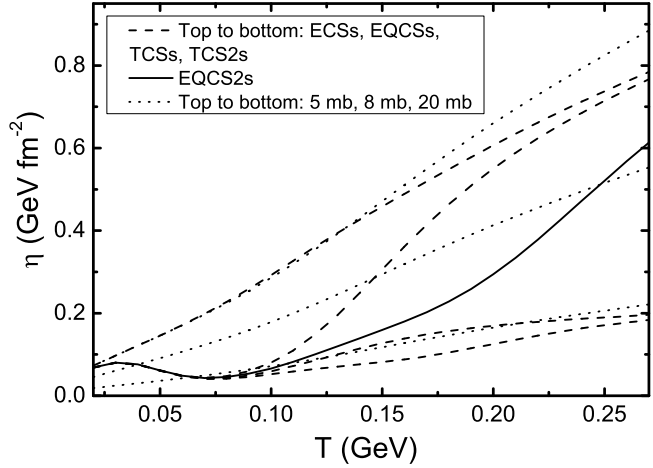,width=8.6cm} \caption{The shear viscosity of
the hadron gas, calculated with the ECSs, EQCSs, TCSs, TCS2s
(dashed line) and the EQCS2s (solid line). One constant cross
section for all hadrons is used in the approximating calculations.
Different values are chosen: $5~mb$, $8~mb$, $20~mb$ (dotted
line). See text for more details. \label{ShearAll} }
\end{center}
\end{figure}

\begin{figure}[h!]
\begin{center}
\epsfig{file=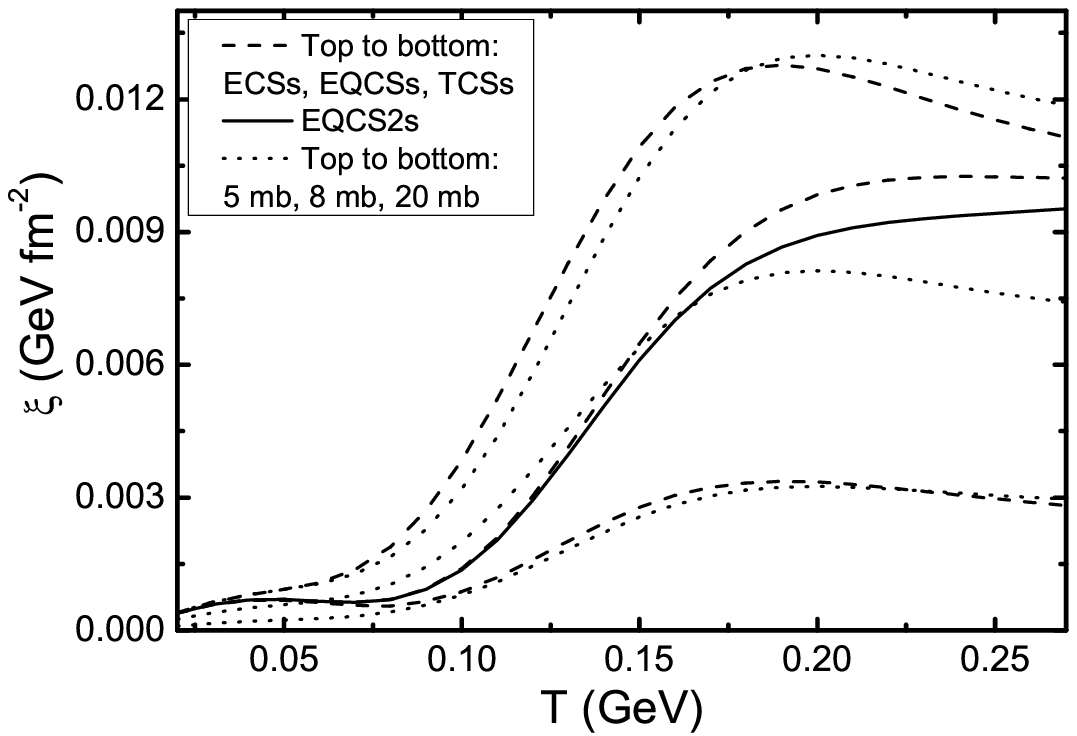,width=8.6cm} \caption{The bulk viscosity of
the hadron gas, calculated with the ECSs, EQCSs, TCSs (dashed
line) and the EQCS2s (solid line). One constant cross section for
all hadrons is used in the approximating calculations. Different
values are chosen: $5~mb$, $8~mb$, $20~mb$ (dotted line). See text
for more details. \label{BulkAll}}
\end{center}
\end{figure}

The results for the shear and the bulk viscosities are shown in
fig. \ref{ShearAll} and fig. \ref{BulkAll} correspondingly. They
are calculated with the different cross sections (with all
corrections and fixes): ECQs, EQCSs, EQCS2s, TCSs, TCS2s. They are
described in Sec. \ref{hardcorsec}. The bulk viscosity with the
TCS2s is not shown because it's smaller only on $5\%$ or less than
the one with the TCSs. Up to 5 (3) test-functions are used in the
calculations of the bulk (shear) viscosity. The maximal errors are
$11\%$ and are less than $4.2\%$ outside the range $40~MeV\leq T
\leq 90~MeV$. The best convergence is for the case of the ECSs
(the errors are less than $2\%$). For quantitative results the
recommended cross sections are the EQCS2s, as is commented in Sec.
\ref{hardcorsec}. For qualitative analysis the TCSs and the TCS2s
are more suitable. Also it's shown that the approximation of one
constant cross section for all hadrons is a good one. For the
shear viscosity with the EQCSs or the EQCS2s it's somewhat worse.
This can be explained in the following way. There are descending
and growing cross sections, and these opposite dependencies
compensate approximately each other, so that at some temperatures
the cross sections manifest themselves approximately as a constant
one. Some EQCSs and EQCS2s have quite steeply descending tails,
which explains relatively fast rises in the shear viscosities,
because of which one constant cross section doesn't provides worse
approximation. An explanation why the bulk viscosity is
approximated well in the same case with one constant cross
sections at some temperatures would be somewhat more complicated.

Also the calculations without resonances (the particles with the
width larger or equal to $0.2~MeV$) are done with the EQCSs to
find out the magnitude of their contributions. After the exclusion
26 particle species remain. The bulk viscosity decreases not more
than in 2.8 times, and the shear viscosity decreases not more than
in 1.6 times (using the TCSs these factors are somewhat smaller).

Note that at $T\approx 160~MeV$ the viscosities calculated with
the TCS2s are approximately 2 times smaller than the viscosities
calculated with the EQCS2s respectively. This reflects the fact
that the contribution to the total number of collisions from the
inelastic processes is approximately the same as from the elastic
plus the quasielastic processes at the freeze-out temperature.

The maximum of the bulk viscosity is, of course, sensitive to the
energy dependence of the cross sections, as can be seen from fig.
\ref{BulkAll}. E. g. if the BB and some MB EQCSs were not fixed,
as is described in Sec. \ref{hardcorsec}, the maximum would be
present at $T\approx 190~MeV$. After the fixes the BB and some MB
EQCSs and EQCS2s have steeply descending tails, and the maximum
shifts towards much higher temperatures. Though for the
qualitative analysis the TCSs and the TCS2s are more suitable
because at $T\gtrsim 160~MeV$ the inelastic processes make not
small contributions. With these cross sections the bulk viscosity
has the maximum at $T\approx 190~MeV$. With one constant cross
section the maximum is present at $T\approx 200~MeV$ with
\cite{Moroz:2011vn} or without charmed and bottomed particles.

\begin{figure}[h!]
\begin{center}
\epsfig{file=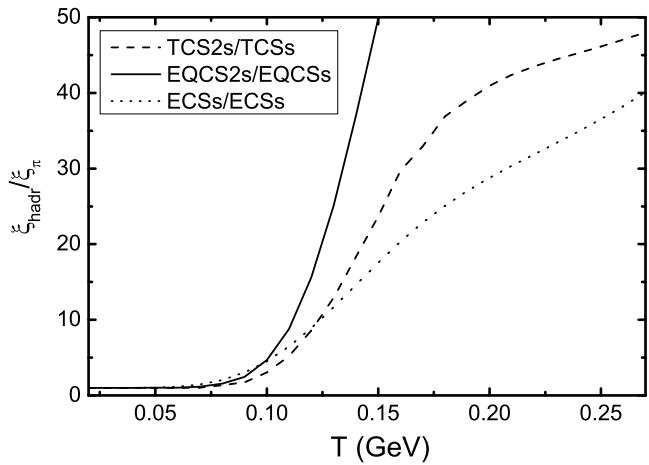,width=8.6cm} \caption{The ratio of the bulk
viscosity of the hadron gas to the one of the pion gas. The
different cross sections are used: TCS2s/TCSs (dashed line),
EQCS2s/EQCSs (solid line), ECSs/ECSs (dotted line). See text for
more details. \label{BulkComparison}}
\end{center}
\end{figure}

\begin{figure}[h!]
\begin{center}
\epsfig{file=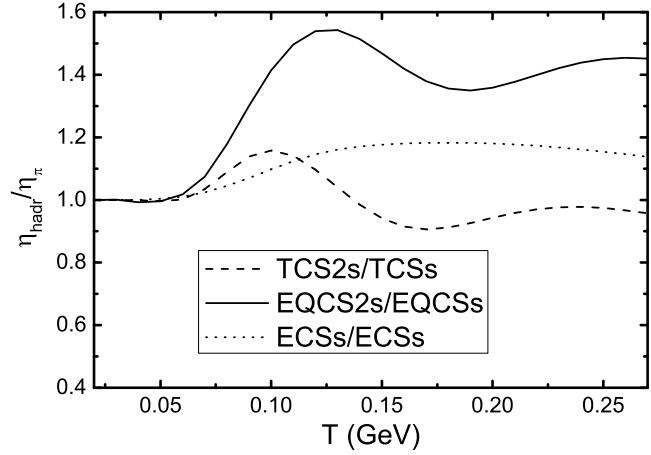,width=8.6cm} \caption{The ratio of the shear
viscosity of the hadron gas to the one of the pion gas. The
different cross sections are used: TCS2s/TCSs (dashed line),
EQCS2s/EQCSs (solid line), ECSs/ECSs (dotted line). See text for
more details. \label{ShearComparison} }
\end{center}
\end{figure}

In several papers the viscosities of the hadron gas were studied
by the ones of the pion gas. This approximations turn out to be
bad while calculating the bulk viscosity\footnote{This discrepancy
could be noticed earlier the results of the \cite{nhngr} and the
\cite{toneev}. Though they required confirmations or
justifications.}. In fig. \ref{BulkComparison} the ratio of the
bulk viscosity of the hadron gas to the one of the pion gas, using
different cross sections, is shown. With all the used cross
sections the deviations are quite large. Purely pion gas implies
the ECSs. The bulk viscosity of the hadron gas with the EQCS2s is
divided on the one of the pion gas with the EQCS. This ratio
reaches 122 at $T=270~MeV$. Considering a closer to the pion gas
approximation, excluding the resonances in the hadron gas, this
ratio becomes $2.3-2.6$ times smaller at $T=120-140~MeV$. The same
factor is equal to $2.0-2.1$ for the used TCS2s and the TCSs for
the hadron gas and the pion gas correspondingly. Though the
resonances should not be excluded. At the same time the
corresponding ratios of the particle number densities and the
energy densities at $T=140~MeV$ are approximately equal to 2 and 3
respectively. Note that the ratios of the viscosities are less
sensitive to the corrections discussed in Sec. \ref{condappl} than
the viscosities themselves. The ratio of the shear viscosity of
the hadron gas to the one of the pion gas is not larger than 1.6,
as can be seen in fig. \ref{ShearComparison}. It can even be
somewhat smaller than 1 if the TCS2s and the TCS2 are used. This
seems to be because the shear viscosity is not much sensitive to
the mass spectrum as the bulk viscosity, and the contributions
from the hadrons other than pions at high temperatures (where pion
numbers don't dominate) come with somewhat larger (on average)
cross sections.

Some simplified explanations of the enlarged bulk viscosity in the
hadron gas (to some extent) and the position of its maximum can be
made. The bulk viscosity is sensitive to particle's masses $m \sim
T$, and the hadron gas mass spectrum provides such masses at
different temperatures. For a fixed $m/T$ and approximately
constant cross section the bulk viscosity grows with the
temperature, as can be seen from the formula (\ref{xi}), using it
as an estimate. Particles with very large masses have relatively
small number densities because of the exponential suppression
$e^{-m/T}$ so that the maximum is set by particles with not the
largest masses at $T \approx 190~MeV$, and a relatively slow
further descending follows at higher temperatures.

\begin{figure}[h!]
\begin{center}
\epsfig{file=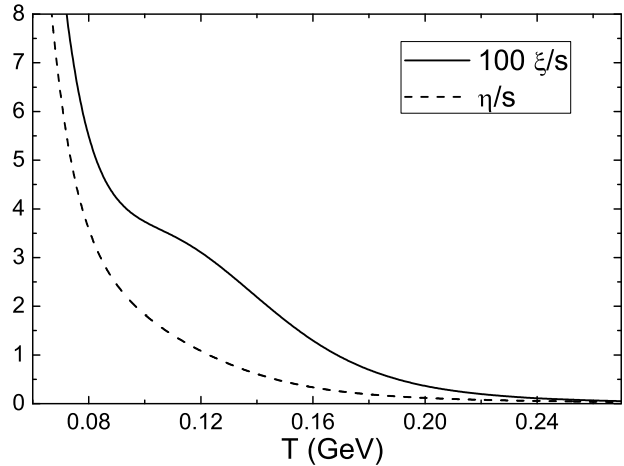,width=8.6cm} \caption{The ratio of the shear
viscosity to the entropy density and the ratio of the bulk
viscosity times 100 to the entropy density in the hadron gas. The
EQCS2s are used. \label{EtasXis} }
\end{center}
\end{figure}

\begin{figure}[h!]
\begin{center}
\epsfig{file=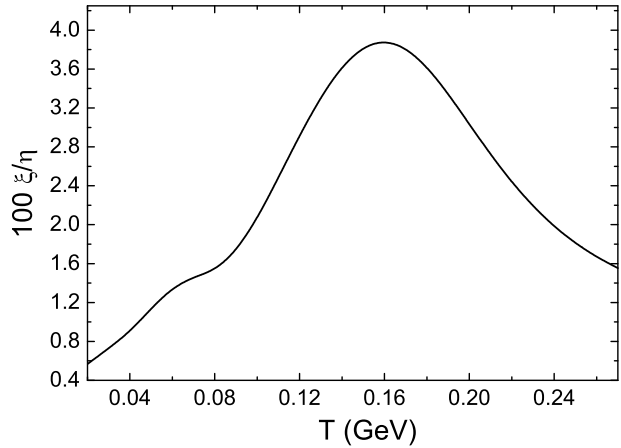,width=8.6cm} \caption{The ratio of the bulk
viscosity times 100 to the shear viscosity in the hadron gas. The
EQCS2s are used. \label{EtaXi} }
\end{center}
\end{figure}

The ratio of the shear viscosity to the entropy density $\eta/s$
and the ratio of the bulk viscosity to the entropy density $\xi/s$
in the hadron gas are shown in fig. \ref{EtasXis}. The EQCS2s are
used. As long as the maximum of the bulk viscosity is not sharp,
the ratio $\xi/s$ doesn't have a maximum and is a descending
function of the temperature. The entropy density is calculated by
the formula (\ref{entrden}) using the ideal gas formulas in the
(\ref{ignk}) and the (\ref{assign2}). The ratio of the bulk
viscosity to the shear viscosity with the EQCS2s is shown in fig.
\ref{EtaXi}.

\begin{figure}[h!]
\begin{center}
\epsfig{file=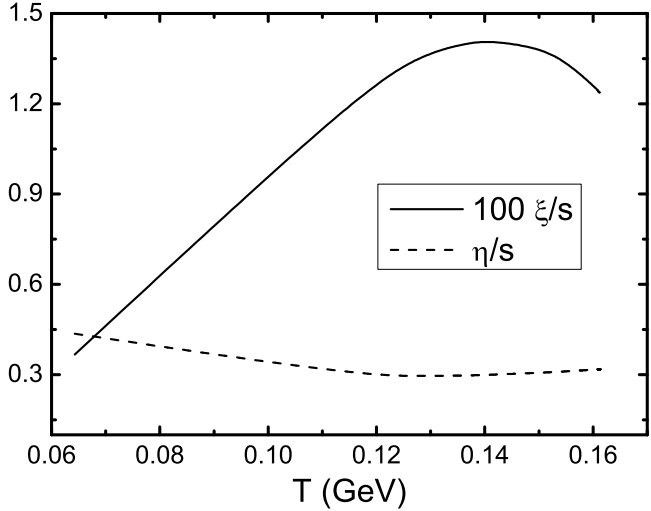,width=8.6cm} \caption{The ratio of the bulk
viscosity times 100 to the entropy density and the ratio of the
shear viscosity to the entropy density as functions of the
freeze-out values of the temperature $T$. All other freeze-out
line variables' values can be found in the
\cite{Gorenstein:2007mw}. The EQCS2s are used.
\label{EtasXisFreezeout} }
\end{center}
\end{figure}

The dependencies from the temperature of the $\eta/s$ and the
$\xi/s$ calculated along the chemical freeze-out line are found
too and are depicted in fig. \ref{EtasXisFreezeout}. The EQCS2s
are used. As was discussed in Sec. \ref{condappl}, the
calculations with large chemical potentials may contain large
deviations, especially in the bulk viscosity, however, in the
hadron gas the contributions from the bosons and the fermions may
cancel substantially. The calculations along the chemical
freeze-out line could have not small deviations for the bulk
viscosity because of the inelastic processes. At the considered
collision energies the strange particle numbers are not described
well by the equilibrium statistical calculations. It's expected
that this is because they don't reach the chemical equilibrium
before the chemical freeze-out takes place. After the introduction
of the strange saturation factors $\gamma_s$
\cite{Becattini:2005xt} the experimental data gets described
better. As long as the considered chemical perturbations are not
quite accurate and are not small they are used in a
phenomenological way, being inserted into all particle number
densities\footnote{The nonequilibrium chemical potential-like
perturbations of the form $T n_{s,k}\ln\gamma_s$ ($n_{s,k}$ is the
number of strange quarks in hadrons of the $k$-th species)
\cite{Becattini:2005xt}, obtained from statistical description and
reflecting suppression of the strange particle numbers, obviously
violate conservation laws to some extent. However, they are used
because of the simplicity in phenomenological estimating
calculations. There are also more elaborated calculations of the
chemical perturbations, see e. g. \cite{Letessier:2005qe}. Also
see \cite{Heinz:2007in} for some discussions.}. Because of all
this the calculations along the chemical freeze-out line are less
reliable than the ones at zero chemical potentials.
All variables' values of the chemical freeze-out line, including
the strangeness saturation factor $\gamma_s$, are conveniently
presented in the \cite{Gorenstein:2007mw}. The convergence of the
calculations (with all the cross section types) is good with the
errors less than $4\%$ for both the viscosities. Also it was
checked how the results change if the chemical equilibrium is
assumed. The entropy density increases up to $36\%$. The shear
viscosity increases no more than on $13\%$. The bulk viscosity
decreases no more than on $5.1\%$ (though with the ECSs the
decrease would be on $44\%$).

\section{ Analytical results }
\subsection{ The single-component gas \label{singcomsec}}
In the single-component gas, using one test-function, the matrix
equations can be easily solved, and the shear (\ref{fineta}) and
the bulk (\ref{finxi2}) viscosities become (indexes "1" of the
particle species are omitted)
 \eq{\label{etasc}
 \eta=\frac1{10}\frac{T}{\sigma(T)}\frac{(\gamma^0)^2}{C^{00}},
 }
 \eq{\label{xisc}
 \xi=\frac{T}{\sigma(T)}\frac{(\alpha^2)^2}{A^{22}}.
 }
In this approximation the explicit closed-form (expressed through
special and elementary functions) relativistic formulas for the
bulk and the shear viscosities were obtained in the
\cite{anderson}. There the parameter ${a=2r}$. In \cite{groot}
(Chap. XI, Sec. 1) they are written through the parameter
${\sigma=2r^2}$.\footnote{It is the differential cross section for
identical particles. The total cross section is ${\int
\frac{d\Omega}2 2r^2=4\pi r^2}$.} The results are
 \eq{\label{etaincor}
 \eta=\frac{15}{64\pi}\frac{T}{r^2}\frac{z^2K_2^2(z)\hat h^2}
 {(5z^2+2)K_2(2z)+(3z^3+49z)K_3(2z)},
 }
 \eq{\label{xi}
 \xi=\frac1{64\pi}\frac{T}{r^2}\frac{z^2K_2^2(z)[(5-3\gamma)\hat h-3\gamma]^2}
 {2K_2(2z)+zK_3(2z)},
 }
where ${\gamma=\frac1{c_\upsilon}+1=\frac{z^2+5\hat h-\hat
h^2}{z^2+5\hat h-\hat h^2-1}}$. Though the correct result for the
shear viscosity is
 \eq{\label{eta}
 \eta=\frac{15}{64\pi}\frac{T}{r^2}\frac{z^2K_2^2(z)\hat h^2}
 {(15z^2+2)K_2(2z)+(3z^3+49z)K_3(2z)}.
 }
This result is in agreement with the result in \cite{prakash,
leeuwen}. To get the (\ref{eta}) and the (\ref{xi}) the collision
brackets in the $C^{00}$ (\ref{C4ind}) and the $A^{22}$
(\ref{A4ind}) can be taken from Appendix \ref{appJ} with
${z_k=z_l=z}$ and the $\gamma^0$ and $\alpha^2$ can be taken from
Appendix \ref{appA}. In the nonrelativistic limit, ${z \gg 1}$,
one gets\footnote{This reproduces the result of Chapman and Enskog
in the nonrelativistic theory for the shear viscosity. The
vanishing value of the bulk viscosity is obtained in the limit
$m\rightarrow \infty$ \cite{landau10} (Sections 8, 10). The result
of the vanishing bulk viscosity of a monoatomic classical gas in
the nonrelativistic theory is attributed to James Clerk Maxwell,
see \cite{Weinberg:1971mx}. }
 \eq{\label{nonrel}
 \eta=\frac{5}{64\sqrt{\pi}}\frac{T}{r^2}\sqrt{z}\left(1+\frac{25}{16}z^{-1}+...\right),
 }
 \eq{
 \xi=\frac{25}{256\sqrt{\pi}}\frac{T}{r^2}z^{-3/2}\left(1-\frac{183}{16}z^{-1}+...\right).
 }
In the ultrarelativistic limit, ${z \ll 1}$, one gets\footnote{The
vanishing value of the bulk viscosity of a monoatomic classical
gas in the ultrarelativistic limit is attributed to I. M.
Khalatnikov, see \cite{landau10}.}
 \eq{
 \eta=\frac{3}{10\pi}\frac{T}{r^2}\left(1+\frac{1}{20}z^2+...\right),
 }
 \eq{
 \xi=\frac1{288\pi}\frac{T}{r^2}z^4\left(1+\left(\frac{49}{12}-6\ln2+6\gamma_E\right)z^2+6z^2\ln z
 +...\right),
 }
where $\gamma_E$ is the Euler's constant, ${\gamma_E\approx
0.577}$.

The perturbation of the distribution function $\varphi$
(\ref{varphi}) can be found too:
 \eq{
 \varphi=\frac1{n\sigma(T)}\left(-(A^0+A^1\tau+A^2\tau^2)\nabla_\mu U^\mu+C^0
 \overset{\circ}{\overline{\pi^\mu \pi^\nu}}
 \overset{\circ}{\overline{\nabla_\mu U_\nu}}\right),
 }
where the $C^0$ is equal to
 \eq{
 C^0=\frac{15}{64\pi}\frac{\sigma(T)}{r^2}\frac{z^2K_2^2(z)\hat h}
 {(15z^2+2)K_2(2z)+(3z^3+49z)K_3(2z)},
 }
and the $A^2$ is equal to
 \eq{
 A^2=\frac1{64\pi}\frac{\sigma(T)}{r^2}\frac{z^2K_2^2(z)[(5-3\gamma)\hat h-3\gamma]}
 {2K_2(2z)+zK_3(2z)}.
 }
The $A^0$ and the $A^1$ are used to satisfy the matching
conditions (\ref{condfit}) and are equal to
 \eq{
 A^0=A^2\frac{a^2a^4-(a^3)^2}{\Delta_A}, \quad
 A^1=A^2\frac{a^2a^3-a^1a^4}{\Delta_A}, \quad
 \Delta_A\equiv a^1 a^3-(a^2)^2,
 }
where the $a^s$ can be found in Appendix \ref{appA}. In the
nonrelativistic limit ${z\gg 1}$ one has
 \eq{
 \varphi=\frac{5\pi e^{z-\hat\mu}}{32\sqrt2 T^3 z^2 r^2}
 \left(-(\tau^2+2z\tau-z^2)\nabla_\mu U^\mu+
 2\overset{\circ}{\overline{\pi^\mu \pi^\nu}}
 \overset{\circ}{\overline{\nabla_\mu U_\nu}}\right).
 }
In the ultrarelativistic limit $z\ll 1$ one has
 \eq{
 \varphi=\frac{\pi e^{-\hat\mu}}{480 T^3 r^2}\left(-5z^2(\tau^2+8\tau-12)\nabla_\mu
 U^\mu+36\overset{\circ}{\overline{\pi^\mu \pi^\nu}}
 \overset{\circ}{\overline{\nabla_\mu U_\nu}}\right).
 }
Note that although the shear viscosity diverges for ${T\rightarrow
\infty}$ the perturbative expansion over the gradients does not
break down because the $\varphi$ does not diverge (it tends to
zero, conversely).

The phenomenological formula, coming from the momentum transfer
considerations in the kinetic-molecular theory, for the shear
viscosity is ${\eta_{ph}\propto l n \langle \abs{\vec p} \rangle}$
(with the coefficient of proportionality of order 1), where
$\langle \abs{\vec p} \rangle$ is the average relativistic
momentum (\ref{avmom}), $l$ is the mean free path. It gives the
correct leading $m$ and $T$ parameter dependence of the
(\ref{eta}) with a quite precise coefficient\footnote{This formula
is justified only for rarified systems where the ideal gas
equation of state is applicable.}. The mean free path can be
estimated as ${l\approx 1/(\sigma_{tot}n)}$ (see Appendix
\ref{appmfp}). Choosing the coefficient of proportionality to
match the nonrelativistic limit one gets \cite{Gorenstein:2007mw}
 \eq{\label{etaph}
 \eta_{ph}=\frac{5}{64\sqrt{\pi}}\frac{\sqrt{mT}}{r^2}\frac{K_{5/2}(m/T)}{K_2(m/T)}.
 }
If the bulk viscosity is expressed as ${\xi_{ph}\propto l n
\langle \abs{\vec p} \rangle}$ the coefficient of proportionality
is not of order 1. In the nonrelativistic limit it is $25/(512
\sqrt2 z^2)$ and in the ultrarelativistic limit it is $z^4/(864
\pi)$. To reproduce these asymptotical dependencies the bulk
viscosity should be proportional to the second power of the
averaged product of the source term $\hat Q$ and the $\tau$ that
is to the $(\alpha^2)^2$.

If a system has no charges, then terms proportional to the $\tau_k$ in the (\ref{Qsource}) are absent, and the
$R$ quantity gets another form. This results in quite different values of the $\alpha_k^r$. In particular, for
the single-component gas in the case ${z\gg 1}$ one gets
 \eq{\label{alfrac1}
 \frac{(\alpha^2)^2|_{q_{11}=0}}{(\alpha^2)^2|_{q_{11}=1}}=\frac{4 z^4}{25}+...,
 }
and in the case ${z\ll 1}$ one gets
 \eq{\label{alfrac2}
 \frac{(\alpha^2)^2|_{q_{11}=0}}{(\alpha^2)^2|_{q_{11}=1}}=4+... .
 }
In both cases these estimates suppose enhancement of the bulk viscosity (\ref{xisc}) if the number-changing
processes are not negligible.

Although constant cross sections are the most simple and the most
universal ones in approximate calculations, let's write down also
formulas for some other simple energy dependencies of cross
sections. Using the collision bracket from Appendix \ref{appJ},
one gets for the cross section $\sigma s/(2 m)^2 = \sigma
\upsilon^2/(2 z)^2$ ($\sigma$ is just a positive dimensional
constant here)
 \eq{\label{eta1}
 \eta_1=\frac{15 T}{32 \pi\sigma}\frac{\hat h^2 z^3 K_2^2(z)}
 {9 z (3 z^2+34) K_2(2 z)+(3 z^4+157 z^2+920) K_3(2 z)},
 }
 \eq{\label{xi1}
 \xi_1=\frac{T}{32 \pi \sigma}\frac{z^3 [(5-3 \gamma) \hat h-3 \gamma]^2 K_2^2(z)}
 {(z^2+20) K_3(2 z)+6 z K_2(2 z)},
 }
and for the cross section $\sigma (2m)^2/s = \sigma (2z)^2
/\upsilon^2$:
 \eq{\label{eta2}
 \eta_2=\frac{15 T}{32 \pi \sigma}\frac{\hat h^2 z^2 K_2^2(z)}
 {(3 z^2-2) K_2(2 z)+z (3 z^2+1) K_3(2 z)},
 }
 \eq{\label{xi2}
 \xi_2=\frac{T}{32 \pi \sigma}\frac{z^2 [(5-3 \gamma) \hat h-3 \gamma]^2 K_2^2(z)}
 {(z K_3(2 z)-2 K_2(2 z))}.
 }
Using the low-energy current algebra isospin averaged $\pi\pi$
differential cross section \cite{prakash, Weinberg:1966kf}
 \eq{\label{sigmaca}
 \sigma_{CA}=\frac13\frac{s}{64\pi^2f_\pi^4}\left[3-8\frac{m_\pi^2}{s}+7\frac{m_\pi^4}{s^2}
 +\left(1-8\frac{m_\pi^2}{s}+16\frac{m_\pi^4}{s^2}\right)\cos^2\Theta\right],
 }
($f_\pi=93~MeV$ is the pion decay constant) and treating all pions
as identical particles at zero chemical potentials, one gets the
formulas
 \eq{\label{etaca}
 \eta_{CA}=\frac{360 \pi f_\pi^4}{T}\frac{\hat h^2 z K_2^2(z)}
 {9 z (93 z^2+1730) K_2(2 z)+(69 z^4+6167 z^2+47104) K_3(2 z)},
 }
 \eq{\label{xica}
 \xi_{CA}=\frac{24 \pi f_\pi^4}{T}\frac{z [(5-3 \gamma) \hat h-3 \gamma]^2 K_2^2(z)}
 {(23 z^2+1024) K_3(2 z)+210 z K_2(2 z)},
 }
which are exactly 2 times larger than the 1-st order calculations
in the \cite{prakash}\footnote{So that it looks like the double
counting factor $1/2$ is lost in the calculations of the
viscosities in the \cite{prakash} (and presumably for the heat
conductivity), and is not lost in the current algebra total cross
section. }. Taking the scattering angle averaged cross section
instead of the (\ref{sigmaca}), one would get approximately the
same result for the viscosities (errors are not more than $4\%$).
In the low temperature limit, $z\gg1$, one gets the following
expansions for the shear viscosities (\ref{eta1}), (\ref{eta2}),
(\ref{etaca}):
 \eq{
 \eta_1=\frac{5}{32 \sqrt{\pi}}\frac{T}{\sigma}\sqrt{z}\left(1-\frac{39}{16}z^{-1}+...\right),
 }
 \eq{
 \eta_2=\frac{5}{32 \sqrt{\pi}}\frac{T}{\sigma}\sqrt{z}\left(1+\frac{89}{16}z^{-1}+...\right),
 }
 \eq{
 \eta_{CA}=\frac{120\pi^{3/2}}{23}\frac{f_\pi^4}{T}z^{-3/2}\left(1-\frac{2049}{368}z^{-1}+...\right),
 }
and in the high temperature limit, $z\ll 1$, respectively:
 \eq{
 \eta_1=\frac{3}{92 \pi}\frac{T}{\sigma}z^2\left(1-\frac2{23} z^2+...\right),
 }
 \eq{
 \eta_2=\frac{6}{\pi}\frac{T}{\sigma} z^{-2}\left(1+\frac1{20}(9 - 4 \gamma_E)z^2
 -\frac1{5}z^2\ln z+...\right),
 }
 \eq{
 \eta_{CA}=\frac{45 \pi}{92}\frac{f_\pi^4}{T}\left(1-\frac{17}{368}z^2+...\right).
 }
In the low temperature limit, $z\gg1$, one gets the following
expansions for the bulk viscosities (\ref{xi1}), (\ref{xi2}),
(\ref{xica}):
 \eq{
 \xi_1=\frac{25}{128 \sqrt{\pi}}\frac{T}{\sigma}z^{-3/2}\left(1-\frac{247}{16}z^{-1}+...\right),
 }
 \eq{
 \xi_2=\frac{25}{128 \sqrt{\pi}}\frac{T}{\sigma}z^{-3/2}\left(1-\frac{119}{16}z^{-1}+...\right),
 }
 \eq{
 \xi_{CA}=\frac{150 \pi^{3/2}}{23}\frac{f_\pi^4}{T}z^{-7/2}\left(1-\frac{6833}{368}z^{-1}+...\right),
 }
and in the high temperature limit, $z\ll 1$, respectively:
 \eq{
 \xi_1=\frac{1}{1440 \pi}\frac{T}{\sigma}z^6\left(1+\frac1{30}(109 + 180 \gamma_E - 180 \ln 2)z^2
 +6 z^2\ln z...\right),
 }
 \eq{
 \xi_2=\frac{1}{36 \pi}\frac{T}{\sigma}z^2\left(1+\frac13(13 + 12 \gamma_E - 18 \ln 2)z^2
 +4 z^2\ln z+...\right),
 }
 \eq{
 \xi_{CA}=\frac{\pi}{96}\frac{f_\pi^4}{T}z^4\left(1+\frac1{24}(89 + 144 \gamma_E - 144 \ln 2)z^2
 +6 z^2\ln z+...\right).
 }
Let's also write down the viscosities with the cross section
$\sigma [1+c_1 \upsilon^2/(2 z)^2+c_2 (2 z)^2/\upsilon^2]$,
unifying the previously considered ones:
 \begin{eqnarray}
 \nonumber \eta_3&=&\frac{15}{32 \pi \sigma} T \hat h^2 z^3 K^2_2(z)/
  [z (2 + 306 c_1 - 2 c_2 + 3 (5 + 9 c_1 + c_2) z^2) K_2(2 z)\\
  &+& (920 c_1 + (49 + 157 c_1 + c_2) z^2 + 3 (1 + c_1 + c_2) z^4) K_3(2 z)],
 \end{eqnarray}
 \eq{
 \xi_3=\frac{T}{32 \pi \sigma} \frac{z^3 [(5-3 \gamma) \hat h-3 \gamma]^2 K^2_2(z)}
 {2 (1 + 3 c_1 - c_2) z K_2(2 z) + (20 c_1 + (1 + c_1 + c_2) z^2) K_3(2 z)},
 }
where $c_1$ and $c_2$ are some dimensionless coefficients such
that $c_1 + c_2 + 1 \geq 0$ and $c_1\geq 0$ to make the cross
section non-negative. The last formulas may have badly convergent
expansions over the $z$ and the $z^{-1}$ for some values of the
$c_1$ and the $c_2$ so that they ain't expanded.

\subsection{ The binary mixture \label{binmixsec}}

The mixture of two species with masses $m_1$, $m_2$ and the
different classical elastic differential constant cross sections
$\sigma^{cl}_{11}$, ${\sigma^{cl}_{12} = \sigma^{cl}_{21}}$,
$\sigma^{cl}_{22}$ is considered in this section. Using the
(\ref{fineta}) with ${n_2=0}$ and solving the matrix equation
(\ref{etamatreq}) one gets for the shear viscosity
 \eq{
 \eta=\frac{T}{10\sigma(T)}\frac1{\Delta_\eta}[(x_{1'}\gamma_1^0)^2C_{2'2'}^{00}-
 2x_{1'}x_{2'}\gamma_1^0\gamma_2^0C_{1'2'}^{00}+(x_{2'}\gamma_2^0)^2C_{1'1'}^{00}],
 }
where ${\Delta_\eta = C_{1'1'}^{00}C_{2'2'}^{00} -
(C_{1'2'}^{00})^2}$. The collision brackets for the
$C_{k'l'}^{00}$ (\ref{C4ind}) can be found in Appendix \ref{appJ}
and the $\gamma_k^0$ can be found in Appendix \ref{appA}.

In the important limiting case when one mass is large ${z_2\gg 1}$
($g_2$ and $\hat \mu_2$ are finite so that ${x_{2'}\ll 1}$) and
another mass is finite one can perform asymptotic expansion of the
special functions. Then one has ${x_{1'} \propto O(1)}$,
${\gamma_1^0 \propto O(1)}$, ${x_{2'} \propto
O(e^{-z_2}z_2^{3/2})}$, ${\gamma_2^0 \propto O(z_2)}$. The
collisions of light and heavy particles dominate over the
collisions of heavy and heavy particles in the $C_{2'2'}^{00}$,
and one has ${[\overset{\circ}{ \overline{\pi^{\mu} \pi^{\nu}} },
\overset{\circ}{ \overline{\pi_{\mu} \pi_{\nu}} }]_{21} \propto
O(z_2)}$, ${C_{2'2'}^{00} \propto O(e^{-z_2}z_2^{5/2})}$. In the
$C_{1'1'}^{00}$ the collisions of light and light particles
dominate, and one gets ${C_{1'1'}^{00} \propto O(1)}$. And
${[\overset{\circ}{ \overline{\pi^{\mu} \pi^{\nu}} },
\overset{\circ}{ \overline{\pi_{1\mu} \pi_{1\nu}} }]_{12} \propto
O(1)}$, ${C_{1'2'}^{00} \propto O(e^{-z_2}z_2^{3/2})}$. In the
shear viscosity the first nonvanishing contribution is the
single-component shear viscosity (\ref{eta}), where one should
take ${r^2=\sigma^{cl}_{11}}$ and ${z=z_1}$. The next correction
is
 \eq{
 \Delta \eta=z_2^{5/2}e^{-z_2}\frac{3 Tg_2 e^{z_1-\hat\mu_1+\hat\mu_2}}
 {64\sqrt{2\pi}(3+3z_1+z_1^2)g_1\sigma^{cl}_{12}}.
 }
The approximate formula \cite{Gorenstein:2007mw}
 \eq{
 \eta=\sum_k\eta_kx_k,
 }
where $\eta_k$ is given by the (\ref{eta}) or the (\ref{etaph})
with mass $m_k$ and cross section $\sigma^{cl}_{kk}$, would give
somewhat different heavy mass power dependence $O(e^{-z_2}z_2^2)$.

Using the (\ref{finxi2}) with ${n_1=1}$ and solving the matrix
equation (\ref{ximatreq2}) one gets for the bulk viscosity
 \eq{
 \xi=\frac{T}{\sigma(T)}\frac{(x_{2'}\alpha_2^1)^2}{A_{2'2'}^{11}}=
 \frac{T}{\sigma(T)}\frac{x_{1'}x_{2'}\alpha_1^1\alpha_2^1}{A_{1'2'}^{11}}.
 }
Using the definition of the $A_{2'2'}^{11}$ (\ref{A4ind}) and the
fact ${[\tau,\tau_1]_{kl} + [\tau,\tau]_{kl} = 0}$ (\ref{br211})
one gets ${A_{2'2'}^{11} = x_{1'}x_{2'}[\tau,\tau]_{12}}$. Using
the (\ref{br2pos}) one gets ${[\tau,\tau]_{12} > 0}$. Then using
${x_{1'}\alpha_1^1 + x_{2'}\alpha_2^1 = 0}$, coming from the
(\ref{lhsencons}), the bulk viscosity can be rewritten as
 \eq{
 \xi=\frac{T}{\sigma(T)}\frac{x_{2'}(\alpha_2^1)^2}{x_{1'}[\tau,\tau]_{12}}
 =\frac{T}{\sigma(T)}\frac{x_{1'}(\alpha_1^1)^2}{x_{2'}[\tau,\tau]_{12}}>0.
 }
The collision bracket $[\tau,\tau]_{12}$ can be found in Appendix
\ref{appJ}, and the $\alpha_k^1$ can be found in Appendix
\ref{appA}.

In the limiting case ${z_2\gg 1}$ one has ${x_{1'} \propto O(1)}$,
${x_{2'} \propto O(e^{-z_2}z_2^{3/2})}$, ${\alpha_1^1 \propto
O(e^{-z_2}z_2^{3/2})}$, ${\alpha_2^1 \propto O(1)}$, ${A_{22}^{11}
\propto A_{12}^{11} \propto O(e^{-z_2}z_2^{1/2})}$,
${[\tau,\tau]_{12} \propto O(z_2^{-1})}$. Then for the bulk
viscosity one gets
 \eq{
 \xi=e^{-z_2}z_2^{5/2}\frac{g_2 T e^{-\hat\mu_1+\hat\mu_2+z_1}[2 z_1^2-5-2\hat h_1^2
 +10 \hat h_1]^2}{128 \sqrt{2 \pi } g_1 \sigma^{cl}_{12}(z_1^2+3 z_1+3)[z_1^2-1-\hat h_1^2+5 \hat h_1]^2}+....
 }

\section{Concluding remarks}

The shear and the bulk viscosities of the hadron gas and the pion
gas were calculated using the UrQMD cross sections.

The physics of the bulk viscosity is very interesting. In
particular, in mixtures it can strongly depend on the mass
spectrum. For instance, at the temperature ${T=120~MeV}$
(${140~MeV}$) and zero chemical potentials the bulk viscosity of
the hadron gas is larger in 8.6-15.6 (14.6-40) times than the bulk
viscosity of the pion gas. The used UrQMD cross sections have
allowed to perform this comparison more accurately.

Also the bulk viscosity can strongly depend on the quantum
statistics corrections, the equation of state and the inelastic
processes, which can be explained by nontrivial form of its source
term(s). It's a future task to find the universal and optimal
criterion for switching on/off the inelastic processes. Numerical
calculations of the bulk viscosity along and around the chemical
freeze-out line which don't involve the approximations of
conserved or not conserved particle numbers (like in the
\cite{Muronga:2003tb}, though the procedure of collisions of
particles introduce some errors itself, which should be kept in
mind) are desirable to get more accurate values of the bulk
viscosity at these points. This is also needed for a better
understanding of the chemical freeze-out itself and connected with
it problems and to get a better description of the deviations from
the chemical equilibrium.

The transport coefficients are connected with fluctuations through
the fluctuation-dissipation theorem. Because of strong dependence
of the bulk viscosity on the equation of state one might expect to
find its maximum at or near a phase transition point. According to
lattice calculations at zero chemical potentials
\cite{Aoki:2006we}, the QCD phase transition is an analytical
crossover. Calculations of the Polyakov loop \cite{Hidaka:2008dr},
the 't Hooft loop \cite{Dumitru:2010mj} and some other
calculations \cite{Ratti:2011au} suggest that hadronic degrees of
freedom survive partially at some temperatures above the critical
one. In the qualitative calculations for the hadron gas with the
UrQMD total cross sections the maximum of the bulk viscosity has
been found at $T\approx 190~MeV$. This value fits in the
transition temperature range $T_c=185-195~MeV$ found from lattice
calculations by the hotQCD collaboration, but calculations of the
Wuppertal-Budapest collaboration with physical quark masses give
the region $T_c=150-170~MeV$ \cite{Borsanyi:2010bp}. According to
the Wuppertal-Budapest collaboration calculations of thermodynamic
functions with physical quark masses \cite{Borsanyi:2010cj}, the
scaled trace of the energy-momentum tensor has its peak at
$T\approx 190~MeV$. In some other lattice calculations with
somewhat different quark masses \cite{Cheng:2009zi,
Petreczky:2009ey} this peak is somewhat sharper and still is
present at $T\approx 190~MeV$. According to the lattice
calculations in the \cite{Borsanyi:2010cj}, the squared speed of
sound in the charge-neutral hadron gas (implying equal to zero and
not developing chemical potentials) has it's minimum at $T\approx
150~MeV$, unlike the ideal gas calculations ($T\approx 190~MeV$)
or the ones in the \cite{toneev} ($T\approx 180~MeV$). So that one
could expect to find the maximum of the bulk viscosity somewhere
in the temperature range $T=150-190~MeV$, presumably closer to the
lowest bound. This range covers the critical temperature values
found from the lattice calculations.

The shear viscosity is less dependent on the mass spectrum, the
quantum statistics corrections, the equation of state and the
inelastic processes. This may be explained by its more trivial
source term(s).

\acknowledgments

The author would like to thank to Prof. Zoltan Fodor for providing
important references. Also the author is grateful to Prof. Shin
Muroya and Dr. Juan Torres-Rincon for discussions.

\appendix

\section{The values of the $\alpha^r_k$, $\gamma_k^r$ and $a_k^r$ \label{appA}}
Their definitions are
 \eq{\label{alphagammaadef}
 \alpha_k^r\equiv(\hat Q_k,\tau_k^r), \quad \gamma_k^r\equiv(\tau_k^r
 \overset{\circ}{\overline{\pi_k^\mu\pi_k^\nu}},
 \overset{\circ}{\overline{\pi_{k\mu}\pi_{k\nu}}}), \quad
 a^s_k\equiv(1,\tau_k^s)_k,
 }
where the round brackets are
 \eq{
 (F,G)_k\equiv\frac1{4\pi z_k^2K_2(z_k)T^2}\int_{p_k} F(p_k)G(p_k)e^{-\tau_k}.
 }
Then one can rewrite the $a_k^s$ as
 \eq{\label{aksdef}
 a_k^s=\frac1{z_k^2K_2(z_k)}\int_{z_k}^\infty d\tau (\tau^2-z_k^2)^{1/2}\tau^s e^{-\tau}.
 }
There is a recurrence relation for the $a^s_k$:
 \eq{
 a^s_k=(s+1)a^{s-1}_k+z_k^2 a^{s-2}_k-(s-2)z_k^2 a^{s-3}_k.
 }
It can be derived from the (\ref{aksdef}) written in the form
 \eq{
 a_k^s=\frac1{z_k^2K_2(z_k)}\int_{z_k}^\infty d\tau
 (\tau^2-z_k^2)^{3/2}\tau^{s-2} e^{-\tau}+z_k^2 a_k^{s-2}.
 }
Then after integration by parts the recurrence relation follow.
Some values of the $a_k^s$ are
 \eq{
 a^0_k=\frac1{z_k^2}(\hat h_k-4),
 }
 \eq{
 a^1_k=1,
 }
 \eq{
 a^2_k=\hat h_k-1,
 }
 \eq{
 a^3_k=3\hat h_k+z_k^2,
 }
 \eq{
 a^4_k=(15+z_k^2)\hat h_k+2z_k^2,
 }
 \eq{
 a^5_k=6(15+z_k^2)\hat h_k+z_k^2(15+z_k^2),
 }
 \eq{
 a^6_k=(630+45z_k^2+z_k^4)\hat h_k+5z_k^2(21+z_k^2).
 }
The $\alpha_k^r$ can be expressed through the $a^r_k$ after the
integration of the (\ref{Qsource}) (or the (\ref{Qsource2}) if
only the elastic collisions are considered) over the momentum,
using the definition (\ref{alphagammaadef}). For systems with only
the elastic collisions some values of the $\alpha_k^r$ are written
below, in agreement with \cite{groot} (Chap. VI, Sec. 3):
 \eq{\label{elal0}
 \alpha_k^0=0,
 }
 \eq{
 \alpha_k^1=\frac{2(c_\upsilon-9)\hat h_k+3\hat h_k^2-3z_k^2}{c_\upsilon}
 =\frac{\gamma_k-\gamma}{\gamma_k-1},
 }
 \eq{
 \alpha_k^2=2\hat h_k-3\frac{c_{\upsilon,k}}{c_\upsilon}-3\frac{\hat
 h_k+1}{c_\upsilon}=(5-3\gamma)\hat
 h_k-3\gamma_k\frac{\gamma-1}{\gamma_k-1},
 }
where the assignments $\gamma$ and $\gamma_k$ from \cite{groot}
are used. They can be expressed through the $c_\upsilon$ and the
$c_{\upsilon,k}$, defined in the (\ref{elquant}), as
 \eq{
 \gamma\equiv \frac1{c_\upsilon}+1, \quad
 \gamma_k\equiv\frac1{c_{\upsilon,k}}+1.
 }
The $\gamma_k^r$ can be rewritten as
 \eq{
 \gamma_k^r=\frac23\frac1{z_k^2K_2(z_k)}\int_{z_k}^\infty d\tau
 (\tau^2-z_k^2)^{5/2}\tau^r e^{-\tau}.
 }
Then it can be rewritten through the $a_k^r$:
 \eq{
 \gamma_k^r=\frac23(a^{r+4}-2z_k^2a^{r+2}+z_k^4a^r).
 }
Some values of the $\gamma_k^r$ are
 \eq{
 \gamma_k^0=10\hat h_k,
 }
 \eq{
 \gamma_k^1=10(6\hat h_k+z_k^2),
 }
 \eq{
 \gamma_k^2=10(7z_k^2+\hat h_k(42+z_k^2)).
 }

\section{The entropy density formula \label{appTherm}}
The Gibbs's potential is defined as
 \eq{\label{Phidef}
 \Phi(P,T)\equiv E(S,V)-ST+PV.
 }
The differential of the energy is defined as
 \eq{
 dE=TdS-PdV+\sum_k\mu_kdN_k=TdS-PdV+\sum_a\mu_adN_a,
 }
where it is rewritten through the independent chemical potentials
and the particle net charges $N_a$. Then the differential of the
$\Phi$ reads:
 \eq{\label{dPhi}
 d\Phi=-SdT+VdP+\sum_a \mu_a dN_a.
 }
Because the $\Phi$ is the function of the intrinsic variables $P$,
$T$ and the extrinsic $N_a$ the only possible form of it in the
thermodynamic limit is
 \eq{\label{Phi}
 \Phi=\sum_a N_a \phi_a(P,T),
 }
where $\phi_a$ are unknown functions. Then from the (\ref{dPhi})
one gets ${\frac{\p \Phi}{\p N_a} = \mu_a}$, which means that
${\phi_a = \mu_a}$. Then substituting the (\ref{Phi}) into the
(\ref{Phidef}) one gets the relation
 \eq{
 \sum_a N_a \mu_a(P,T)=E(S,V)-ST+PV.
 }
Being written for local infinitesimal volume it transforms into
the expression
 \eq{
 \sum_a n_a \mu_a=\epsilon-sT+P,
 }
from where the entropy density $s$ can be found:
 \eq{\label{entrden}
 s=\frac{\epsilon+P}{T}-\sum_a n_a \hat \mu_a.
 }

\section{The calculation of the collision brackets \label{appJ}}
The momentum parametrization and the most transformations of the
12-dimensional integrals used below are taken from \cite{groot}
(Chap. XI and XIII). Let's start from some assignments. The full
momentum is
 \eq{
 P^\mu = p^\mu_k+p^\mu_{1l}={p'}^\mu_k+{p'}^\mu_{1l}={P'}^\mu.
 }
The "relative" momentums before collision $Q^\mu$ and after
collision ${Q'}^\mu$ are defined as
 \eq{
 Q^\mu=\Delta_P^{\mu\nu}(p_{k\nu}-p_{1l\nu}), \quad
 {Q'}^\mu=\Delta_P^{\mu\nu}({p'}_{k\nu}-{p'}_{1l\nu}),
 }
with the assignment
 \eq{
 \Delta_P^{\mu\nu}=g^{\mu\nu}-\frac{P^\mu P^\nu}{P^2},
 }
where $P^2\equiv P^\mu P_\mu$. The covariant cosine of the
scattering angle can be expressed through the $Q^\mu$ and the
${Q'}^\mu$ as
 \eq{
 \cos\Theta=-\frac{Q \cdot Q'}{\sqrt{-Q^2}\sqrt{-{Q'}^2}},
 }
where $\cdot$ denotes convolution of 4-vectors. One also has ${Q^2
= {Q'}^2}$ and
 \eq{\label{Q2}
 Q^2=4m_k^2-(1+\alpha_{kl})^2P^2=-\left(P^2-M_{kl}^2\right)\left[1-\frac{M_{kl}^2}{P^2}
 \left(1-\frac{4\mu_{kl}}{M_{kl}}\right)\right],
 }
where
 \eq{\label{alphakl}
 M_{kl}\equiv m_k+m_l, \quad \mu_{kl}\equiv \frac{m_k
 m_l}{m_k+m_l}, \quad  \alpha_{kl}\equiv\frac{m_k^2-m_l^2}{P^2}=\mathrm{sign}(m_k-m_l)
 \sqrt{1-\frac{4\mu_{kl}}{M_{kl}}}\frac{M_{kl}^2}{P^2}.
 }
The function $\mathrm{sign}(x)$ is equal to 1, if ${x>0}$ and
equal to $-1$, if ${x<0}$. Note that not all $P^\mu$ and $Q^\mu$
are independent:
 \eq{
 P^\mu Q_\mu=0, \quad P^\mu {Q'}_\mu=0.
 }
To come from the variables ${(p_k^\mu,p_{1l}^\mu)}$ to the
variables ${(P^\mu,Q^\mu)}$ in the measure of integration first
one has to come from the ${(p_k^\mu,p_{1l}^\mu)}$ to the
${(p_k^\mu+p_{1l}^\mu,p_k^\mu-p_{1l}^\mu)}$ (the determinant is
equal to $16$) and then shift the relative momentum
$p_k^\mu-p_{1l}^\mu$ on the $\alpha_{kl}P^\mu$. Analogically for
the ${({p'}_k^\mu,{p'}_{1l}^\mu)}$ and the
${({P'}^\mu,{Q'}^\mu)}$. The inverse relations for the
$p_k^\mu,p_{1l}^\mu,{p'}_k^\mu,{p'}_{1l}^\mu$ through the
$P^\mu,Q^\mu,{Q'}^\mu$ are
 \eq{
 p_k^\mu=\frac12(1+\alpha_{kl})P^\mu+\frac12{Q}^\mu,
 }
 \eq{
 p_{1l}^\mu=\frac12(1-\alpha_{kl})P^\mu-\frac12{Q}^\mu,
 }
 \eq{
 {p'}_k^\mu=\frac12(1+\alpha_{kl})P^\mu+\frac12{Q'}^\mu,
 }
 \eq{
 {p'}_{1l}^\mu=\frac12(1-\alpha_{kl})P^\mu-\frac12{Q'}^\mu.
 }

There is a need to calculate the following integrals
 \begin{eqnarray}\label{Jint0}
  \nonumber &~& J_{kl}^{(a,b,d,e,f|q,r)} \equiv \frac{\gamma_{kl}}{T^6(4\pi)^2z_k^2z_l^2K_2(z_k)K_2(z_l)
  \sigma(T)} \int_{p_k,p_{1l},{p'}_k,{p'}_{1l}} e^{-P\cdot U/T}(1+\alpha_{kl})^q\\
  &~& \times(1-\alpha_{kl})^r \left(\frac{P^2}{T^2}\right)^a
  \left(\frac{P\cdot U}{T}\right)^b\left(\frac{Q\cdot U}{T}\right)^d\left(\frac{Q'\cdot U}{T}\right)^e
  \left(\frac{-Q\cdot {Q'}}{T^2}\right)^f W_{kl}.
 \end{eqnarray}
Let's start from the case of constant cross sections. After
nontrivial transformations, described in more details in
\cite{groot}, one arrives at
 \begin{eqnarray}\label{Jint}
  \nonumber &~&J_{kl}^{(a,b,d,e,f|q,r)}=\frac{\pi(d+e+1)!!\sigma^{(d,e,f)}_{1kl}}{z_k^2z_l^2K_2(z_k)K_2(z_l)}
  \sum_{q_1=0}^q\sum_{r_1=0}^r\sum_{k_2=0}^{\frac{d+e}2+f+1}\sum_{k_3=0}^{\frac{d+e}2+f+1}
  \sum_{h=0}^{[b/2]}(z_k+z_l)^{2(q_1+r_1+k_2+k_3)} \\
  &~&\times\left(\frac{z_k-z_l}{z_k+z_l}\right)^{q_1+r_1+2k_3}
  (-1)^{r_1+k_2+k_3+h}(2h-1)!! \binom{b}{2h}\binom{q}{q_1}\binom{r}{r_1}
  \binom{\frac{d+e}2+f+1}{k_2}\\
  \nonumber &~& \times\binom{\frac{d+e}2+f+1}{k_3} I\left(2(a+f-q_1-r_1-k_2-k_3)+3,
  b+\frac{d+e}2-h+1,z_k+z_l\right),
 \end{eqnarray}
where
 \eq{
 \sigma_{1kl}^{(d,e,f)}=\frac{\sigma^{cl}_{kl}}{\sigma(T)}
 \sum_{g=0}^{\min(d,e)} \sigma^{(f,g)} K(d,e,g),
 }
where ${\sigma^{cl}_{kl} = \gamma_{kl}\sigma_{kl}}$ is the
classical elastic differential constant cross section. The
$\sigma^{(f,g)}$ is equal to the real, nonzero and non-diverging
value (for any non-negative integer $g$)
 \eq{\label{sigmafg}
 \sigma^{(f,g)}=\frac{2g+1}{2}\int_{-1}^1 dx x^f P_g(x)=(2g+1) \frac{f!}{(f-g)!!(f+g+1)!!},
 }
if the difference ${f-g}$ is even and ${g \leq f}$. Above the
$P_{\text{g}}(x)$ is the Legendre polynomial. The $K(d,e,g)$ is
equal to the real, the nonzero and non-diverging quantity (for any
non-negative integer $g$)
 \eq{
 K(d,e,g)=\frac{d! e!}{(d-g)!! (d+g+1)!! (e-g)!! (e+g+1)!!},
 }
if ${g \leq \min(d,e)}$ and both the ${d-g}$ and the ${e-g}$ are
even (which also implies that ${d+e}$ is even). The ${[...]}$
denotes the integer part. The integral $I$ is
 \eq{\label{IIntdef}
 I(r,n,x)\equiv x^{r+n+1}\int_1^\infty du u^{r+n}K_n(xu).
 }
Also there is the following frequently used combination of the $J$
integrals
 \eq{
 {J'}_{kl}^{(a,b,d,e,f|q,r)}\equiv\sum_{u=0}^f (-1)^u\binom{f}{u}
 (2z_k)^{2(f-u)}J_{kl}^{(a+k,b,d+e,0,0|q+2u,r)}-J_{kl}^{(a,b,d,e,f|q,r)}.
 }
The first term in the difference is obtained by the replacement of
the $Q'$ on the $Q$ everywhere except for the $W_{kl}$. Using this
fact, the $J'$ can be rewritten in the form
 \begin{eqnarray}\label{Jpint}
  \nonumber &~&{J'}_{kl}^{(a,b,d,e,f|q,r)}=\frac{\pi(d+e-1)!!
  \sigma^{(d,e,f)}_{kl}}{z_k^2z_l^2K_2(z_k)K_2(z_l)}
  \sum_{q_1=0}^q\sum_{r_1=0}^r\sum_{k_2=0}^{\frac{d+e}2+f+1}\sum_{k_3=0}^{\frac{d+e}2+f+1}
  \sum_{h=0}^{[b/2]}(z_k+z_l)^{2(q_1+r_1+k_2+k_3)} \\
  &~&\times\left(\frac{z_k-z_l}{z_k+z_l}\right)^{q_1+r_1+2k_3}
  (-1)^{r_1+k_2+k_3+h}(2h-1)!! \binom{b}{2h}\binom{q}{q_1}\binom{r}{r_1}
  \binom{\frac{d+e}2+f+1}{k_2}\\
  \nonumber &~& \times\binom{\frac{d+e}2+f+1}{k_3}
  I\left(2(a+f-q_1-r_1-k_2-k_3)+3, b+\frac{d+e}2-h+1,z_k+z_l\right),
 \end{eqnarray}
where
 \begin{eqnarray}\label{sigmadef}
  \nonumber \sigma^{(d,e,f)}_{kl}&=&\frac{\sigma^{cl}_{kl}}{\sigma(T)}(d+e+1)\left(K(d+e,0,0)\sigma^{(0,0)}
  -\sum_{g=0}^{\min(d,e)}K(d,e,g)\sigma^{(f,g)}\right)\\
  &=&\frac{\sigma^{cl}_{kl}}{\sigma(T)}\left(1-(d+e+1)\sum_{g=0}^{\min(d,e)}K(d,e,g)\sigma^{(f,g)}\right).
 \end{eqnarray}
There is a recurrence relation for the integral $I$
(\ref{IIntdef}) \cite{groot} (Chap. XI, Sec. 1):
 \eq{\label{recrel}
 I(r,n,x)=(r-1)(r+2n-1)I(r-2,n,x)+(r-1)x^{r+n-1}K_n(x)+x^{r+n}K_{n+1}(x).
 }
For the calculations one needs only the integrals $I(r,n,x)$ with
the positive values of the $n$ and the odd values of the $r$. If
${r \geq -2n+1}$, the $I$ integrals can be expressed through the
Bessel functions $K_n(x)$, using the (\ref{recrel}), when ${r=1}$
or ${r=-2n+1}$. Then using the recurrence relation for the
$K_n(x)$ \cite{luke}
 \eq{\label{Krecrel}
 K_{n+1}(x)=K_{n-1}(x)+\frac{2n}{x}K_n(x),
 }
the final result can be expressed through a couple of Bessel
functions. If ${r\leq-2n-1}$, then the recurrence relation
(\ref{recrel}) becomes singular if one tries to express the
$I(r,n,x)$ through the $I(-2n+1,n,x)$. Using the (\ref{recrel}),
the $I$ integrals with ${r \leq -2n-1}$ can be expressed through
the integrals $G(n,x)$
 \eq{\label{Gdef}
 G(n,x)\equiv I(-2n-1,n,x)=x^{-n}\int_1^\infty du u^{-n-1}K_n(xu).
 }
There is a recurrence relation for the $G(n,x)$:
 \eq{\label{Grecrel}
 G(n,x)=-\frac1{2n}(G(n-1,x)-x^{-n} K_n(x)).
 }
It can be easily proved by the integration by parts of the
(\ref{Gdef}) and using the following relation for the $K_n(x)$
\cite{luke}
 \eq{\label{dKdx}
 \frac{\p }{\p x}K_n(x)=-\frac{n}{x} K_n(x)-K_{n-1}(x).
 }
It is found that collision brackets have the simplest form if they
are expressed through $G(n,x)$ with ${n=3}$ or ${n=2}$ and the
Bessel functions $K_3(x)$ and $K_2(x)$ or $K_2(x)$ and $K_1(x)$.
It was chosen to take ${G(x) \equiv G(3,x)}$ and $K_3(x)$,
$K_2(x)$. The $G(x)$ can be expressed through the Meijer function
\cite{meijer}
 \eq{
 G(x)=\frac{1}{32}
 G_{1,3}^{3,0}\left((x/2)^2\left|
 \begin{array}{c}
  1 \\
  -3,0,0
 \end{array}\right.
 \right).
 }
The needed scalar collision brackets can be expressed through the
$J'$ as
 \eq{
 [\tau^r,\tau^s_1]_{kl}=\frac1{2^{r+s}}\sum_{u=1}^r\sum_{\upsilon=1}^s
 (-1)^\upsilon \binom{r}{u} \binom{s}{\upsilon} {J'}_{kl}^{(0,r+s-u-\upsilon,u,
 \upsilon,0|r-u,s-\upsilon)},
 }
 \eq{
 [\tau^r,\tau^s]_{kl}=\frac1{2^{r+s}}\sum_{u=1}^r\sum_{\upsilon=1}^s
 \binom{r}{u} \binom{s}{\upsilon} {J'}_{kl}^{(0,r+s-u-\upsilon,u,
 \upsilon,0|r+s-u-\upsilon,0)},
 }
and the needed tensorial collision brackets can be expressed as
 \begin{eqnarray}
  \nonumber &~& [\tau^r\overset{\circ}{\overline{\pi^{\mu} \pi^{\nu}}},
  \tau_1^s\overset{\circ}{\overline{\pi_{1\mu} \pi_{1\nu}}}]_{kl}=
  \frac1{2^{r+s+4}}\sum_{n_1=0}^r\sum_{n_2=0}^s \binom{s}{n_2}\binom{r}{n_1}(-1)^{s-n_2}
  ({J'}_{kl}^{(2,n_1+n_2,r-n_1,s-n_2,0|2+n_1,2+n_2)} \\
  \nonumber &+&2{J'}_{kl}^{(1,n_1+n_2,r-n_1,s-n_2,1|1+n_1,1+n_2)}+{J'}_{kl}^{(0,n_1+n_2,r-n_1,s-n_2,2|n_1,n_2)}) \\
  \nonumber &-&\frac1{2^{r+s+3}}\sum_{n_1=0}^{r+1}\sum_{n_2=0}^{s+1} \binom{s+1}{n_2}\binom{r+1}{n_1}(-1)^{s+1-n_2}
  ({J'}_{kl}^{(1,n_1+n_2,r+1-n_1,s+1-n_2,0|1+n_1,1+n_2)} \\
  \nonumber &+&{J'}_{kl}^{(0,n_1+n_2,r+1-n_1,s+1-n_2,1|n_1,n_2)})+\frac23[\tau^{r+2},\tau^{s+2}_1]_{kl}
  +\frac13 z_l^2 [\tau^{r+2},\tau^{s}_1]_{kl} \\
  &+&\frac13 z_k^2 [\tau^{r},\tau^{s+2}_1]_{kl}- \frac13 z_k^2 z_l^2
  [\tau^{r},\tau^{s}_1]_{kl},
 \end{eqnarray}
 \begin{eqnarray}
  \nonumber &~& [\tau^r\overset{\circ}{\overline{\pi^{\mu} \pi^{\nu}}},
  \tau^s\overset{\circ}{\overline{\pi_{\mu} \pi_{\nu}}}]_{kl}=
  \frac1{2^{r+s+4}}\sum_{n_1=0}^r\sum_{n_2=0}^s \binom{s}{n_2}\binom{r}{n_1}
  ({J'}_{kl}^{(2,n_1+n_2,r-n_1,s-n_2,0|4+n_1+n_2,0)} \\
  \nonumber &-&2{J'}_{kl}^{(1,n_1+n_2,r-n_1,s-n_2,1|2+n_1+n_2,0)}+{J'}_{kl}^{(0,n_1+n_2,r-n_1,s-n_2,2|n_1+n_2,0)}) \\
  \nonumber &-&\frac1{2^{r+s+3}}\sum_{n_1=0}^{r+1}\sum_{n_2=0}^{s+1} \binom{s+1}{n_2}\binom{r+1}{n_1}
  ({J'}_{kl}^{(1,n_1+n_2,r+1-n_1,s+1-n_2,0|2+n_1+n_2,0)} \\
  \nonumber &-&{J'}_{kl}^{(0,n_1+n_2,r+1-n_1,s+1-n_2,1|n_1+n_2,0)})+\frac23[\tau^{r+2},\tau^{s+2}]_{kl}
  +\frac13 z_k^2 [\tau^{r+2},\tau^{s}]_{kl} \\
  &+&\frac13 z_k^2 [\tau^{r},\tau^{s+2}]_{kl}-\frac13  z_k^4[\tau^{r},\tau^{s}]_{kl}.
 \end{eqnarray}
Below some lowest orders collision brackets are presented with the
following notations:
 \begin{eqnarray}
  \nonumber \widetilde K_1 &\equiv& \frac{K_3(z_k+z_l)}{K_2(z_k)K_2(z_l)},
  \quad \widetilde K_2\equiv \frac{K_2(z_k+z_l)}{K_2(z_k)K_2(z_l)},
  \quad \widetilde K_3\equiv \frac{G(z_k+z_l)}{K_2(z_k)K_2(z_l)}, \\
  Z_{kl} &\equiv& z_k+z_l, \quad z_{kl} \equiv z_k-z_l.
 \end{eqnarray}
For the scalar collision brackets one has:
 \eq{\label{br211}
 -[\tau,\tau_1]_{kl}=[\tau,\tau]_{kl}=\frac{\sigma^{cl}_{kl}}{\sigma(T)}\frac{\pi}{2z_k^2z_l^2Z_{kl}^2}
 (P_{s1}^{(1,1)}\widetilde K_1+P_{s2}^{(1,1)}\widetilde K_2+P_{s3}^{(1,1)}\widetilde K_3),
 }
where
 \eq{
 P_{s1}^{(1,1)}=-2 Z_{kl} (z_{kl}^4+4 z_{kl}^2 Z_{kl}^2-2 Z_{kl}^4),
 }
 \eq{
 P_{s2}^{(1,1)}=z_{kl}^4 (3 Z_{kl}^2+8)+32 z_{kl}^2 Z_{kl}^2+8 Z_{kl}^4,
 }
 \eq{
 P_{s3}^{(1,1)}=-3 z_{kl}^4 Z_{kl}^6,
 }
and
 \eq{
 [\tau,\tau_1^2]_{kl}=[\tau^2,\tau_1]_{lk}=\frac{\sigma^{cl}_{kl}}{\sigma(T)}\frac{\pi}{4z_k^2z_l^2Z_{kl}^2}
 (P_{s11}^{(1,2)}\widetilde K_1+P_{s12}^{(1,2)}\widetilde K_2+P_{s13}^{(1,2)}\widetilde K_3),
 }
where
 \eq{
 P_{s11}^{(1,2)}=2 Z_{kl} (z_{kl}^5 Z_{kl}+8 z_{kl}^4+16 z_{kl}^3 Z_{kl}
 +32 z_{kl}^2 Z_{kl}^2+16 z_{kl} Z_{kl}^3-40 Z_{kl}^4),
 }
 \begin{eqnarray}
  \nonumber P_{s12}^{(1,2)}&=&-z_{kl}^5 Z_{kl} (Z_{kl}^2+8)-8 z_{kl}^4 (Z_{kl}^2+8)
  -16 z_{kl}^3 Z_{kl} (Z_{kl}^2+8)\\ &+&16 z_{kl}^2 Z_{kl}^2 (Z_{kl}^2-16)
  +8 z_{kl} Z_{kl}^3 (Z_{kl}^2-16)-8 Z_{kl}^4 (Z_{kl}^2+8),
 \end{eqnarray}
 \eq{
 P_{s13}^{(1,2)}=z_{kl}^5 Z_{kl}^7,
 }
and
 \eq{
 [\tau,\tau^2]_{kl}=[\tau^2,\tau]_{kl}=\frac{\sigma^{cl}_{kl}}{\sigma(T)}\frac{\pi}{4z_k^2z_l^2Z_{kl}^2}
 (P_{s21}^{(1,2)}\widetilde K_1+P_{s22}^{(1,2)}\widetilde K_2+P_{s23}^{(1,2)}\widetilde K_3),
 }
where
 \eq{
 P_{s21}^{(1,2)}=2 Z_{kl} (z_{kl}^5 Z_{kl}-8 z_{kl}^4+16 z_{kl}^3 Z_{kl}
 -32 z_{kl}^2 Z_{kl}^2+16 z_{kl} Z_{kl}^3+40 Z_{kl}^4),
 }
 \begin{eqnarray}
  \nonumber P_{s22}^{(1,2)}&=&-z_{kl}^5 Z_{kl} (Z_{kl}^2+8)+
  8 z_{kl}^4 (Z_{kl}^2+8)-16 z_{kl}^3 Z_{kl} (Z_{kl}^2+8)\\
  &-&16 z_{kl}^2 Z_{kl}^2 (Z_{kl}^2-16)+8 z_{kl} Z_{kl}^3 (Z_{kl}^2-16)
  +8 Z_{kl}^4 (Z_{kl}^2+8),
 \end{eqnarray}
 \eq{
 P_{s23}^{(1,2)}=z_{kl}^5 Z_{kl}^7,
 }
and
 \eq{
 [\tau^2,\tau_1^2]_{kl}=\frac{\sigma^{cl}_{kl}}{\sigma(T)}\frac{\pi}{24z_k^2z_l^2Z_{kl}^2}
 (P_{s11}^{(2,2)}\widetilde K_1+P_{s12}^{(2,2)}\widetilde K_2+P_{s13}^{(2,2)}\widetilde K_3),
 }
where
 \begin{eqnarray}
  \nonumber P_{s11}^{(2,2)}&=&-2 Z_{kl} [z_{kl}^6 (Z_{kl}^2+2)+6 z_{kl}^4
  (11 Z_{kl}^2-32)-72 z_{kl}^2 Z_{kl}^2 (Z_{kl}^2+8)\\ &+&24 Z_{kl}^4 (Z_{kl}^2+96)],
 \end{eqnarray}
 \begin{eqnarray}
  \nonumber P_{s12}^{(2,2)}&=&z_{kl}^6 (Z_{kl}^4+10 Z_{kl}^2+16)-6 z_{kl}^4
  (Z_{kl}^4-56 Z_{kl}^2+256)\\ &+&144 z_{kl}^2 Z_{kl}^2 (5 Z_{kl}^2-32)
  -48 Z_{kl}^4 (13 Z_{kl}^2+32),
 \end{eqnarray}
 \eq{
 P_{s13}^{(2,2)}=-z_{kl}^4 Z_{kl}^6 [z_{kl}^2 (Z_{kl}^2-6)-6 Z_{kl}^2],
 }
and
 \eq{
 [\tau^2,\tau^2]_{kl}=\frac{\sigma^{cl}_{kl}}{\sigma(T)}\frac{\pi}{24z_k^2z_l^2Z_{kl}^2}
 (P_{s21}^{(2,2)}\widetilde K_1+P_{s22}^{(2,2)}\widetilde K_2+P_{s23}^{(2,2)}\widetilde K_3),
 }
where
 \begin{eqnarray}
  \nonumber P_{s21}^{(2,2)}&=&-2 Z_{kl} [z_{kl}^6 (Z_{kl}^2+2)-36 z_{kl}^5
  Z_{kl}+18 z_{kl}^4 (Z_{kl}^2+16)+96 z_{kl}^3 Z_{kl} (Z_{kl}^2-10)\\
  &+&24 z_{kl}^2 Z_{kl}^2 (Z_{kl}^2+56)-48 z_{kl} Z_{kl}^3
  (Z_{kl}^2+20)-24 Z_{kl}^4 (Z_{kl}^2+100)],
 \end{eqnarray}
 \begin{eqnarray}
  \nonumber P_{s22}^{(2,2)}&=&z_{kl}^6 (Z_{kl}^4+10 Z_{kl}^2+16)+
  12 z_{kl}^5 Z_{kl} (Z_{kl}^2-24)-6 z_{kl}^4 (Z_{kl}^4-72 Z_{kl}^2-384)\\
  \nonumber &-&192 z_{kl}^3 Z_{kl} (Z_{kl}^2+40)-48 z_{kl}^2 Z_{kl}^2 (13 Z_{kl}^2-224)
  +96 z_{kl} Z_{kl}^3 (7 Z_{kl}^2-80)\\ &+&48 Z_{kl}^4 (13 Z_{kl}^2+48),
 \end{eqnarray}
 \eq{
 P_{s23}^{(2,2)}=-z_{kl}^4 Z_{kl}^6 [z_{kl}^2 (Z_{kl}^2-6)+12 z_{kl} Z_{kl}-6 Z_{kl}^2].
 }
And for the tensor collision brackets one has:
 \eq{
 [\overset{\circ}{\overline{\pi^{\mu} \pi^{\nu}}},
 \overset{\circ}{\overline{\pi_{1\mu} \pi_{1\nu}}}]_{kl}=
 \frac{\sigma^{cl}_{kl}}{\sigma(T)}\frac{\pi}{72z_k^2z_l^2Z_{kl}^2}(P_{T11}^{(0,0)}\widetilde K_1
 +P_{T12}^{(0,0)}\widetilde K_2+P_{T13}^{(0,0)}\widetilde K_3),
 }
where
 \begin{eqnarray}
  \nonumber P_{T11}^{(0,0)}&=&-2 Z_{kl} [z_{kl}^6 (5 Z_{kl}^2-8)+
  24 z_{kl}^4 (Z_{kl}^2-16)-144 z_{kl}^2 Z_{kl}^2 (Z_{kl}^2+8)\\
  &+&48 Z_{kl}^4 (Z_{kl}^2+72)],
 \end{eqnarray}
 \begin{eqnarray}
  \nonumber P_{T12}^{(0,0)}&=&z_{kl}^6 (5 Z_{kl}^4-40 Z_{kl}^2-64)-
  24 z_{kl}^4 (5 Z_{kl}^4+8 Z_{kl}^2+128)\\&+&576 z_{kl}^2 Z_{kl}^2
  (Z_{kl}^2-16)-192 Z_{kl}^4 (5 Z_{kl}^2+16),
 \end{eqnarray}
 \eq{
 P_{T13}^{(0,0)}=-5 z_{kl}^4 Z_{kl}^6 [z_{kl}^2 (Z_{kl}^2-24)-24 Z_{kl}^2],
 }
and
 \eq{
 [\overset{\circ}{\overline{\pi^{\mu} \pi^{\nu}}},
 \overset{\circ}{\overline{\pi_{\mu} \pi_{\nu}}}]_{kl}=
 \frac{\sigma^{cl}_{kl}}{\sigma(T)}\frac{\pi}{72z_k^2z_l^2Z_{kl}^2}(P_{T21}^{(0,0)}\widetilde K_1
 +P_{T22}^{(0,0)}\widetilde K_2
 +P_{T23}^{(0,0)}\widetilde K_3),
 }
where
 \begin{eqnarray}
  \nonumber P_{T21}^{(0,0)}&=&2 Z_{kl} [z_{kl}^6 (8-5 Z_{kl}^2)+72 z_{kl}^4 (3 Z_{kl}^2-8)
  -480 z_{kl}^3 Z_{kl} (Z_{kl}^2-4) \\ &-&336 z_{kl}^2 Z_{kl}^2 (Z_{kl}^2+8)+240 z_{kl} Z_{kl}^3
  (Z_{kl}^2+8)+192 Z_{kl}^4 (Z_{kl}^2+67)],
 \end{eqnarray}
 \begin{eqnarray}
  \nonumber P_{T22}^{(0,0)}&=&z_{kl}^6 (5 Z_{kl}^4-40 Z_{kl}^2-64)+240 z_{kl}^5
  Z_{kl}^3-24 z_{kl}^4 (5 Z_{kl}^4+48 Z_{kl}^2-192)\\
  \nonumber &+&1920 z_{kl}^3 Z_{kl}
  (Z_{kl}^2-8)-192 z_{kl}^2 Z_{kl}^2 (17 Z_{kl}^2-112)+1920 z_{kl} Z_{kl}^3
  (Z_{kl}^2-8)\\ &+&768 Z_{kl}^4 (5 Z_{kl}^2+6),
 \end{eqnarray}
 \eq{
 P_{T23}^{(0,0)}=-5 z_{kl}^4 Z_{kl}^6 [z_{kl}^2 (Z_{kl}^2-24)+48 z_{kl} Z_{kl}-24 Z_{kl}^2].
 }
If ${z_k=z_l}$, then the $G(x)$ function is eliminated everywhere
and the collision brackets simplify considerably.

The constant cross sections are not the only possible ones
resulting in the analytical expressions for the collision brackets
and, hence, for the transport coefficients. The analytical
expressions through the Bessel and the Meijer functions (possibly,
relatively simple ones) can be obtained also using some
non-constant cross sections. For example, this is possible in the
case when cross sections are proportional to integer powers of the
$P^2$ and/or the $\cos\Theta$. As one can see from the
(\ref{Jint0}), an integer power of the $P^2$ would result just in
shift of the index $a$ which can be easily taken into account in
the (\ref{Jint}) or the (\ref{Jpint}). A power of the $\cos\Theta$
would result in shift of the first index of the $\sigma^{(f,g)}$
(\ref{sigmafg}), and one would get
 \eq{
 \sigma^{(d,e,f|f')}_{kl}=\frac{\sigma^{cl}_{kl}}{\sigma(T)}\left((f'+1)^{-1}
  -(d+e+1)\sum_{g=0}^{\min(d,e)}K(d,e,g)\sigma^{(f+f',g)}\right),
 }
instead of the $\sigma^{(d,e,f)}_{kl}$ (\ref{sigmadef}). So that
the $J$ or the $J'$ integrals just get a different factor.
Half-integer powers\footnote{Actually all powers of the $P^2$ can
also be taken into account in the same way if all mutual
differences between all powers of all the $P^2$ terms are
integers.} of the $P^2$ can also be taken into account in the same
way as the integer powers of the $P^2$ however one would need to
introduce one more special function through which to express the
$I$ integrals with even $r$ using the recurrence relation
(\ref{recrel}). The convenient choice of this function is found to
be
 \eq{
 G_2(n,x)\equiv I(-2n+2,n,x)=x^{-n+3}\int_1^\infty du u^{-n+2}K_n(xu).
 }
Then one can express these functions through the function (the
$n=3$ is found to be the convenient choice)
 \eq{
 G_2(x)\equiv G_2(3,x)=\frac{\pi}{6}-\frac{1}{4}
 G_{1,3}^{2,1}\left((x/2)^2\left|
 \begin{array}{c}
  1 \\
  -\frac32,\frac32,0
 \end{array}\right.
 \right),
 }
using the recurrence relation
 \eq{
 G_2(n,x)=\frac1{2n-3}[K_n(x)x^{-n+3}-G_2(n-1,x)].
 }
Its derivation is the same as the derivation of the
(\ref{Grecrel}). Using the above mentioned prescriptions one can
calculate collision brackets for a quite large class of cross
sections. Few examples of simple energy dependencies for cross
sections are considered below.

Using the cross sections $\sigma_{kl}P^2/(T^2Z_{kl}^2) \equiv
\sigma_{kl}\upsilon^2/Z_{kl}^2$, growing with the energy, one
finds for the scalar collision brackets ($\sigma_{kl}$ is just a
positive dimensional constant here, and ${\sigma^{cl}_{kl} =
\gamma_{kl}\sigma_{kl}}$)
 \eq{
 -[\tau,\tau_1]_{kl}=[\tau,\tau]_{kl}=\frac{\sigma^{cl}_{kl}}{\sigma(T)}\frac{8 \pi }{z_k^2 z_l^2 Z_{kl}^2}
 (P_{s1}^{(1,1|1)}\widetilde K_1+P_{s2}^{(1,1|1)}\widetilde K_2),
 }
where
 \eq{
 P_{s1}^{(1,1|1)}=\frac{1}{4} Z_{kl} (z_{kl}^4-2 z_{kl}^2 Z_{kl}^2+Z_{kl}^4+48 Z_{kl}^2),
 }
 \eq{
 P_{s2}^{(1,1|1)}=-z_{kl}^4-z_{kl}^2 Z_{kl}^2+2 Z_{kl}^4,
 }
and
 \eq{
 [\tau,\tau_1^2]_{kl}=[\tau^2,\tau_1]_{lk}=-\frac{\sigma^{cl}_{kl}}{\sigma(T)}\frac{4 \pi }{z_k^2 z_l^2 Z_{kl}^3}
 (P_{s11}^{(1,2|1)}\widetilde K_1+P_{s12}^{(1,2|1)}\widetilde K_2),
 }
where
 \eq{
 P_{s11}^{(1,2|1)}=Z_{kl} (z_{kl}^5+2 z_{kl}^4 Z_{kl}+4 z_{kl}^3 Z_{kl}^2-
 10 z_{kl}^2 Z_{kl}^3-5 z_{kl} Z_{kl}^4+8 Z_{kl}^3 (Z_{kl}^2+30)),
 }
 \begin{eqnarray}
  \nonumber P_{s12}^{(1,2|1)}&=&\frac{1}{2} [-z_{kl}^5 (Z_{kl}^2+8)+z_{kl}^4 Z_{kl} (Z_{kl}^2-16)
  +2 z_{kl}^3 Z_{kl}^2 (Z_{kl}^2-16)\\
  &-&2 z_{kl}^2 Z_{kl}^3 (Z_{kl}^2+8)-z_{kl} Z_{kl}^4 (Z_{kl}^2+8)+Z_{kl}^5 (Z_{kl}^2+80)],
 \end{eqnarray}
and
 \eq{
 [\tau,\tau^2]_{kl}=[\tau^2,\tau]_{kl}=\frac{\sigma^{cl}_{kl}}{\sigma(T)}\frac{4 \pi }{z_k^2 z_l^2 Z_{kl}^3}
 (P_{s21}^{(1,2|1)}\widetilde K_1+P_{s22}^{(1,2|1)}\widetilde K_2),
 }
where
 \eq{
 P_{s21}^{(1,2|1)}=Z_{kl} (-z_{kl}^5+2 z_{kl}^4 Z_{kl}-4 z_{kl}^3 Z_{kl}^2-10 z_{kl}^2 Z_{kl}^3+5 z_{kl} Z_{kl}^4+8 Z_{kl}^3 (Z_{kl}^2+30)),
 }
 \begin{eqnarray}
  \nonumber P_{s22}^{(1,2|1)}&=&\frac{1}{2} [z_{kl}^5 (Z_{kl}^2+8)+z_{kl}^4 Z_{kl} (Z_{kl}^2-16)
  -2 z_{kl}^3 Z_{kl}^2 (Z_{kl}^2-16)\\
  &-&2 z_{kl}^2 Z_{kl}^3 (Z_{kl}^2+8)+z_{kl} Z_{kl}^4 (Z_{kl}^2+8)+Z_{kl}^5 (Z_{kl}^2+80)],
 \end{eqnarray}
and
 \eq{
 [\tau^2,\tau_1^2]_{kl}=-\frac{\sigma^{cl}_{kl}}{\sigma(T)}\frac{\pi }{6 z_k^2 z_l^2 Z_{kl}^4}
 (P_{s11}^{(2,2|1)}\widetilde K_1+P_{s12}^{(2,2|1)}\widetilde K_2+P_{s13}^{(2,2|1)}\widetilde K_3),
 }
where
 \begin{eqnarray}
  \nonumber P_{s11}^{(2,2|1)}&=&2 [-z_{kl}^6 Z_{kl} (5 Z_{kl}^2+4)+18 z_{kl}^4 Z_{kl}^3
  (Z_{kl}^2-2)\\
  &-&18 z_{kl}^2 Z_{kl}^5 (Z_{kl}^2+70)+6 Z_{kl}^5 (Z_{kl}^4+222 Z_{kl}^2+5120)],
 \end{eqnarray}
 \begin{eqnarray}
  \nonumber P_{s12}^{(2,2|1)}&=&z_{kl}^6 (-Z_{kl}^4+44 Z_{kl}^2+32)+36 z_{kl}^4 Z_{kl}^2
  (3 Z_{kl}^2+8)\\
  &-&36 z_{kl}^2 Z_{kl}^4 (11 Z_{kl}^2-8)+12 Z_{kl}^6 (19 Z_{kl}^2+856),
 \end{eqnarray}
 \eq{
 P_{s13}^{(2,2|1)}=z_{kl}^6 Z_{kl}^8,
 }
and
 \eq{
 [\tau^2,\tau^2]_{kl}=\frac{\sigma^{cl}_{kl}}{\sigma(T)}\frac{\pi }{6 z_k^2 z_l^2 Z_{kl}^4}
 (P_{s21}^{(2,2|1)}\widetilde K_1+P_{s22}^{(2,2|1)}\widetilde K_2+P_{s23}^{(2,2|1)}\widetilde K_3),
 }
where
 \begin{eqnarray}
  \nonumber P_{s21}^{(2,2|1)}&=&2 Z_{kl} [z_{kl}^6 (5 Z_{kl}^2+4)+12 z_{kl}^5 Z_{kl}
  (Z_{kl}^2-10)-6 z_{kl}^4 Z_{kl}^2 (Z_{kl}^2-46)\\
  \nonumber &-&24 z_{kl}^3 Z_{kl}^3 (Z_{kl}^2+20)-6 z_{kl}^2 Z_{kl}^4 (Z_{kl}^2+182)+12 z_{kl}
  Z_{kl}^5 (Z_{kl}^2+98)\\
  &+&6 Z_{kl}^4 (Z_{kl}^4+226 Z_{kl}^2+5440)],
 \end{eqnarray}
 \begin{eqnarray}
  \nonumber P_{s22}^{(2,2|1)}&=&z_{kl}^6 (Z_{kl}^4-44 Z_{kl}^2-32)+24 z_{kl}^5 Z_{kl}
  (Z_{kl}^2+40)+12 z_{kl}^4 Z_{kl}^2 (5 Z_{kl}^2-184)\\
  \nonumber &-&48 z_{kl}^3 Z_{kl}^3 (7 Z_{kl}^2-80)-12 z_{kl}^2 Z_{kl}^4 (19 Z_{kl}^2+184)
  +24 z_{kl} Z_{kl}^5 (13 Z_{kl}^2+40)\\
  &+&12 Z_{kl}^6 (19 Z_{kl}^2+904),
 \end{eqnarray}
 \eq{
 P_{s23}^{(2,2|1)}=-z_{kl}^6 Z_{kl}^8.
 }
And for the tensor collision brackets one has:
 \eq{
 [\overset{\circ}{\overline{\pi^{\mu} \pi^{\nu}}},
 \overset{\circ}{\overline{\pi_{1\mu} \pi_{1\nu}}}]_{kl}=
 -\frac{\sigma^{cl}_{kl}}{\sigma(T)}\frac{2 \pi }{9 z_k^2 z_l^2 Z_{kl}^4}(P_{T11}^{(0,0|1)}\widetilde K_1
 +P_{T12}^{(0,0|1)}\widetilde K_2+P_{T13}^{(0,0|1)}\widetilde K_3),
 }
where
 \begin{eqnarray}
  \nonumber P_{T11}^{(0,0|1)}&=&2 [2 z_{kl}^6 Z_{kl} (Z_{kl}^2-1)+9 z_{kl}^4 Z_{kl}^3
  (Z_{kl}^2-2)\\
  &-&9 z_{kl}^2 Z_{kl}^5 (Z_{kl}^2+22)+3 Z_{kl}^5 (Z_{kl}^4+126 Z_{kl}^2+2240)],
 \end{eqnarray}
 \begin{eqnarray}
  \nonumber P_{T12}^{(0,0|1)}&=&z_{kl}^6 (-5 Z_{kl}^4-14 Z_{kl}^2+16)-18 z_{kl}^4 Z_{kl}^2
  (3 Z_{kl}^2-8)\\
  &-&18 z_{kl}^2 Z_{kl}^4 (5 Z_{kl}^2-8)+6 Z_{kl}^6 (13 Z_{kl}^2+376),
 \end{eqnarray}
 \eq{
 P_{T13}^{(0,0|1)}=5 z_{kl}^6 Z_{kl}^8,
 }
and
 \eq{
 [\overset{\circ}{\overline{\pi^{\mu} \pi^{\nu}}},
 \overset{\circ}{\overline{\pi_{\mu} \pi_{\nu}}}]_{kl}=
 \frac{\sigma^{cl}_{kl}}{\sigma(T)}\frac{2 \pi }{9 z_k^2 Z_{kl}^4 z_l^2}(P_{T21}^{(0,0|1)}\widetilde K_1
 +P_{T22}^{(0,0|1)}\widetilde K_2+P_{T23}^{(0,0|1)}\widetilde K_3),
 }
where
 \begin{eqnarray}
  \nonumber P_{T21}^{(0,0|1)}&=&2 Z_{kl} [-2 z_{kl}^6 (Z_{kl}^2-1)+15 z_{kl}^5 Z_{kl}
  (Z_{kl}^2-4)+6 z_{kl}^4 Z_{kl}^2 (Z_{kl}^2+23)\\
  \nonumber &-&30 z_{kl}^3 Z_{kl}^3 (Z_{kl}^2+8)-3 z_{kl}^2 Z_{kl}^4 (7 Z_{kl}^2+614)+15 z_{kl}
  Z_{kl}^5 (Z_{kl}^2+68)\\
  &+&6 Z_{kl}^4 (2 Z_{kl}^4+377 Z_{kl}^2+8480)],
 \end{eqnarray}
 \begin{eqnarray}
  \nonumber P_{T22}^{(0,0|1)}&=&z_{kl}^6 (5 Z_{kl}^4+14 Z_{kl}^2-16)-60 z_{kl}^5 Z_{kl}
  (Z_{kl}^2-8)+6 z_{kl}^4 Z_{kl}^2 (29 Z_{kl}^2-184)\\
  \nonumber &-&240 z_{kl}^3 Z_{kl}^3 (Z_{kl}^2-8)-6 z_{kl}^2 Z_{kl}^4 (85 Z_{kl}^2+184)
  +60 z_{kl} Z_{kl}^5 (5 Z_{kl}^2+8)\\
  &+&6 Z_{kl}^6 (67 Z_{kl}^2+2824)),
 \end{eqnarray}
 \eq{
 P_{T23}^{(0,0|1)}=-5 z_{kl}^6 Z_{kl}^8.
 }
Using the descending cross sections $\sigma_{kl}
Z_{kl}^2/\upsilon^2$ one finds for the scalar collision brackets
 \eq{
 -[\tau,\tau_1]_{kl}=[\tau,\tau]_{kl}=\frac{\sigma^{cl}_{kl}}{\sigma(T)}\frac{\pi }{8 z_k^2 z_l^2}
 (P_{s1}^{(1,1|-1)}\widetilde K_1+P_{s2}^{(1,1|-1)}\widetilde K_2+P_{s3}^{(1,1|-1)}\widetilde K_3),
 }
where
 \eq{
 P_{s1}^{(1,1|-1)}=2 Z_{kl} (-z_{kl}^4+8 z_{kl}^2+8 Z_{kl}^2),
 }
 \eq{
 P_{s2}^{(1,1|-1)}=z_{kl}^4 (Z_{kl}^2-8)-8 z_{kl}^2 (3 Z_{kl}^2+8)-64 Z_{kl}^2,
 }
 \eq{
 P_{s3}^{(1,1|-1)}=z_{kl}^2 Z_{kl}^4 (-z_{kl}^2 (Z_{kl}^2-24)+24 Z_{kl}^2),
 }
and
 \eq{
 [\tau,\tau_1^2]_{kl}=[\tau^2,\tau_1]_{lk}=-\frac{\sigma^{cl}_{kl}}{\sigma(T)}\frac{\pi }{64 z_k^2 z_l^2}
 (P_{s11}^{(1,2|-1)}\widetilde K_1+P_{s12}^{(1,2|-1)}\widetilde K_2+P_{s13}^{(1,2|-1)}\widetilde K_3),
 }
where
 \eq{
 P_{s11}^{(1,2|-1)}=Z_{kl} (-2 z_{kl}^5 Z_{kl}+32 z_{kl}^4+64 z_{kl}^3 Z_{kl}
 +512 z_{kl}^2+256 z_{kl} Z_{kl}+512 Z_{kl}^2),
 }
 \begin{eqnarray}
  \nonumber P_{s12}^{(1,2|-1)}&=&z_{kl}^5 Z_{kl} (Z_{kl}^2-16)-16 z_{kl}^4 (Z_{kl}^2+8)
  -32 z_{kl}^3 Z_{kl} (Z_{kl}^2+8)\\
  &-&256 z_{kl}^2 (Z_{kl}^2+8)-128 z_{kl} Z_{kl} (Z_{kl}^2+8)+128 Z_{kl}^2 (Z_{kl}^2-16),
 \end{eqnarray}
 \eq{
 P_{s13}^{(1,2|-1)}=z_{kl}^3 Z_{kl}^5 (-z_{kl}^2 (Z_{kl}^2-32)+16 z_{kl} Z_{kl}+32 Z_{kl}^2),
 }
and
 \eq{
 [\tau,\tau^2]_{kl}=[\tau^2,\tau]_{kl}=\frac{\sigma^{cl}_{kl}}{\sigma(T)}\frac{\pi }{64 z_k^2 z_l^2}
 (P_{s21}^{(1,2|-1)}\widetilde K_1+P_{s22}^{(1,2|-1)}\widetilde K_2+P_{s23}^{(1,2|-1)}\widetilde K_3),
 }
where
 \eq{
 P_{s21}^{(1,2|-1)}=2 Z_{kl} (z_{kl}^5 Z_{kl}+16 z_{kl}^4-32 z_{kl}^3 Z_{kl}
 +256 z_{kl}^2-128 z_{kl} Z_{kl}+256 Z_{kl}^2),
 }
 \begin{eqnarray}
  \nonumber P_{s22}^{(1,2|-1)}&=&z_{kl}^5 Z_{kl} (Z_{kl}^2-16)-16 z_{kl}^4 (Z_{kl}^2+8)
  +32 z_{kl}^3 Z_{kl} (Z_{kl}^2+8)\\
  &-&256 z_{kl}^2 (Z_{kl}^2+8)+128 z_{kl} Z_{kl}(Z_{kl}^2+8)+128 Z_{kl}^2 (Z_{kl}^2-16),
 \end{eqnarray}
 \eq{
 P_{s23}^{(1,2|-1)}=z_{kl}^3 Z_{kl}^5 (z_{kl}^2 (Z_{kl}^2-32)+16 z_{kl} Z_{kl}-32 Z_{kl}^2),
 }
and
 \eq{
 [\tau^2,\tau_1^2]_{kl}=-\frac{\sigma^{cl}_{kl}}{\sigma(T)}\frac{\pi }{3840 z_k^2 z_l^2}
 (P_{s11}^{(2,2|-1)}\widetilde K_1+P_{s12}^{(2,2|-1)}\widetilde K_2+P_{s13}^{(2,2|-1)}\widetilde K_3),
 }
where
 \begin{eqnarray}
  \nonumber P_{s11}^{(2,2|-1)}&=&2 Z_{kl} [z_{kl}^6 (7 Z_{kl}^2-76)-60 z_{kl}^4 (5 Z_{kl}^2-32)\\
  &-&11520 z_{kl}^2 (Z_{kl}^2-6)+3840 Z_{kl}^2 (Z_{kl}^2+18)],
 \end{eqnarray}
 \begin{eqnarray}
  \nonumber P_{s12}^{(2,2|-1)}&=&z_{kl}^6 (-7 Z_{kl}^4+188 Z_{kl}^2+320)+60 z_{kl}^4 (5 Z_{kl}^4+80 Z_{kl}^2-256)\\
  &+&23040 z_{kl}^2 (Z_{kl}^2-24)+7680 Z_{kl}^2 (7 Z_{kl}^2-72),
 \end{eqnarray}
 \eq{
 P_{s13}^{(2,2|-1)}=z_{kl}^4 Z_{kl}^6 (z_{kl}^2 (7 Z_{kl}^2-300)-300 Z_{kl}^2),
 }
and
 \eq{
 [\tau^2,\tau^2]_{kl}=\frac{\sigma^{cl}_{kl}}{\sigma(T)}\frac{\pi }{3840 z_k^2 z_l^2}
 (P_{s21}^{(2,2|-1)}\widetilde K_1+P_{s22}^{(2,2|-1)}\widetilde K_2+P_{s23}^{(2,2|-1)}\widetilde K_3),
 }
where
 \begin{eqnarray}
  \nonumber P_{s21}^{(2,2|-1)}&=&2 Z_{kl} [-z_{kl}^6 (7 Z_{kl}^2-76)-360 z_{kl}^5 Z_{kl}
  +60 z_{kl}^4 (5 Z_{kl}^2+64)-11520 z_{kl}^3 Z_{kl}\\
  &-&3840 z_{kl}^2 (Z_{kl}^2-22)+7680 z_{kl} Z_{kl} (Z_{kl}^2-10)+3840 Z_{kl}^2 (Z_{kl}^2+22)],
 \end{eqnarray}
 \begin{eqnarray}
  \nonumber P_{s22}^{(2,2|-1)}&=&z_{kl}^6 (7 Z_{kl}^4-188 Z_{kl}^2-320)+120 z_{kl}^5 Z_{kl}(3 Z_{kl}^2+16)\\
  \nonumber &-&60 z_{kl}^4 (5 Z_{kl}^4+48 Z_{kl}^2+512)-3840 z_{kl}^3 Z_{kl} (Z_{kl}^2-24)\\
  &-&7680 z_{kl}^2 (7 Z_{kl}^2+88)+15360 z_{kl} Z_{kl} (Z_{kl}^2+40)+7680 Z_{kl}^2 (7 Z_{kl}^2-88),
 \end{eqnarray}
 \begin{eqnarray}
  \nonumber P_{s23}^{(2,2|-1)}&=&z_{kl}^3 Z_{kl}^5 [z_{kl}^3 (300 Z_{kl}-7 Z_{kl}^3)
  -120 z_{kl}^2 (3 Z_{kl}^2-32)\\
  &+&60 z_{kl} Z_{kl} (5 Z_{kl}^2-32)+3840 Z_{kl}^2].
 \end{eqnarray}
And for the tensor collision brackets one has:
 \eq{
 [\overset{\circ}{\overline{\pi^{\mu} \pi^{\nu}}},
 \overset{\circ}{\overline{\pi_{1\mu} \pi_{1\nu}}}]_{kl}=
 -\frac{\sigma^{cl}_{kl}}{\sigma(T)}\frac{\pi }{5760 z_k^2 z_l^2}(P_{T11}^{(0,0|-1)}\widetilde K_1
 +P_{T12}^{(0,0|-1)}\widetilde K_2+P_{T13}^{(0,0|-1)}\widetilde K_3),
 }
where
 \begin{eqnarray}
  \nonumber P_{T11}^{(0,0|-1)}&=&2 Z_{kl} [z_{kl}^6 (7 Z_{kl}^2-76)-60 z_{kl}^4 (5 Z_{kl}^2-32)
  -11520 z_{kl}^2 (Z_{kl}^2-6)\\
  &+&3840 Z_{kl}^2 (Z_{kl}^2+18)],
 \end{eqnarray}
 \begin{eqnarray}
  \nonumber P_{T12}^{(0,0|-1)}&=&z_{kl}^6 (-7 Z_{kl}^4+188 Z_{kl}^2+320)+60 z_{kl}^4 (5 Z_{kl}^4
  +80 Z_{kl}^2-256)\\
  &+&23040 z_{kl}^2 (Z_{kl}^2-24)+7680 Z_{kl}^2 (7 Z_{kl}^2-72),
 \end{eqnarray}
 \eq{
 P_{T13}^{(0,0|-1)}=z_{kl}^4 Z_{kl}^6 (z_{kl}^2 (7 Z_{kl}^2-300)-300 Z_{kl}^2),
 }
and
 \eq{
 [\overset{\circ}{\overline{\pi^{\mu} \pi^{\nu}}},
 \overset{\circ}{\overline{\pi_{\mu} \pi_{\nu}}}]_{kl}=
 \frac{\sigma^{cl}_{kl}}{\sigma(T)}\frac{\pi }{5760 z_k^2 z_l^2}(P_{T21}^{(0,0|-1)}\widetilde K_1
 +P_{T22}^{(0,0|-1)}\widetilde K_2+P_{T23}^{(0,0|-1)}\widetilde K_3),
 }
where
 \begin{eqnarray}
  \nonumber P_{T21}^{(0,0|-1)}&=&2 Z_{kl} [-z_{kl}^6 (7 Z_{kl}^2-76)-1800 z_{kl}^5 Z_{kl}
  +60 z_{kl}^4 (5 Z_{kl}^2-32)\\
  &-&3840 z_{kl}^2 (7 Z_{kl}^2-22)+19200 z_{kl} Z_{kl} (Z_{kl}^2-4)+7680 Z_{kl}^2 (2 Z_{kl}^2+11)],
 \end{eqnarray}
 \begin{eqnarray}
  \nonumber P_{T22}^{(0,0|-1)}&=&z_{kl}^6 (7 Z_{kl}^4-188 Z_{kl}^2-320)
  +600 z_{kl}^5 Z_{kl} (3 Z_{kl}^2-16)\\
  \nonumber &-&60 z_{kl}^4 (5 Z_{kl}^4-240 Z_{kl}^2-256)-38400 z_{kl}^3 Z_{kl}^3
  +7680 z_{kl}^2 (17 Z_{kl}^2-88)\\
  &-&76800 z_{kl} Z_{kl} (Z_{kl}^2-8)+7680 Z_{kl}^2 (13 Z_{kl}^2-88),
 \end{eqnarray}
 \begin{eqnarray}
  \nonumber P_{T23}^{(0,0|-1)}&=&z_{kl}^3 Z_{kl}^5 [z_{kl}^3 (300 Z_{kl}-7 Z_{kl}^3)
  -600 z_{kl}^2 (3 Z_{kl}^2-64)\\
  &+&300 z_{kl} Z_{kl} (Z_{kl}^2-64)+38400 Z_{kl}^2].
 \end{eqnarray}
Also averaged cross sections with the considered energy
dependencies can be written down easily. Below there are only
averaged energy insertions:
 \eq{
  \langle\upsilon^2/Z_{kl}^2\rangle = Z_{kl}^{-2}
  \frac{J_{kl}^{(1,0,0,0,0|0,0)}}{J_{kl}^{(0,0,0,0,0|0,0)}}
  =\frac{Z_{kl} ((-z_{kl}^2+Z_{kl}^2+24) K_3(Z_{kl})+4 Z_{kl} K_2(Z_{kl}))}{Z_{kl}
  (Z_{kl}^2-z_{kl}^2) K_3(Z_{kl})+4 z_{kl}^2 K_2(Z_{kl})},
 }
 \begin{eqnarray}
  \nonumber \langle Z_{kl}^2/\upsilon^2 \rangle &=& Z_{kl}^2
  \frac{J_{kl}^{(-1,0,0,0,0|0,0)}}{J_{kl}^{(0,0,0,0,0|0,0)}}\\
  &=&\frac{12 z_{kl}^2 Z_{kl}^4 G(Z_{kl})+Z_{kl} (Z_{kl}^2-z_{kl}^2) K_3(Z_{kl})
  -4 (z_{kl}^2+Z_{kl}^2) K_2(Z_{kl})}{Z_{kl} (Z_{kl}^2-z_{kl}^2) K_3(Z_{kl})+4 z_{kl}^2 K_2(Z_{kl})}.
 \end{eqnarray}
Using these averaged insertions one could reproduce correct
asymptotic temperature dependencies of the viscosities however
with a somewhat different coefficient in the high-temperature
limit.

For complicated forms of cross sections one can calculate
numerically the 1- or 2-dimensional integrals (corresponding to
the unknown energy and angular dependencies). As long as
implemented UrQMD cross sections don't depend on the scattering
angle (being treated as averaged ones over it) one has to
calculate only 1-dimensional integrals numerically. For this
purpose the $J$ integrals can be represented in the form
 \begin{eqnarray}\label{Jint2}
  \nonumber &~&J_{kl}^{(a,b,d,e,f|q,r)}=\frac{\pi(d+e+1)!!}{z_k^2z_l^2K_2(z_k)K_2(z_l)}
  \sum_{h=0}^{[b/2]}(-1)^h(2h-1)!! \binom{b}{2h} Z_{kl}^{2 (a + f) + b + (d + e)/2 - h + 5} \\
  \nonumber &~&\times \int_1^\infty du \sigma^{(d,e,f)}_{1kl}(Z_{kl}u) (u^2 + z_{kl}/Z_{kl})^q
  (u^2 - z_{kl}/Z_{kl})^r u^{2 (a - f - q - r) + b - h - 3(d + e)/2}\\
  &~&\times (u^2 - 1)^{(d + e)/2 + f + 1} (u^2 - z_{kl}^2/Z_{kl}^2)^{(d + e)/2 + f + 1}
  K_{(d + e)/2 + b - h + 1}(Z_{kl}u),
 \end{eqnarray}
where the $\sigma_{1kl}^{(d,e,f)}(Z_{kl}u)$ is generalized to
 \eq{
 \sigma_{1kl}^{(d,e,f)}(Z_{kl}u)=\frac{\sigma^{cl}_{kl}(Z_{kl}u)}{\sigma(T)}
 \sum_{g=0}^{\min(d,e)} \sigma^{(f,g)} K(d,e,g), \quad
 \sigma^{cl}_{kl}(Z_{kl}u)\equiv \gamma_{kl}\sigma_{kl}(Z_{kl}u).
 }
And the $J'$ integrals can be represented in the form
 \begin{eqnarray}\label{Jpint2}
  \nonumber &~&{J'}_{kl}^{(a,b,d,e,f|q,r)}=\frac{\pi(d+e-1)!!}{z_k^2z_l^2K_2(z_k)K_2(z_l)}
  \sum_{h=0}^{[b/2]}(-1)^h(2h-1)!! \binom{b}{2h} Z_{kl}^{2 (a + f) + b + (d + e)/2 - h + 5} \\
  \nonumber &~&\times \int_1^\infty du \sigma^{(d,e,f)}_{kl}(Z_{kl}u) (u^2 + z_{kl}/Z_{kl})^q
  (u^2 - z_{kl}/Z_{kl})^r u^{2 (a - f - q - r) + b - h - 3(d + e)/2}\\
  &~&\times (u^2 - 1)^{(d + e)/2 + f + 1} (u^2 - z_{kl}^2/Z_{kl}^2)^{(d + e)/2 + f + 1}
  K_{(d + e)/2 + b - h + 1}(Z_{kl}u),
 \end{eqnarray}
where the $\sigma^{(d,e,f)}_{kl}(Z_{kl}u)$ is generalized to
 \eq{\label{sigmadef2}
 \sigma^{(d,e,f)}_{kl}(Z_{kl}u)=\frac{\sigma^{cl}_{kl}(Z_{kl}u)}{\sigma(T)}\left(1-(d+e+1)
 \sum_{g=0}^{\min(d,e)}K(d,e,g)\sigma^{(f,g)}\right).
 }
To calculate collision brackets faster one can bring all
integrated expressions under one integral and simplify the
integrand.

\section{ The collision rates and the mean free paths \label{appmfp}}

The quantity $\frac{W_{k'l'}} {p_{k'}^0 p_{1l'}^0 {p'}_{k'}^0
{p'}_{1l'}^0} d^3{p'}_{k'} d^3{p'}_{1l'}$, which enters in the
elastic collision integral (\ref{ckelgroot}), represents the
probability of scattering per unit time times unit volume for two
particles which had momentums $\vec p_{k'}$ and $\vec p_{1l'}$
before scattering and momentums in the ranges ${(\vec {p'}_{k'},
\vec {p'}_{k'} + d\vec {p'}_{k'})}$ and ${(\vec {p'}_{1l'}, \vec
{p'}_{1l'} + d\vec {p'}_{1l'})}$ after the scattering. The
quantity $ g_{k'} \frac{d^3p_{k'}}{ (2\pi)^3 } f_{k'}$ represents
the number of particles per unit volume, which have momentums in
the range ${(\vec p_{k'}, \vec p_{k'} + d\vec p_{k'})}$. The
number of collisions of particles of the $k'$-th species with
particles of the $l'$-th species per unit time per unit volume is
then\footnote{It represents some sum over all possible collisions.
In the case of the same species one factor $\gamma_{k'l'}$ just
cancels the double counting in momentum states after scattering
and another factor $\gamma_{k'l'}$ also reflects the fact that
scattering takes place for ${{n_{k'}\choose 2} \approx
\frac12n_{k'}^2}$ pairs of undistinguishable particles in a given
unit volume.}
 \eq{\label{totratekl}
 \widetilde R^{el}_{k'l'}\equiv g_{k'}g_{l'}\frac{\gamma_{k'l'}^2}{(2\pi)^6}\int
 \frac{d^3p_{k'}}{p_{k'}^0}\frac{d^3p_{1l'}}{p_{1l'}^0}\frac{d^3p'_{k'}}{{p'}_{k'}^0}
 \frac{d^3p'_{1l'}}{{p'}_{1l'}^0}f_{k'}^{(0)}f_{1l'}^{(0)}W_{k'l'}.
 }
To get the corresponding number of collisions of particles of the
$k'$-th species with particles of the $l'$-th species per unit
time \emph{per particle of the $k'$-th species}, $R^{el}_{k'l'}$,
one has to divide the (\ref{totratekl}) on the
$\gamma_{k'l'}n_{k'}$ (recall that ${n_{k'} \propto g_{k'}}$ by
definition), which is the number of particles of the $k'$-th
species per unit volume divided on the number of particles of the
$k'$-th species taking part in the given type of reaction (2 for
binary elastic collisions, if particles are identical, and 1
otherwise). This rate can be directly obtained averaging the
collision rate with fixed momentum $p_k$ of the $k$-th particle
species
 \eq{
 \mc R^{el}_{kl'}\equiv g_{l'}\gamma_{kl'}\int
 \frac{d^3p_{1l'}}{(2\pi)^3}d^3p'_{k}d^3p'_{1l'}f_{1l'}^{(0)}
 \frac{W_{kl'}}{p_{k}^0p_{1l'}^0{p'}_{k}^0{p'}_{1l'}^0},
 }
over the momentum with the probability distribution
$\frac{d^3p_{k}} {(2\pi)^3} \frac{f_{k}}{n_k}$ (and spin states
which is trivial):
 \eq{
 R^{el}_{k'l'}\equiv g_{k'}g_{l'}\frac{\gamma_{k'l'}}{(2\pi)^6n_{k'}}\int
 \frac{d^3p_{k'}}{p_{k'}^0} \frac{d^3p_{1l'}}{p_{1l'}^0}
 \frac{d^3p'_{k'}}{{p'}_{k'}^0}\frac{d^3p'_{1l'}}{{p'}_{1l'}^0}f_{k'}^{(0)}
 f_{1l'}^{(0)}W_{k'l'}=\frac{\widetilde R^{el}_{k'l'}}{\gamma_{k'l'}n_{k'}}.
 }
So that to get the mean rate of the elastic collisions per
particle of the $k'$-th species with all particles in the system
one can just integrate the sum of the gain terms in the collision
integral (\ref{ckelgroot}) over $\frac{d^3p_k}{(2\pi)^3p_k^0n_k}$
and average it over spin:
 \eq{
 R_{k'}^{el}\equiv \sum_{l'} R_{k'l'}^{el}.
 }
One can express the $\widetilde R_{k'l'}^{el}$ through the
$J_{kl}^{(0, 0, 0, 0, 0| 0, 0)}$ integrals from Appendix
\ref{appJ} as
 \eq{\label{Rklel}
 \widetilde R_{k'l'}^{el}=\gamma_{k'l'}\sigma(T)n_{k'} n_{l'}
 J_{k'l'}^{(0, 0, 0, 0, 0| 0, 0)}.
 }
For simplicity let's consider constant cross sections in what
follows. Then the (\ref{Rklel}) becomes
 \eq{
 \widetilde R_{k'l'}^{el}=g_{k'}g_{l'}\gamma_{k'l'}\frac{2\sigma^{cl}_{k'l'} T^6}{\pi^3}
  [(z_{k'}-z_{l'})^2 K_2(z_{k'}+z_{l'})+z_{k'} z_{l'}(z_{k'}+z_{l'})
  K_3(z_{k'}+z_{l'})],
 }
where $\sigma^{cl}_{k'l'}$ is the classical elastic differential
constant cross section of scattering of a particle of the $k'$-th
species on particles of the $l'$-th species. For the case of large
temperature or when both masses are small, ${z_{k'} \ll 1}$ and
${z_{l'} \ll 1}$, one has expansion
 \eq{
 \widetilde R_{k'l'}^{el}=g_{k'}g_{l'}\gamma_{k'l'}
 \frac{4 \sigma^{cl}_{k'l'} T^6}{\pi^3}
 \left(1-\frac14(z_{k'}^2+z_{l'}^2)+...\right).
 }
For the case of small temperature or when both masses are large,
${z_{k'}\gg 1}$ and ${z_{l'}\gg 1}$, one has expansion
 \eq{
 \widetilde R_{k'l'}^{el}=g_{k'}g_{l'}\gamma_{k'l'}\frac{\sqrt{2}\sigma^{cl}_{k'l'} T^6 z_{k'} z_{l'}
 \sqrt{z_{k'}+z_{l'}} e^{-z_{k'}-z_{l'}}}{\pi^{5/2}}\left(1+\frac{8 z_{k'}^2+19 z_{k'} z_{l'}
 +8 z_{l'}^2}{8 z_{k'} z_{l'} (z_{k'}+z_{l'})}+...\right).
 }
For the case when only one mass is large, ${z_{l'}\gg 1}$, one has
somewhat different expansion
 \eq{
 \widetilde R_{k'l'}^{el}=g_{k'}g_{l'}\gamma_{k'l'}\frac{\sqrt{2} \sigma^{cl}_{k'l'} T^6 (z_{k'}+1) z_{l'}^{3/2}
 e^{-z_{k'}-z_{l'}}}{\pi^{5/2}}\left(1+\frac{4 z_{k'}^2+15 z_{k'}+15}{8 z_{k'}+8}z_{l'}^{-1}...\right).
 }
The $\sigma(T) J_{k'l'}^{(0, 0, 0, 0, 0| 0, 0)}$ in the
(\ref{Rklel}) can be replaced in the limit of high temperatures
with $4\pi \sigma^{cl}_{k'l'} \langle \abs{\vec v_{k'}}\rangle$
and in the limit of low temperatures with $4\pi \sigma^{cl}_{k'l'}
\langle \abs{\vec v_{k'}}\rangle \sqrt{1+m_{k'}/m_{l'}}= 4\pi
\sigma^{cl}_{k'l'} \langle \abs{\vec v_{rel,k'l'}}\rangle$, where
$\langle \abs{\vec v_{k'}}\rangle$ is the mean modulus of
particle's velocity of the $k'$-th species,
 \eq{\label{avvel}
 \langle \abs{\vec v_{k'}}\rangle=\frac{\int d^3p_{k'} \frac{|\vec p_{k'}|}{p_{k'}^0}
 f^{(0)}_{k'}(p_{k'})}{\int d^3p_{k'} f^{(0)}_{k'}(p_{k'})}
 =\frac{2 e^{-z_{k'}}(1+z_{k'})}{z_{k'}^2K_2(z_{k'})}
 =\sqrt{\frac{8}{\pi z_{k'}}}\frac{K_{3/2}(z_{k'})}{K_{2}(z_{k'})},
 }
and $\langle \abs{\vec v_{rel,k'l'}}\rangle$ is the mean modulus
of the relative velocity, which coincides with the $\langle
\abs{\vec v_{k'}}\rangle$ for high temperatures. Then the
resultant collision rate $R^{el}_{k'}$ would reproduce simple
nonrelativistic collision rates know in the kinetic-molecular
theory. To get the (approximate) mean free time one has just to
invert the $R^{el}_{k'}$:
 \eq{
 t^{el}_{k'}=\frac{1}{R^{el}_{k'}}.
 }
The (approximate) mean free path $l^{el}_{k'}$ can be obtained
after multiplication of it on the $\langle \abs{\vec
v_{k'}}\rangle$:
 \eq{\label{lel}
 l^{el}_{k'}=\frac{\langle \abs{\vec v_{k'}}\rangle}{R^{el}_{k'}}.
 }
For the single-component gas one gets
 \eq{\label{mfpsc}
 l^{el}_{1'}=\frac{\langle \abs{\vec v_{1'}}\rangle}{R^{el}_{1'1'}}=
 \frac{\pi e^{-z_{1}} (z_{1}+1)}{g_{1}4 \sigma^{cl}_{11} T^3 z_{1}^3 K_3(2 z_{1})}.
 }
The nonrelativistic limit of the (\ref{mfpsc}) with the ${g_1=1}$
coincides with the same limit of the formula
 \eq{
 l^{el}_{1}=\frac{\langle \abs{\vec v_1}\rangle}{4\pi\sigma^{cl}_{11}n_1
 \langle \abs{\vec v_{rel}}\rangle}=\frac{1}{4\pi\sigma^{cl}_{11}n_1\sqrt2},
 }
which is the mean free path formula coming from the
nonrelativistic kinetic-molecular theory obtained by Maxwell. The
ultrarelativistic limit of the (\ref{mfpsc}) with the ${g_1=1}$
coincides with the same limit of the formula
 \eq{
 l_1^{el}=\frac{1}{4\pi\sigma^{cl}_{11}n_1}.
 }

Analogically one can introduce inelastic rates $R^{inel}_k$ of any
inelastic processes to occur for the $k'$-th particles species.
Then the mean free time $t^{inel}_{k'}$ in what any inelastic
process occurs for the particles of the $k'$-th species can be
introduced as
 \eq{\label{tkinel}
 t^{inel}_{k'}=\frac{1}{R^{inel}_{k'}}.
 }
The mean free path for the particles of the $k'$-th species is
obtained through the rate ${R^{el}_{k'} + R^{inel}_{k'}}$ and can
be written as
 \eq{\label{mfp}
 l_{k'}=\frac{\langle \abs{\vec v_{k'}}\rangle}{R^{el}_{k'}+R^{inel}_{k'}}.
 }

\end{document}